%% file: main.tex
\documentclass[letterpaper,12pt]{article}
\usepackage{subfiles}

\usepackage{tabularx}
\usepackage{amsmath}
\usepackage{amssymb}
\usepackage[margin=1in,letterpaper]{geometry}
\usepackage{cite}
\usepackage[final]{hyperref}
\hypersetup{
    colorlinks,
    linkcolor={red!50!black},
    citecolor={blue!50!black},
    urlcolor={blue!80!black}
}
\usepackage{float}
\usepackage{subfig}
\usepackage{tablefootnote}
\usepackage{graphicx}
\usepackage{epstopdf}
\usepackage{placeins}
\usepackage{graphicx}
\usepackage{gensymb}
\graphicspath{ {main.tex} }
\usepackage{mathtools}

\usepackage[utf8]{inputenc}
\usepackage{setspace}
\usepackage{braket}
\usepackage[font=footnotesize,skip=4pt]{caption}
\usepackage[nice]{nicefrac}
\usepackage{color,soul}
\usepackage[table]{xcolor}
\usepackage{upgreek}
\usepackage{longtable}
\usepackage[printonlyused,withpage]{acronym}
\usepackage{url}
\usepackage[pdf]{pstricks} 

\usepackage[ruled,vlined]{algorithm2e}
\usepackage{algpseudocode}
\usepackage{algorithmicx}

\usepackage[version=3]{mhchem}

\newcommand{\dd}[2]{\frac{\mathrm{d}#1}{\mathrm{d}#2}}

\newcommand{\yI}{[O^{T^+}_1]}
\newcommand{\yII}{[O^{T^+}_2]}
\newcommand{\yIII}{[O^{T^-}_1]}
\newcommand{\yIV}{[O^{T^-}_2]}
\newcommand{\yV}{[T^-O^{T^+}_1]}

\newcommand{\yVI}{[T^-O^{T^+}_2]}
\newcommand{\yVII}{[T^-O^{T^+}_1 O^{T^+}_2]}
\newcommand{\yVIII}{[T^+T^-]}
\newcommand{\yIX}{[T^+O^{T^-}_1]}
\newcommand{\yX}{[T^+O^{T^-}_2]}

\newcommand{\yXI}{[T^+O^{T^-}_1 O^{T^-}_2]}
\newcommand{\yXII}{[T^-]}
\newcommand{\yXIII}{[T^+]}
\newcommand{\yXIV}{[O^{T^+}_1 O^{T^-}_1]}
\newcommand{\yXV}{[O^{T^+}_2 O^{T^-}_2]}

\newcommand{\kplusoItminus}{k^+_{O^{T^-}_1}}
\newcommand{\kplusoIItminus}{k^+_{O^{T^-}_2}}
\newcommand{\kminusoItminus}{k^-_{O^{T^-}_1}}
\newcommand{\kminusoIItminus}{k^-_{O^{T^-}_2}}
\newcommand{\kplusoItplus}{k^+_{O^{T^+}_1}}
\newcommand{\kplusoIItplus}{k^+_{O^{T^+}_2}}
\newcommand{\kminusoItplus}{k^-_{O^{T^+}_1}}
\newcommand{\kminusoIItplus}{k^-_{O^{T^+}_2}}
\newcommand{\ktplus}{k^+_{T}}
\newcommand{\ktminus}{k^-_{T}}
\newcommand{\kl}{k_L}
\newcommand{\kplusoI}{k^+_{O_1^{T^\pm}}}
\newcommand{\kplusoII}{k^+_{O_2^{T^\pm}}}
\newcommand{\kminusoI}{k^-_{O_1^{T^\pm}}}
\newcommand{\kminusoII}{k^-_{O_2^{T^\pm}}}

\definecolor{r}{RGB}{232,178,178}
\definecolor{b}{RGB}{178,178,232}
\definecolor{p}{RGB}{162,124,178}
\definecolor{w}{RGB}{255,255,255}
\definecolor{g}{RGB}{100,100,100}

\usepackage{subfiles}

\begin{document}

\onehalfspacing
\linespread{1.25}

\input{title}

\newpage

\tableofcontents
\newpage

\newpage

\input{1-Introduction}

\input{2-ProtocellDesign}

\input{3-ChargeTransfer}

\input{4-Replication}

\input{5-ConnectingMetabolism}

\input{6-Discussion}

\input{7-Conclusion}

\section*{Acknowledgements}
Part of the computation done for this project was performed on the UCloud interactive HPC system, which is managed by the eScience Center at the University of Southern Denmark. 
KRT and SR were in part supported by the Donostia International Physics Center (DIPC), San Sebastian, Spain. 
AK was partly supported by the
European Union’s Horizon 2020 research and innovation programme under the Marie Skłodowska-Curie Grant Agreement
No. 101068029, and by Grant 62828 from the John Templeton
Foundation. The opinions expressed in this publication are those of the author(s) and do not necessarily
reflect the views of the John Templeton Foundation.


\bibliographystyle{unsrt}
\bibliography{bibliography.bib}

\end{document}

%% file: title.tex
\begin{titlepage}
\begin{center}
\renewcommand*{\thefootnote}{\fnsymbol{footnote}}

\LARGE


\textbf{Metabolism, information, and viability in a simulated physically-plausible protocell}

\vspace{1 cm}
\small
By: \textbf{Kristoffer R. Thomsen$^{1}$\footnote{krt.workmail@gmail.com}, Artemy Kolchinsky$^{2,3}$\footnote{artemyk@gmail.com}, Steen Rasmussen$^{1,4,5}$}\footnote{steensantafe@gmail.com}\\
$^{1}$ University of Southern Denmark, 5230 Odense, Denmark\\
$^{2}$ICREA-Complex Systems Lab, Universitat Pompeu Fabra, 08003 Barcelona, Spain\\
$^{3}$ Universal Biology Institute, The University of Tokyo, 7-3-1 Hongo, Bunkyo-ku, Tokyo 113-0033, Japan\\
$^{4}$ Santa Fe Institute, New Mexico 87501, USA\\
$^{5}$ European Centre for Living Technology, 30123 Venice, Italy.\\






\input{0-Abstract}

\end{center}

\renewcommand*{\thefootnote}{\arabic{footnote}}
\setcounter{footnote}{0}
\end{titlepage}

%% file: 0-Abstract.tex
\begin{abstract}
Critical experimental design issues connecting energy transduction and inheritable information within a protocell are explored and elucidated. The protocell design utilizes a photo-driven energy transducer (a ruthenium complex) to turn resource molecules into building blocks, in a manner that is modulated by a combinatorial DNA-based co-factor.
This co-factor molecule serves as part of an electron relay for the energy transduction mechanism, where the charge-transfer rates depend on the sequence that contains an oxo-guanine. The co-factor also acts as a store of inheritable information due to its ability to replicate non-enzymatically through template-directed ligation. Together, the energy transducer and the co-factor act as a metabolic catalyst that 
produces co-factor DNA building blocks as well as fatty acids (from picolinium ester and modified DNA oligomers), where the fatty acids self-assemble into vesicles on which exterior surface both the co-factor (DNA) and the energy transducer are anchored with hydrophobic tails. Here we use simulations to study how the co-factor sequence determines its `fitness' as reflected by charge transfer and replication rates. To estimate the impact on the protocell, we compare these rates with previously measured metabolic rates from a similar system where the charge transfer is directly between the ruthenium complex and the oxo-guanine (without DNA replication and charge transfer). 
Replication and charge transfer turn out to have different and often opposing sequence requirements. Functional information of the co-factor molecules is used to probe the feasibility of randomly picking co-factor sequences from a limited population of co-factors molecules, where a good co-factor can enhance both metabolic biomass production and its own replication rate.  

\end{abstract}

%% file: 1-Introduction.tex
\section{Introduction}\label{sec:intro}

In this work, we use simulations to explore simple molecular mechanisms that, when combined, can function as protocellular metabolisms and inheritable information systems. Both of these functionalities are critical to create `living' materials from `nonliving' materials. We may define a system to be `alive' \cite{rasmussen2008protocells} if it can use free energy to metabolize resources into building blocks, such that the system can grow and divide to restart its `life-cycle'. Furthermore, if metabolism is regulated in part by inheritable information that can change between generations, then selection and simple evolution is possible. In this work, we define a protocell as a minimal physicochemical system that satisfies this operational definition, and we determine under which conditions such a protocell could be `viable' with a combined metabolic-informational system. 

Our bottom-up protocell approach builds on top of a diversity of related designs and methods that have steadily narrowed the gap between nonliving and living matter, e.g., see team efforts by Szostak et al.~(2016)\cite{adamala2016collaboration}, Yomo et al.~(2016)\cite{furubayashi2016packet}, Cronin et al.~(2017)\cite{kitson20163d}, Mann et al.~(2017)\cite{qiao2017predatory,rodriguez2017phagocytosis}, Rasmussen et al.~(2016)\cite{imai2022vesicles}, Adamala et al.~(2022)\cite{sato2022expanding}, the Schwille et al. and the MaxSynBio consortium (2018)\cite{schwille2018maxsynbio,jahnke2019programmable}, and perhaps the most ambitious, the ongoing Dutch Synthetic Cell project (2017)\cite{zwart2019primal,pols2019synthetic} headed by Dogterom. These bottom-up approaches are complementary to top-down approaches that modify existing living cells, as well as synthetic biology approaches that employ the biocatalytic machinery of contemporary life, e.g. see Venter team efforts (2006-now)~\cite{glass2006essential}.

Many experimental, theoretical and simulation studies in the literature have addressed challenges in the development of bottom-up protocells, including issues regarding container growth and division~\cite{villani2014growth,shirt2015emergent}, container functionalities~\cite{sato2022expanding}, formation of autocatalytic metabolic sets~\cite{farmer1986autocatalytic,vasas2012evolution,tanaka2014structure}, information replication~\cite{adamala2013nonenzymatic,sievers1998self,lincoln2009self}, thermodynamics of replication~\cite{fellermann2015non,fellermann2016toward,corominas2019thermodynamics,england2013statistical,kolchinsky2024thermodynamic}, protocell integration~\cite{imai2022vesicles,adamala2013nonenzymatic,rasmussen2003bridging}, inter-protocell interactions~\cite{qiao2017predatory,rodriguez2017phagocytosis} as well as coupling and synchronization of growth among the different protocell processes~\cite{carletti2008sufficient,rocheleau2007emergence}. A comparative discussion of many of these different approaches can be found in Rasmussen et al. (2008) \cite{rasmussen2008roadmap}. 

In all bottom-up approaches, the greatest challenge is to integrate metabolism, information, and container into a functional and autonomous whole that can survive in some environment. Therefore, examining and understanding how novel functionalities emerge in complex physicochemical systems is crucial. This paper reports a continuation of ongoing work exploring how to assemble fully autonomous protocells in the lab with a metabolism design that contains an integrated energy transducer and a combinatorial co-factor \cite{rasmussen2003bridging, rasmussen2004transitions, fellermann2007life, rocheleau2007emergence, declue2009nucleobase, maurer2011interactions, cape2012phototriggered, rasmussen2016generating, engelhardt2020thermodynamics, bornebusch2021reaction, thomsen2022energy}.

In Section 2, we present and discuss the key features of the proposed protocell design, as well as the functionalities essential for the following analysis. In Sections 3 and 4, we use simulations, which are motivated both by empirical observations and first-principles theory, to explore co-factor charge transfer and replication, which is the main focus of this paper. We conduct a simulation analysis of these two processes as they are not yet implemented in the laboratory; they are the two `missing links' in a full laboratory implementation of our protocell design \cite{rasmussen2016generating}. The simulations in Sections 3 and 4 may therefore be seen as preparations for future experiments. In Section 5, we compare and contrast the results from Sections 3 and 4, to rank the co-factor sequence efficiency for charge transfer and replication together with a combination of the two processes. Finally, we apply functional information \cite{szostak2003functional,hazen2007functional} to our findings to get a sense of how likely it is to pick a `good' co-factor at random from a limited DNA sequence ensemble. Section 6 critically discusses our findings, and in Section 7 we draw the conclusions from our study.

%% file: 2-ProtocellDesign.tex
\section{Protocell functionalities and design} \label{sec:Flint_design}
\begin{figure}[H]
\centering
\includegraphics[width=0.4\textwidth]{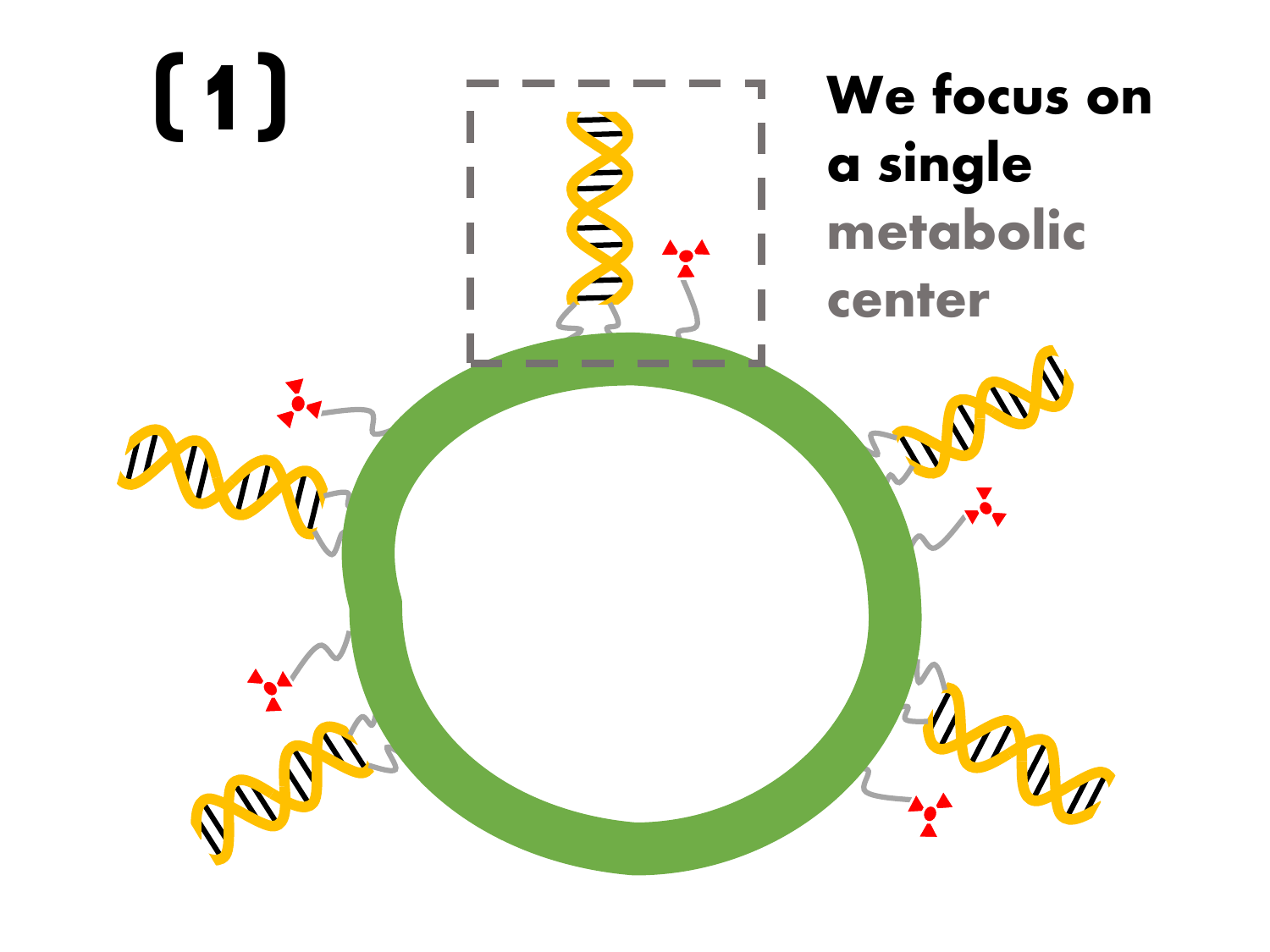}
\includegraphics[width=0.55\textwidth]{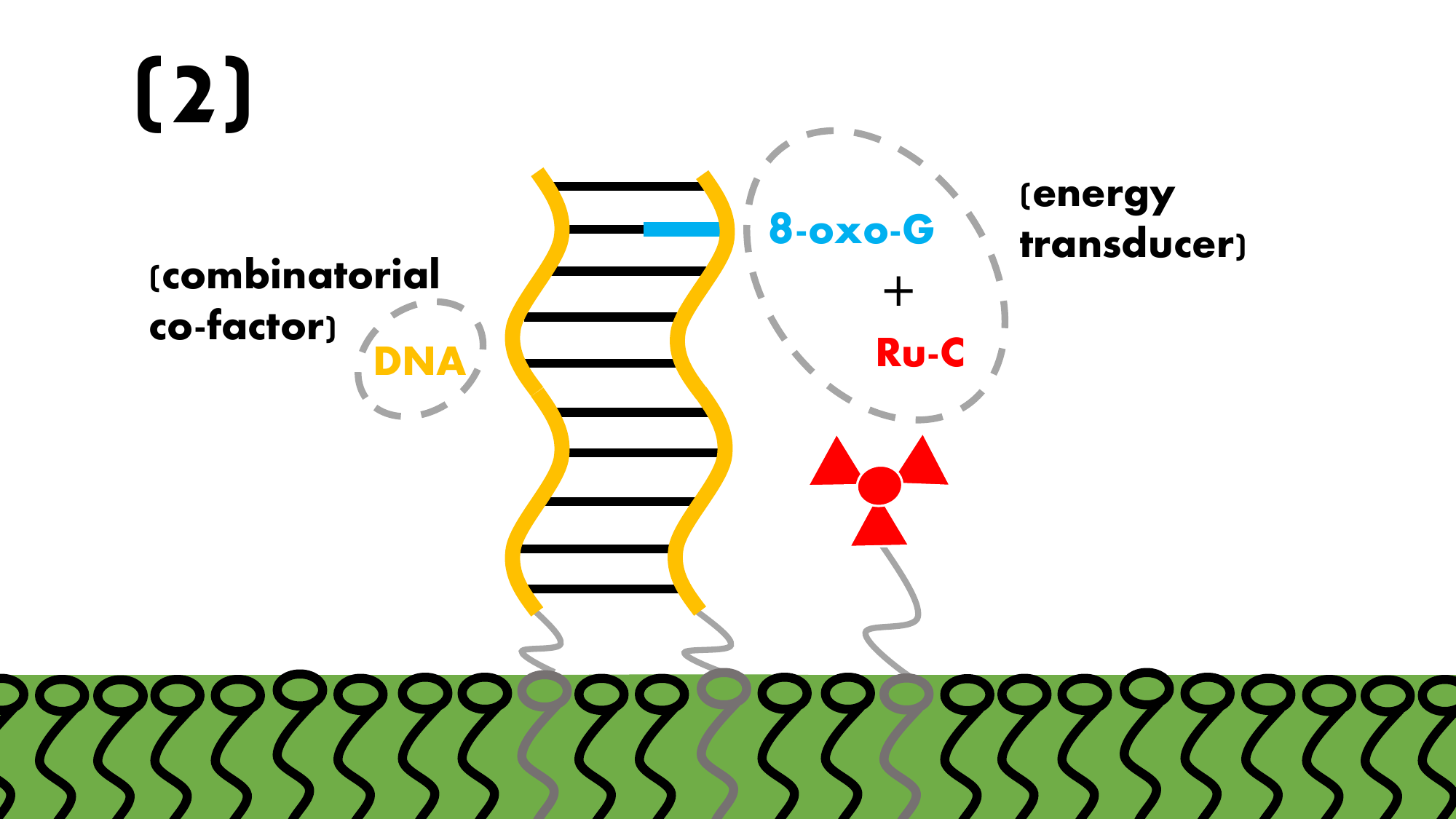}
\caption{Panel (1) shows a full protocell. Panel (2) depicts the metabolic system for the protocell that is composed of a combined energy transducer (ruthenium complex) and a co-factor (DNA) that modulates the energy transduction process. The overall metabolic efficiency depends on the combinatorial properties of the DNA co-factor; different nucleobase composition and sequences alters the efficiency.}
\label{fig:metabolism}
\end{figure}

The protocell is built around a metabolism that is directly coupled to an informational system and a container, all placed in an environment with appropriate access to resources and free energy. The design is ``systemic'' in the sense that the metabolic, information and container components are designed to mutually support each other \cite{rasmussen2003bridging}.
The metabolic system utilizes light driven ruthenium tris(bipyridine) [Ru(bpy)$_3$] complexes (further referred  to as Ru-C) as energy transducers and DNA as co-factors. Both the energy transducer and co-factor are tethered with hydrophopbic anchors to the exterior surface of the fatty acid vesicles, which act as 2D containers; see \autoref{fig:metabolism}. 

\begin{figure}[H]
\includegraphics[width=0.5\textwidth]{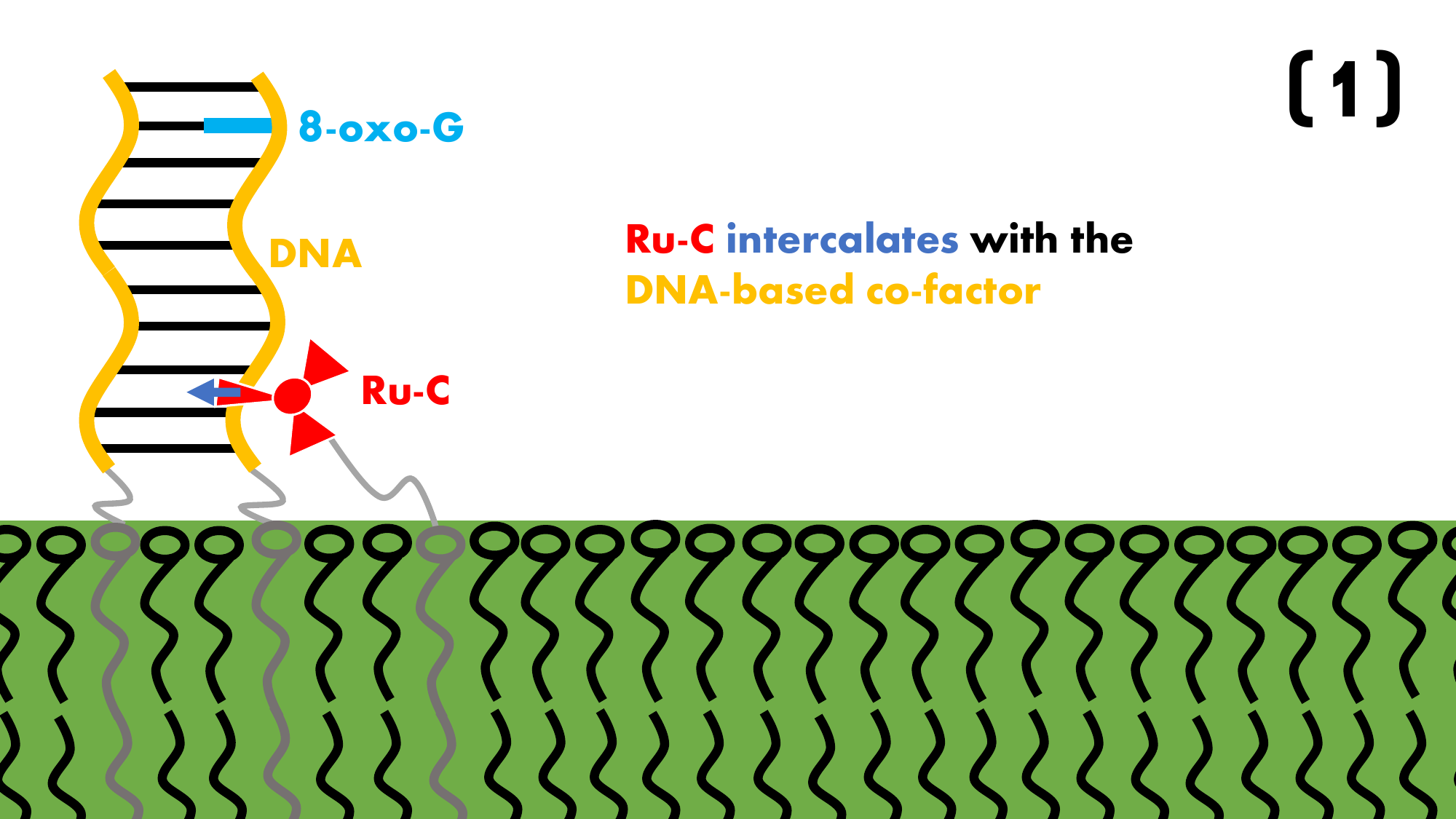}
\includegraphics[width=0.5\textwidth]{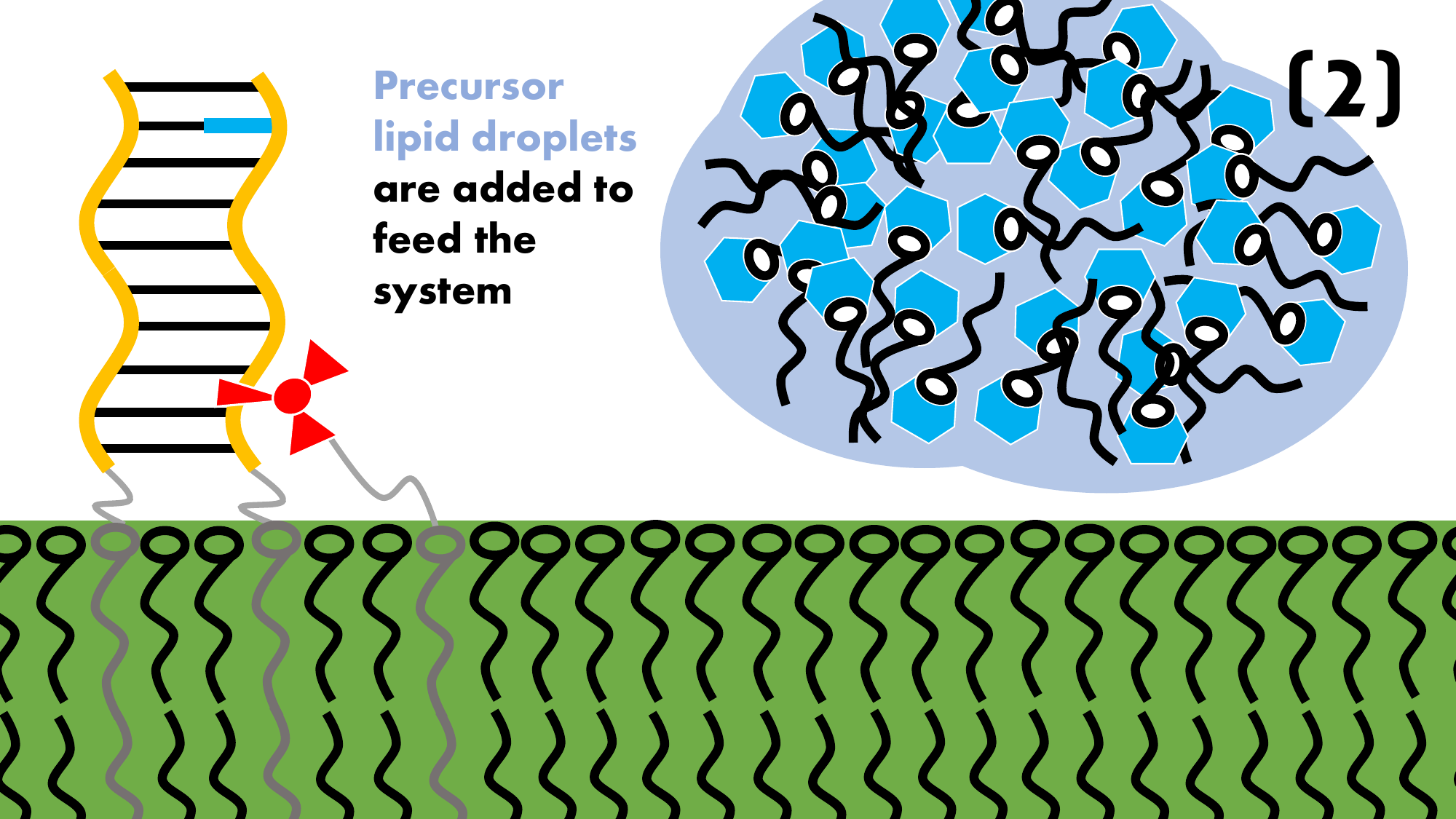}
\includegraphics[width=0.5\textwidth]{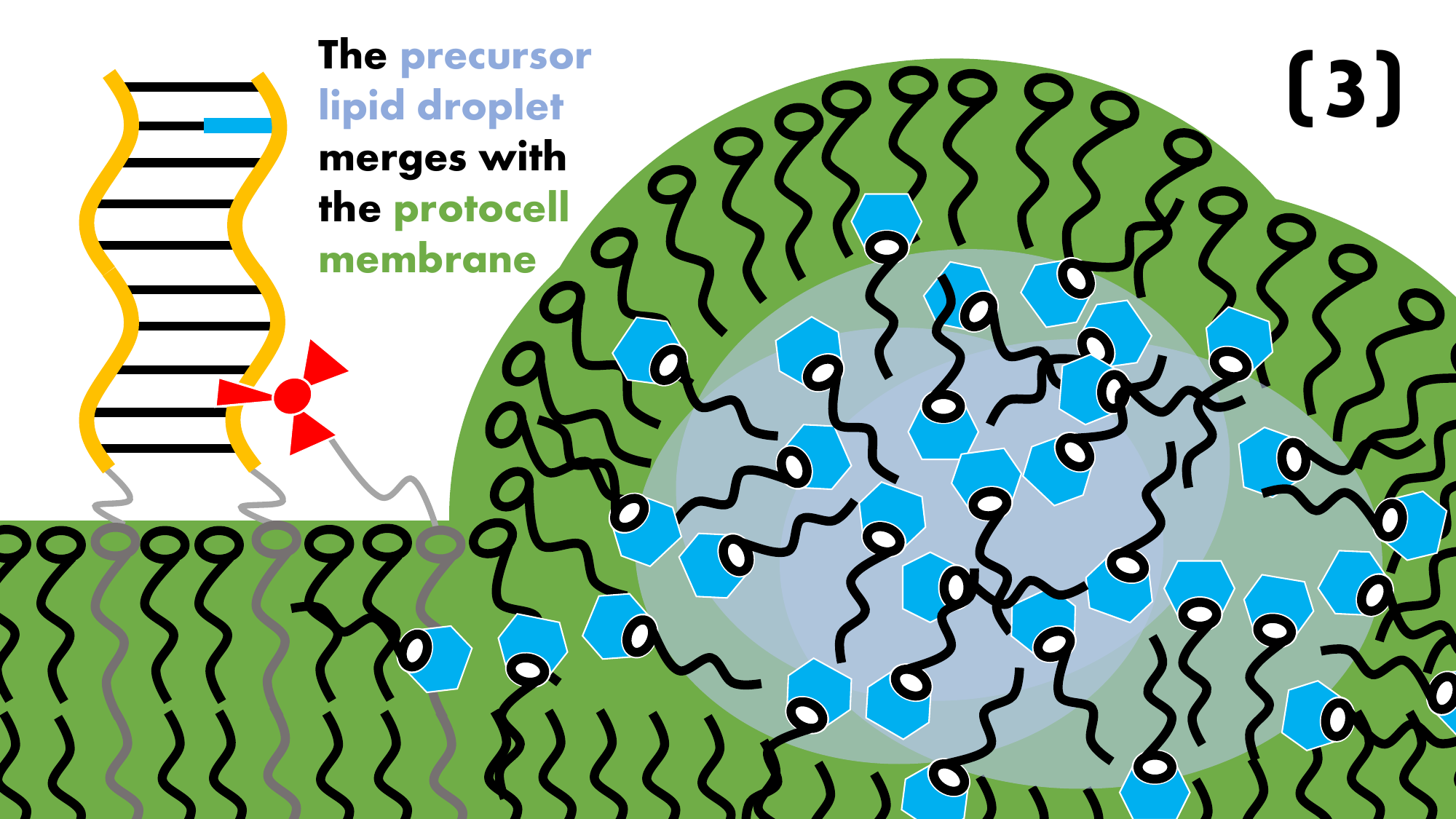}
\includegraphics[width=0.5\textwidth]{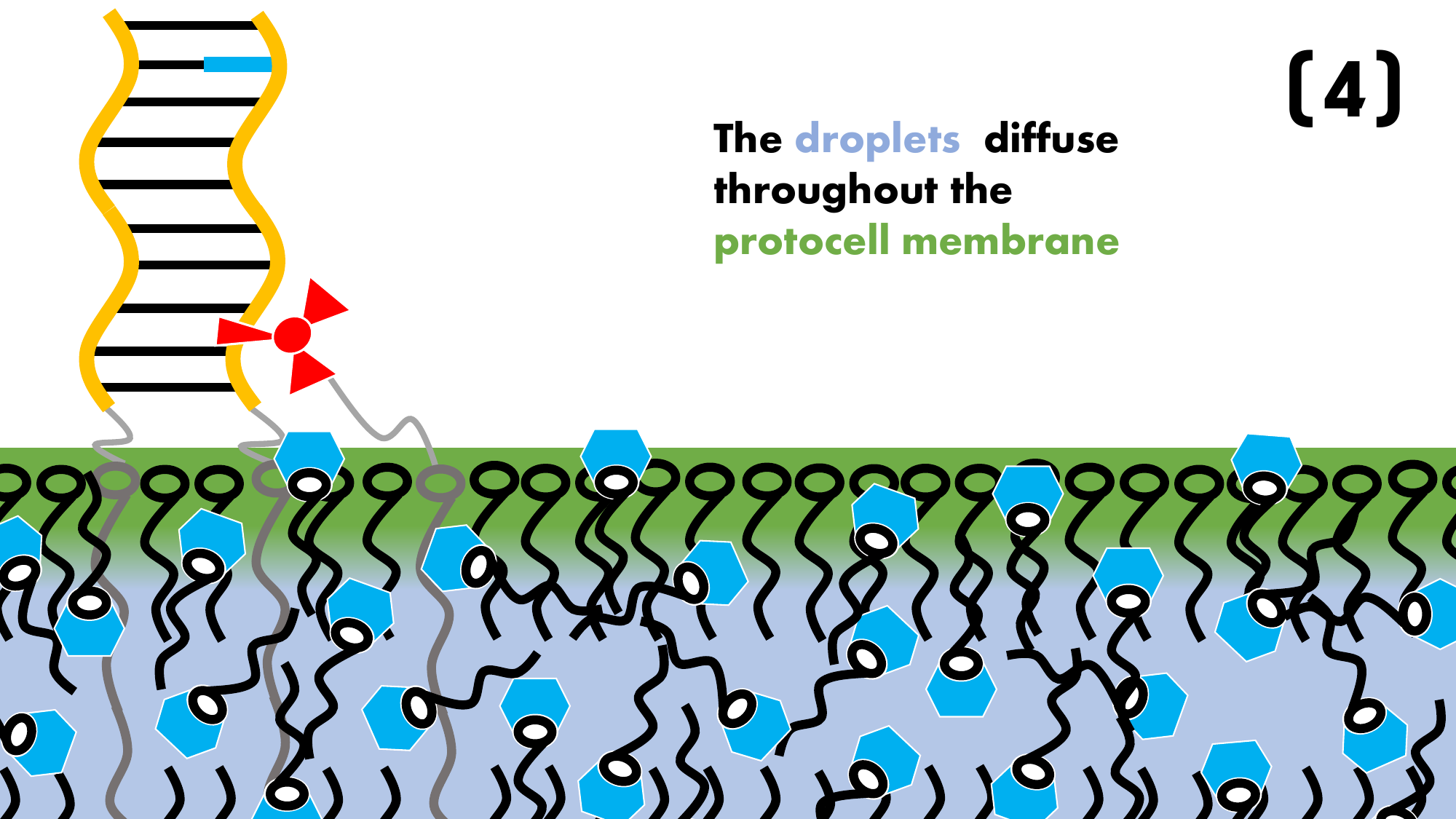}
\caption{Membrane feeding: Panel (1) depicts the initial condition as discussed in \autoref{fig:metabolism}, where the ruthenium complex is intercalated into the DNA duplex. Panel (2) shows that hydrophobic picolinium ester droplets are fed to the portocells via the environment. Due to their hydrophobisity, these droplets are incorporated into the fatty acid membrane where they initially tend to form globular membrane structures, see panel (3) \cite{maurer2011interactions}. Over time, some of the picolinium ester molecules tend to spread and organize within the membrane, Panel (4).}
\label{fig:membrane_feeding}
\end{figure}

\begin{figure}[H]
\includegraphics[width=0.5\textwidth]{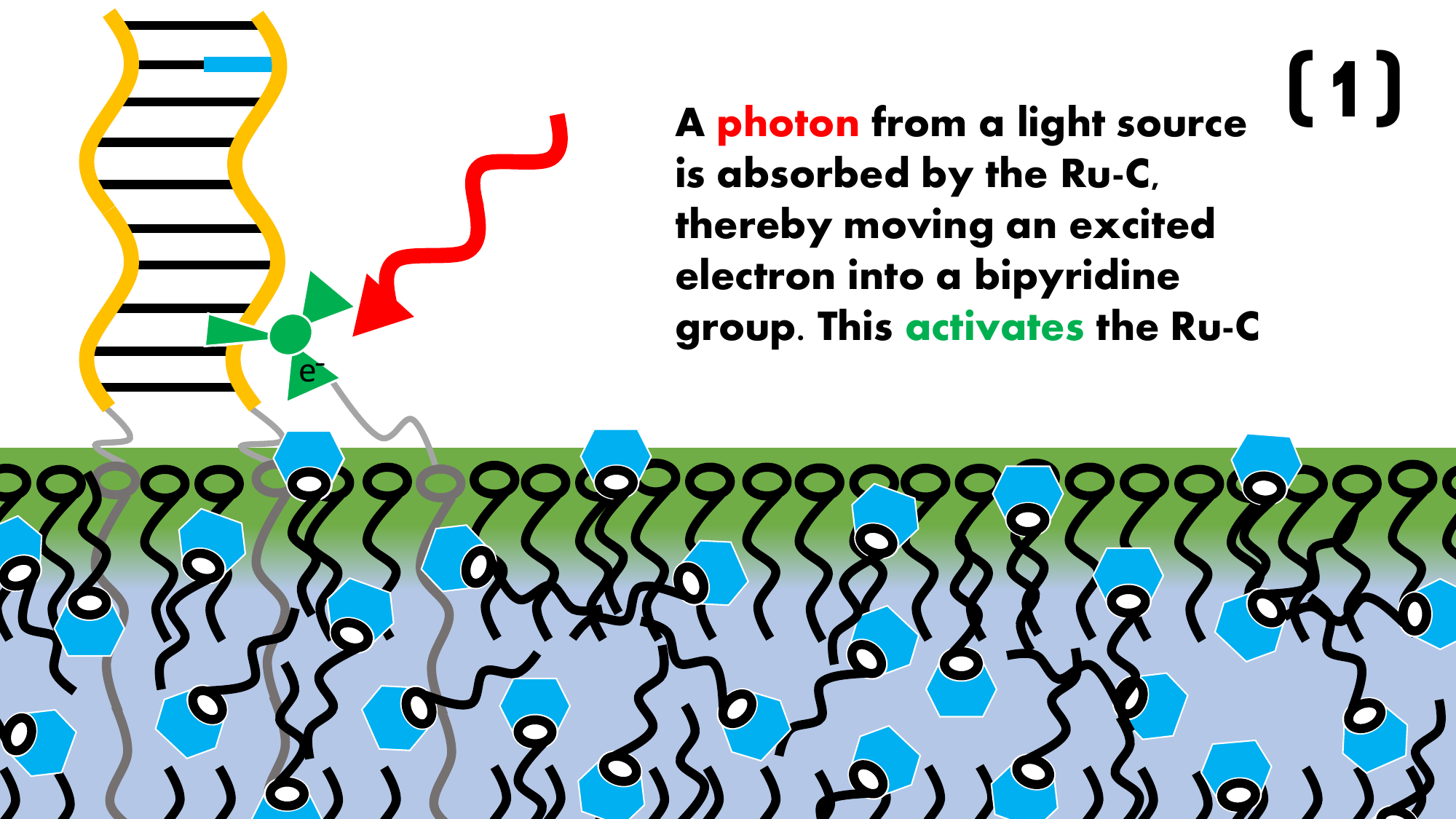}
\includegraphics[width=0.5\textwidth]{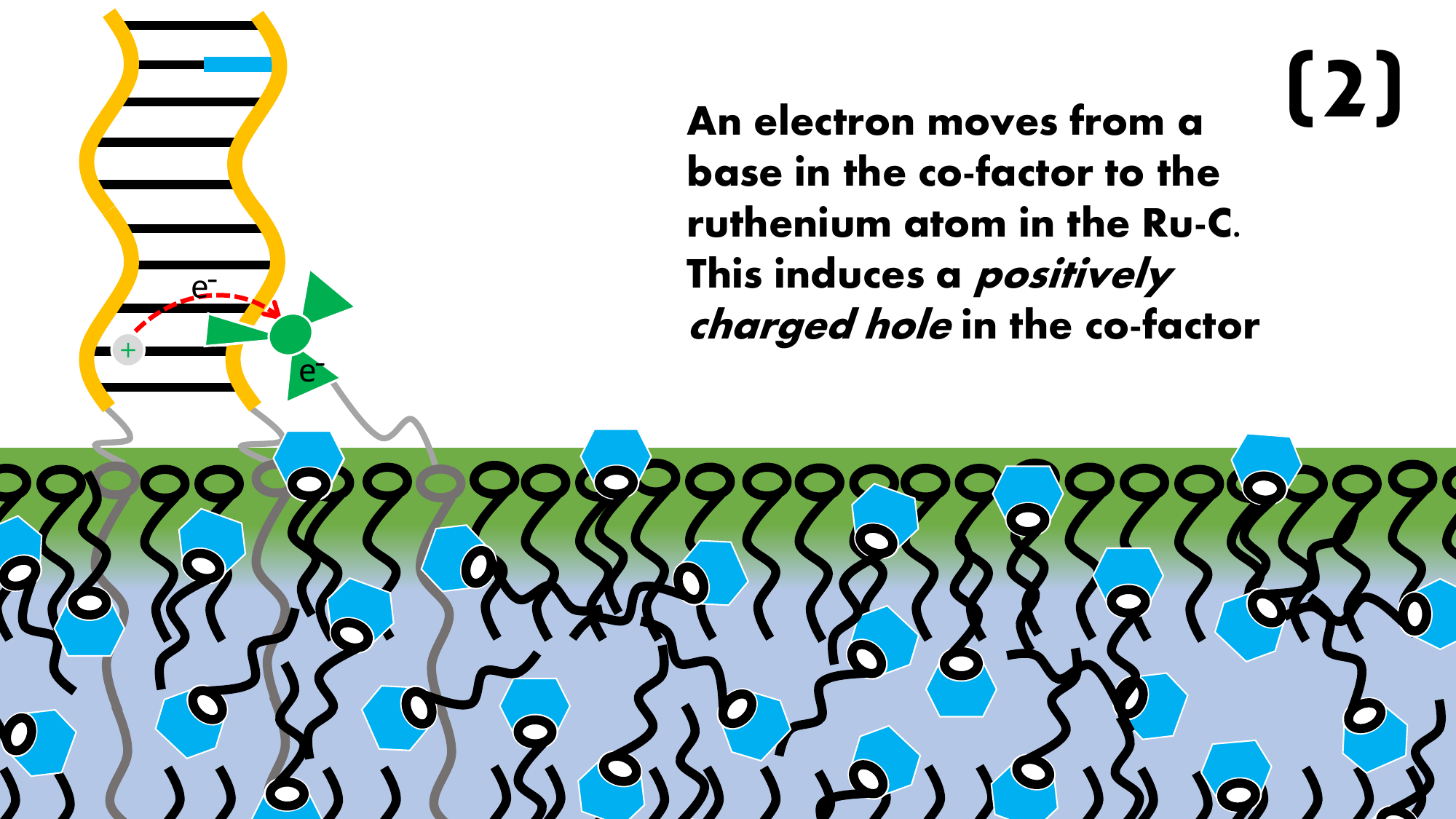}
\includegraphics[width=0.5\textwidth]{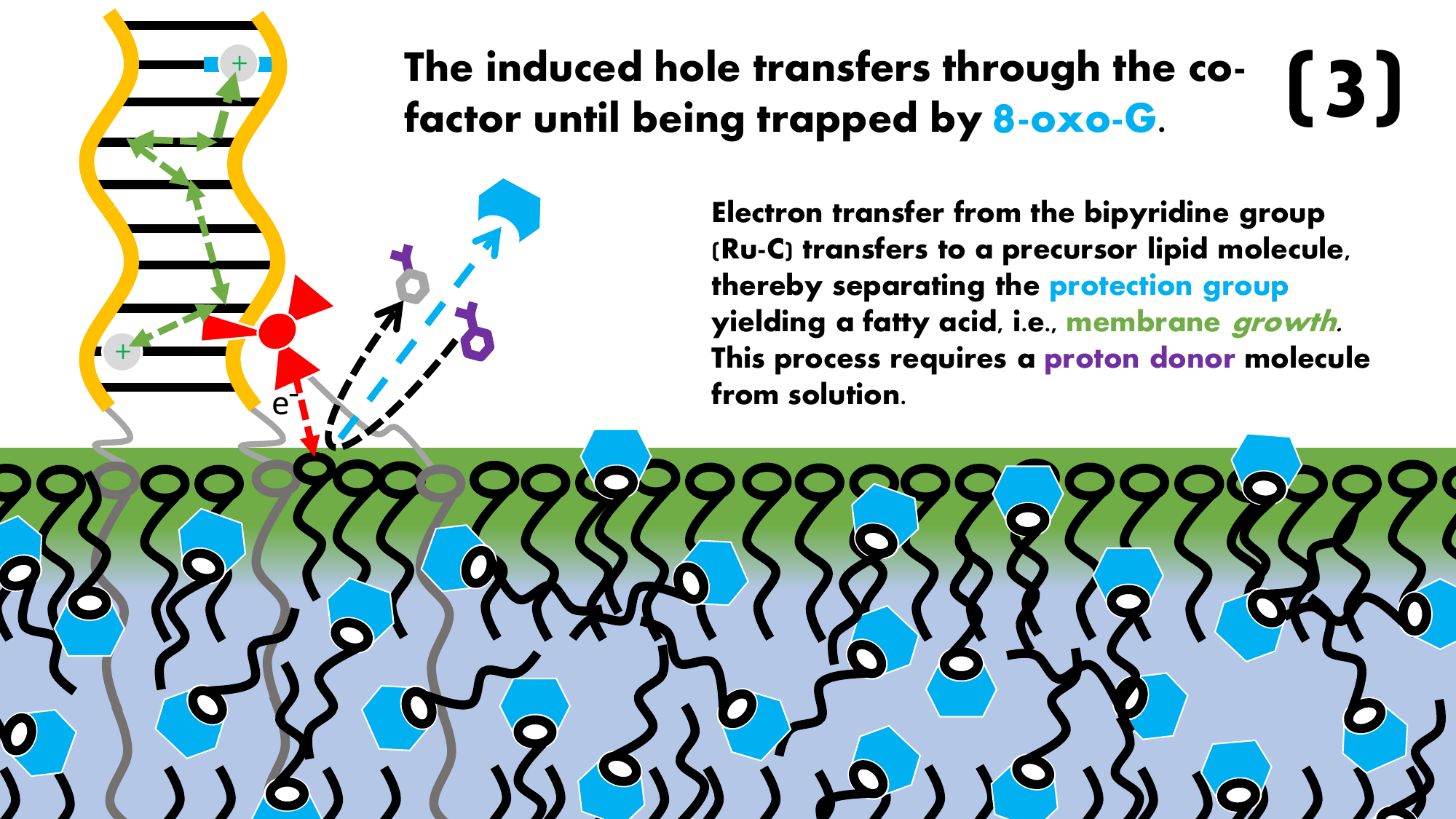}
\includegraphics[width=0.5\textwidth]{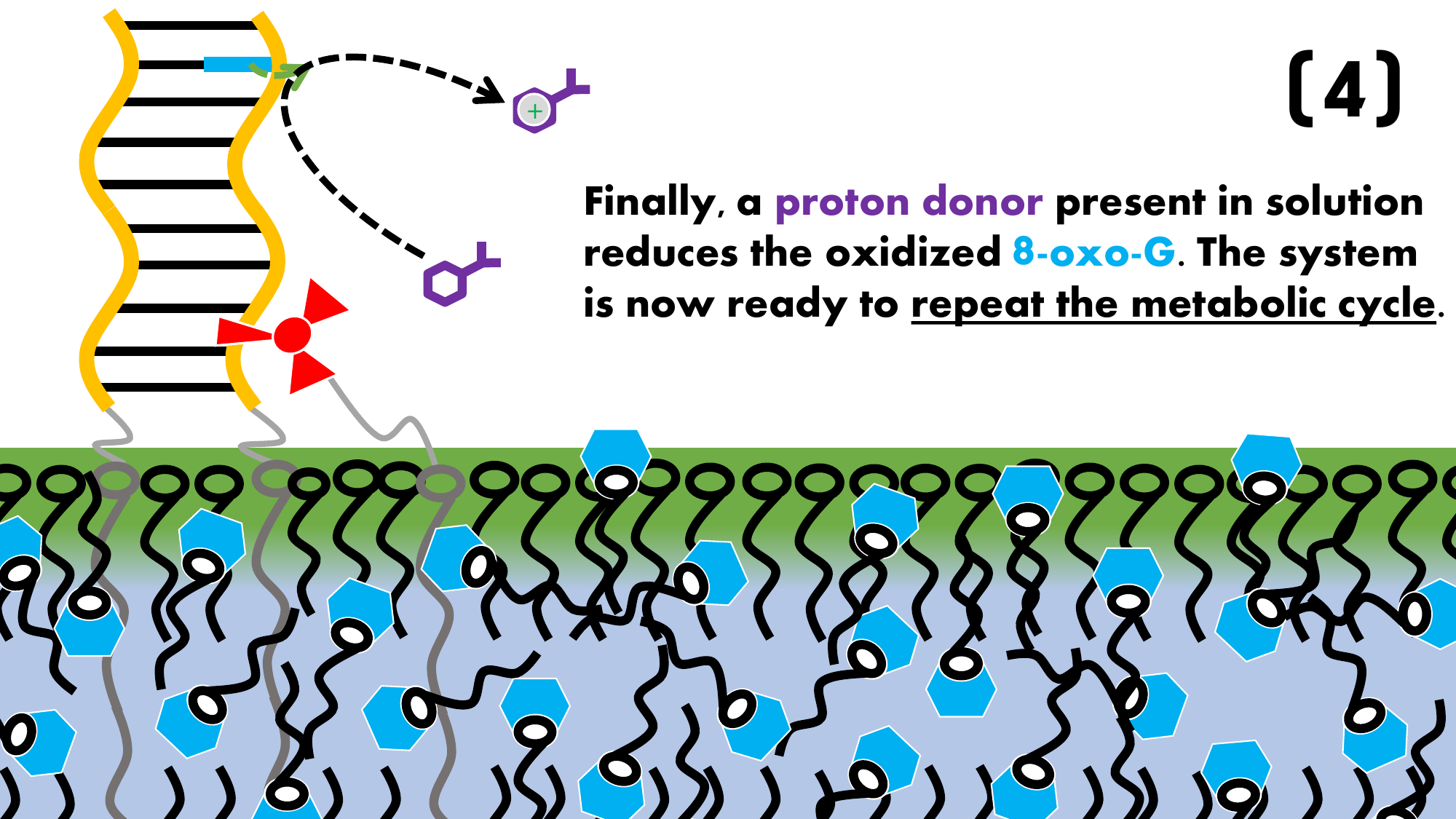}
\caption{Metabolic membrane formation: Panel (1) depicts a photo-induced excitation of the ruthenium complex where an electron from the ruthenium atom jumps out onto one of the bipyridine rings. As a result, an electron from a nearby guanine molecule may now replace the missing ruthenium atom electron; see Panel (2). Panel (3) depicts (i) how the electron hole (the missing electron) within the DNA duplex can diffuse along the DNA duplex due to the $\pi$ stacking of the base pairs. It should be noted that DNA hole transfer is sensitively dependent on the details of the DNA duplexes involved, which is the topic of Section 3 of this paper. Further, (ii) the originally excited electron may now jump from the bipyridine ring onto a nearby picolinium ester molecule, causing a fragmentation that results in a newly formed fatty acid and a free picolil group. Panel (4) depicts (i) trapping of the diffusing hole by the 8-oxo-guanine, resulting in the release of a hydrogen atom and replacement by a hydrogen donor (ii), so that the 8-oxo-guanine is restored and ready to receive another hole (donate another electron). For more details on the ruthenium chemistry involved, see \cite{declue2009nucleobase} and \cite{bornebusch2021reaction}}
\label{fig:charge_transfer}
\end{figure}

\begin{figure}[H]
\includegraphics[width=0.5\textwidth]{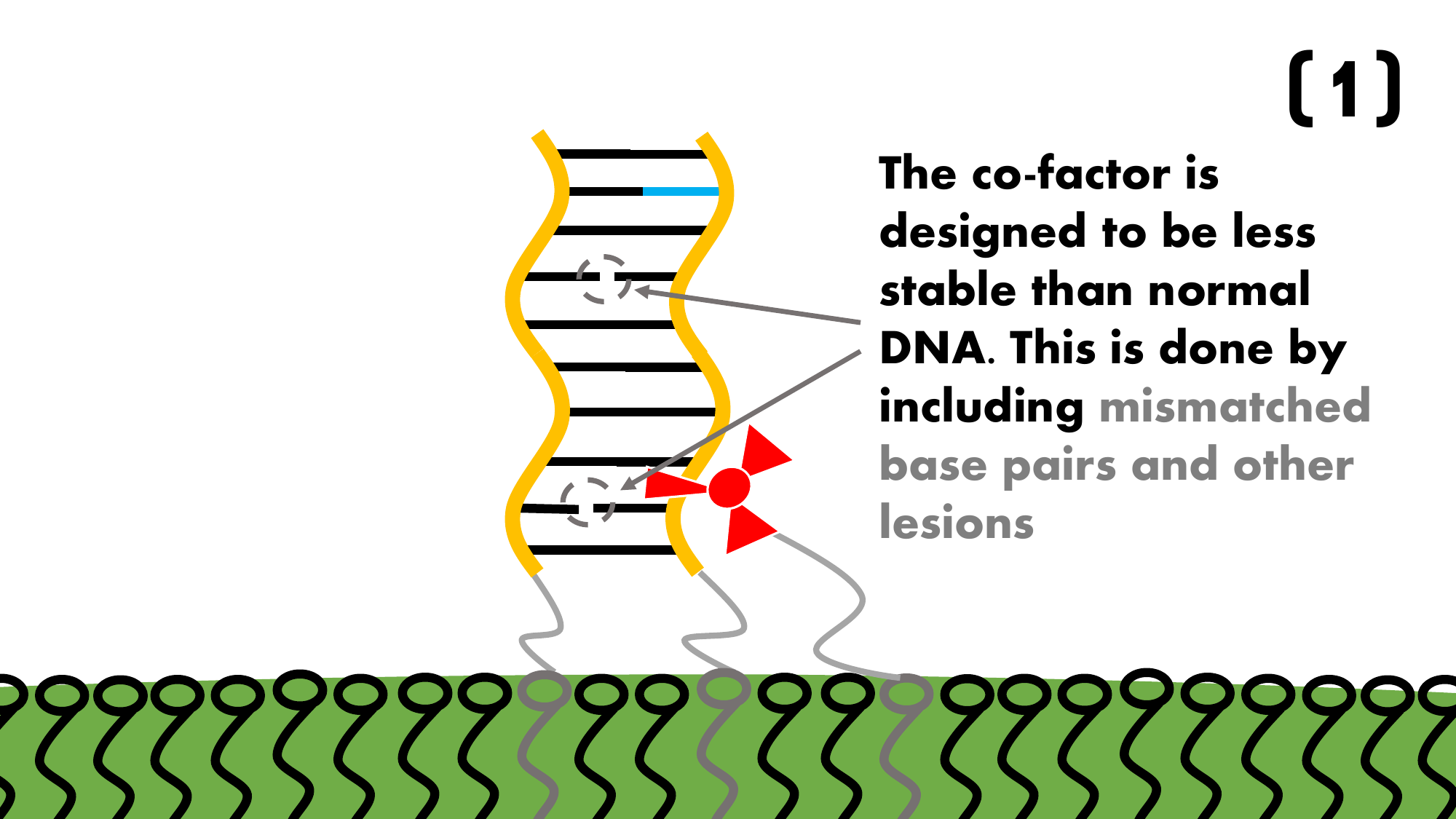}
\includegraphics[width=0.5\textwidth]{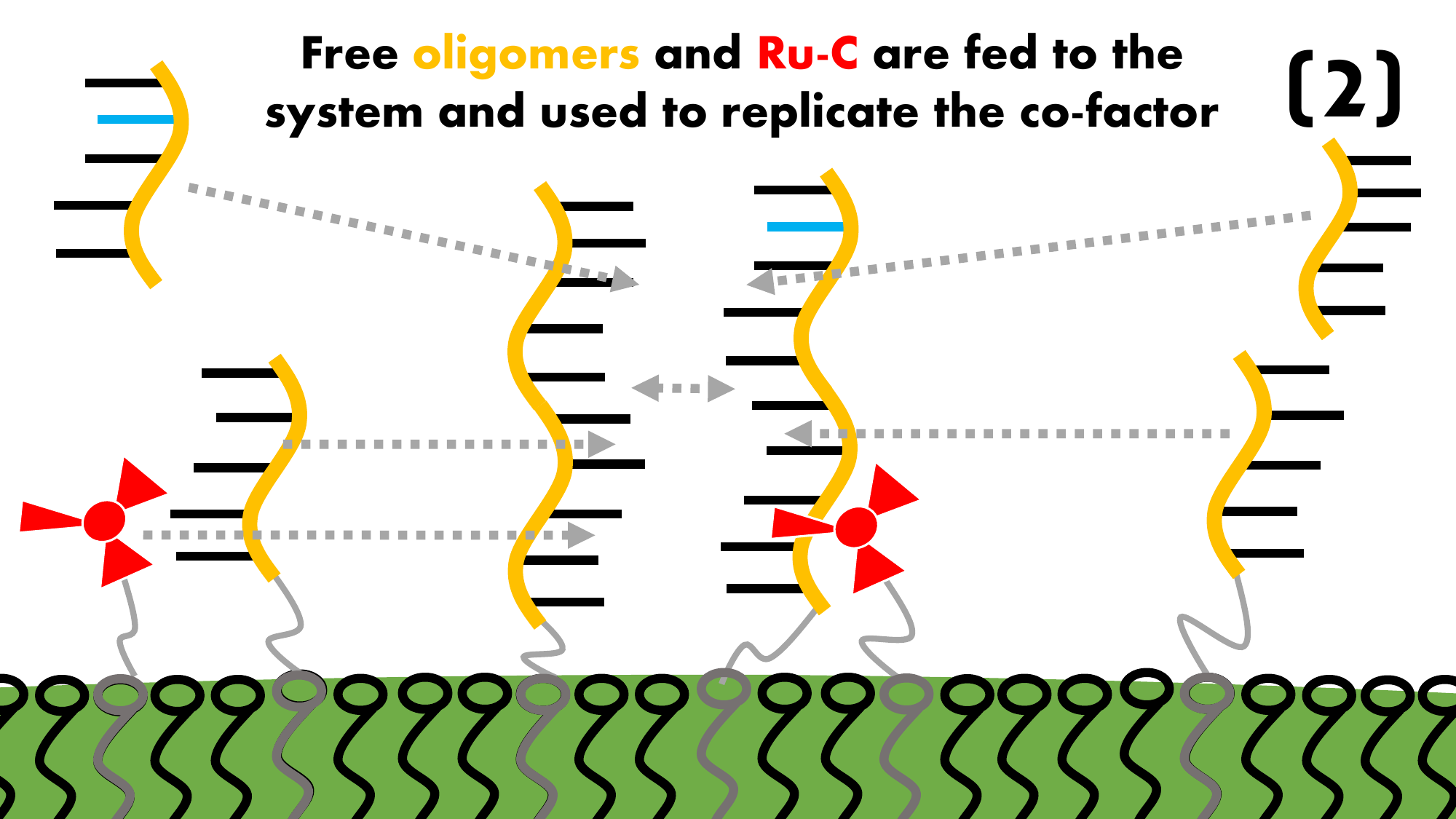}
\includegraphics[width=0.5\textwidth]{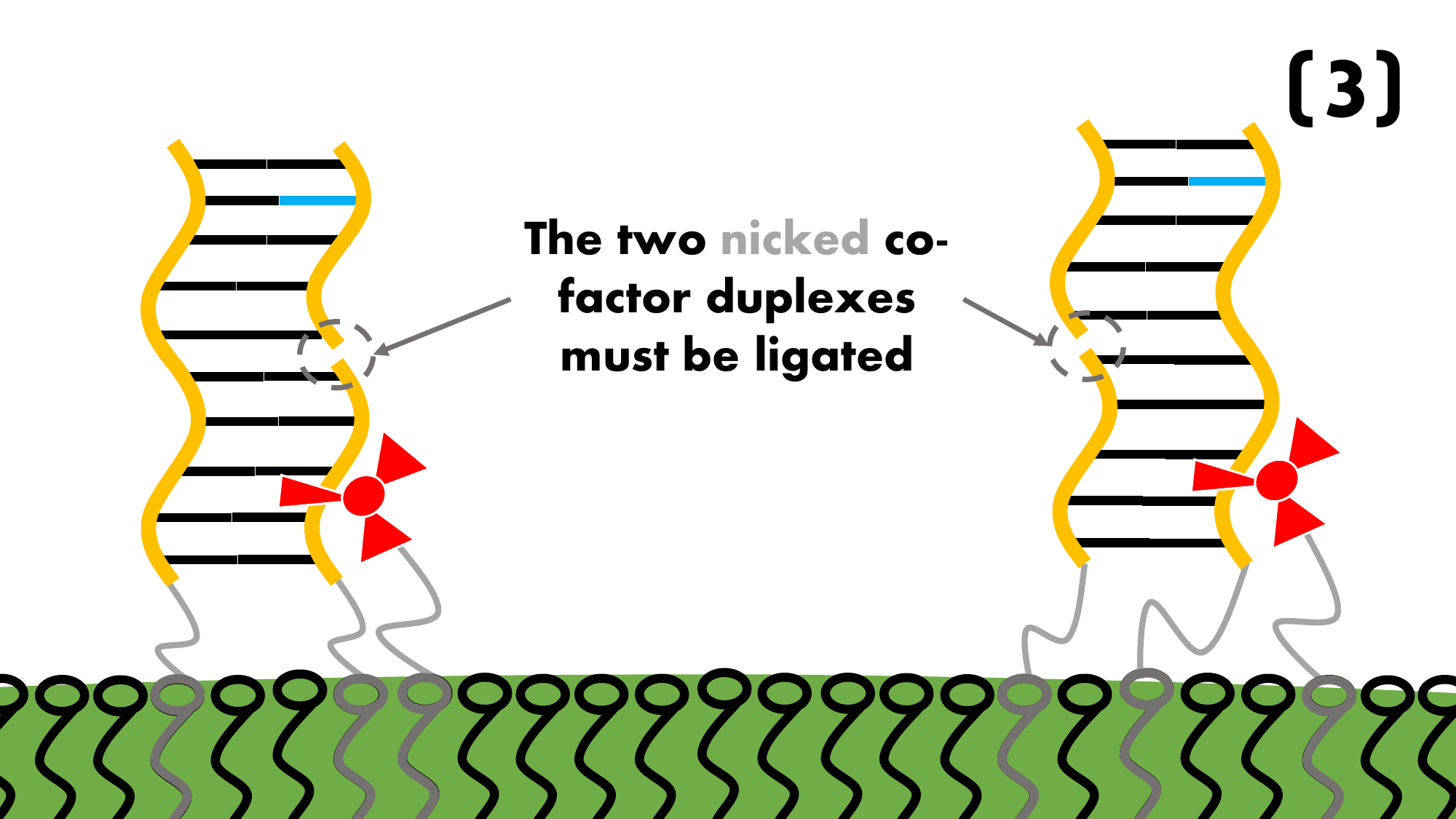}
\includegraphics[width=0.5\textwidth]{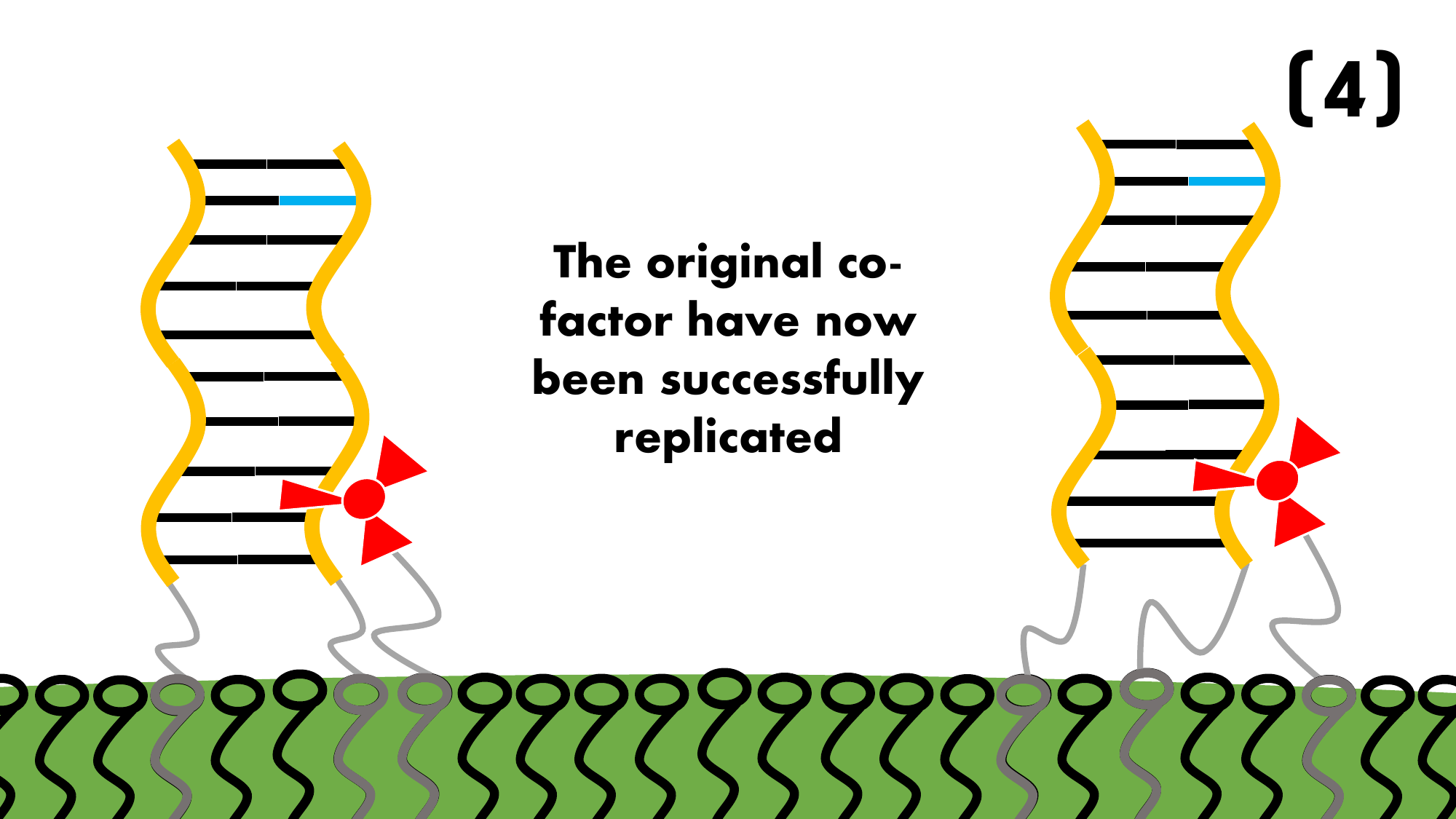}
\caption{Co-factor replication: Panel (1) depicts the same initial conditions as shown in \autoref{fig:membrane_feeding} and  \autoref{fig:charge_transfer}, although an extra detail is added as the involved DNA strands must contain lesions such as mismatches, abasic sites or bulges to destabilize the duplex. Panel (2) depicts (i) feeding of four oligomers and ruthenium complexes from the environment. The ruthenium complex and two of the oligomers have anchors. (ii) Due to the dynamic equilibrium between hybridized and dehybridized oligomers and the full strands, templating between the original strand, its complementary strand, and the newly added oligomers is possible. It should be noted that only DNA sequences with particular properties (destabilizing mismatches, abasic sites, or bulges) can replicate in this manner, which is the topic for Section 4 of this paper. In Panel (3), a full hybridization between the oligomers and the templates has occurred, where ligation can occur directly as the 3' end is activated by imidazole in the situation where no oligomer protection group is present at the 5' end. A hypothetical oligomer protection group could be removed if the protection group bond were disrupted by the traveling hole (positive charge) inside the DNA duplex. Oligomer deprotection is previously demonstrated by an excess electron (not a hole) \cite{cape2012phototriggered} where the activated electron is also provided by ruthenium complex photo-activation. Panel (4) depicts the resulting replication of the co-factor.}
\label{fig:info_replication}
\end{figure}

The combined energy transducer and cofactor, a two-component metabolic system, drives redox reactions on precursors of both amphiphiles (picolinium ester) and information molecules (protected DNA oligomers). In this way, it produces self-assembling vesicles containing building blocks, decanoic acid, and functional co-factor DNA information templates. Thus, our protocell design is in several ways different from modern life: it uses non-biological energy transducers; it has both the energy transducers and the information (co-factor) molecules anchored on the outside of a vesicle (and not inside in the vesicle lumen); it does not have a DNA translation machinery; it uses no enzymes; and it uses simple fatty acids and not phospholipids for containers. A detailed illustration of the key steps in the protocellular processes are found in \autoref{fig:membrane_feeding}, \autoref{fig:charge_transfer}, and \autoref{fig:info_replication}.

The two-component catalytic metabolism works as follows: A ruthenium-based complex (Ru-C) is activated by visible light and converts light energy (photons) into chemical energy (electrons) via two electron transfer processes (i) involving the DNA duplex with a modified nucleobase 8-oxoguanine (oxo-G) and (ii) the presence of a resource precursor molecule that has an easily breakable ester bond. (i) The activated and intercalated Ru-C receives an electron from a nearby guanine within the DNA duplex, which results in an electron hole in the $\pi$ stacked duplex. This electron hole diffuses up and down the DNA $\pi$ stack and eventually reaches the 8-oxo-G where it is absorbed due to the properties of the 8-oxo-G electron donor. As a result of this electron donation, the 8-oxo-G loses a hydrogen atom, which is replenished by dihydrophenyl glycine as a hydrogen donor \cite{declue2009nucleobase}. (ii) The light-activated electron that stems from the ruthenium atom is located at one of the three bipyridine molecules in the ruthenium complex. This electron may now jump to one of the picolinium ester molecules within the membrane and as a result deprotect its picolil group to form a new decanoic acid molecule, which serves as the building block of the fatty acid membrane. To ensure charge balance, a dihydrophenyl glycine is also needed as a hydrogen donor in this process \cite{declue2009nucleobase}. 

A detailed analysis of the DNA charge (electron hole) transfer properties is the topic of Section 3. 

DNA is used as a combinatorial co-factor in the metabolism because of its well known replication property. As the DNA base composition and sequence critically impact the metabolic efficiency, a protocellular inheritance is obtained by replication of the co-factor. Due to the well-known product inhibition in template-directed ligation replication \cite{Kiedrowski_1993}, we apply DNA sequences that are capable of self-replication via isothermal lesion-induced DNA amplification (LIDA) \cite{alladin2015achieving}. A detailed analysis of the co-factor replication properties is the topic for Section 4 of this paper. 

Note that the oxo-G is located above the ligation site to ensure that charge transfer can occur only in conjunction with DNA replication that requires hybridization and ligation into a duplex strand. In that way, the replication process becomes directly linked to the metabolic process. On the contrary, if the oxo-G is located near the vesicle surface at the lower end of the DNA strand, direct charge transfer may become possible between the ruthenium complex and the oxo-G without necessitating DNA hybridization and ligation. Thus, the metabolism would depend only on the presence of an anchored precursor oligomer containing oxo-G. For simplicity, in the following, we require the oxo-G to be located above the ligation site. However, a weaker linkage between replication and metabolism might still be possible with an oxo-G located below (but close to) the ligation site, as charge transfer would also likely require hybridization (but not ligation) and thus an initiation of the replication process. 

To summarize: The combinatorial DNA co-factor system serves as part of an electron relay that modulates the energy transduction efficiency, which directly depends on the charge transfer rates of the co-factor DNA information molecules. Further, non-enzymatic replication of the combinatorial DNA co-factor and added ruthenium complexes ensure continued metabolic functionality in the next generation. Note that both DNA charge transfer and non-enzymatic DNA replication strongly depends on the DNA sequence composition. As we shall see in the following Section, the DNA charge transfer and replication properties have very different sequence requirements, which makes it nontrivial to identify appropriate sequences that can support both functionalities. 

\begin{figure}[H]
\includegraphics[width=0.5\textwidth]{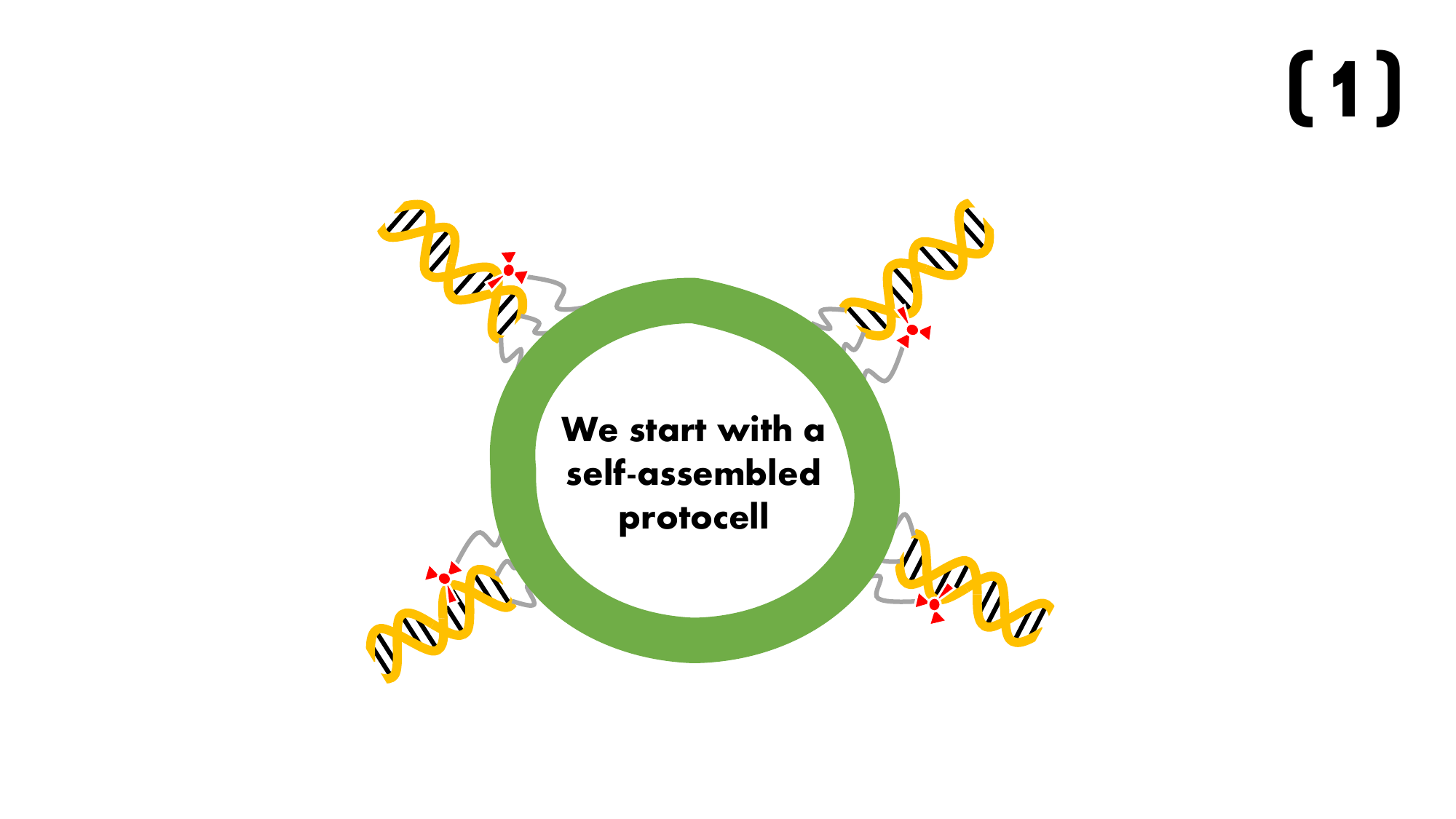}
\includegraphics[width=0.5\textwidth]{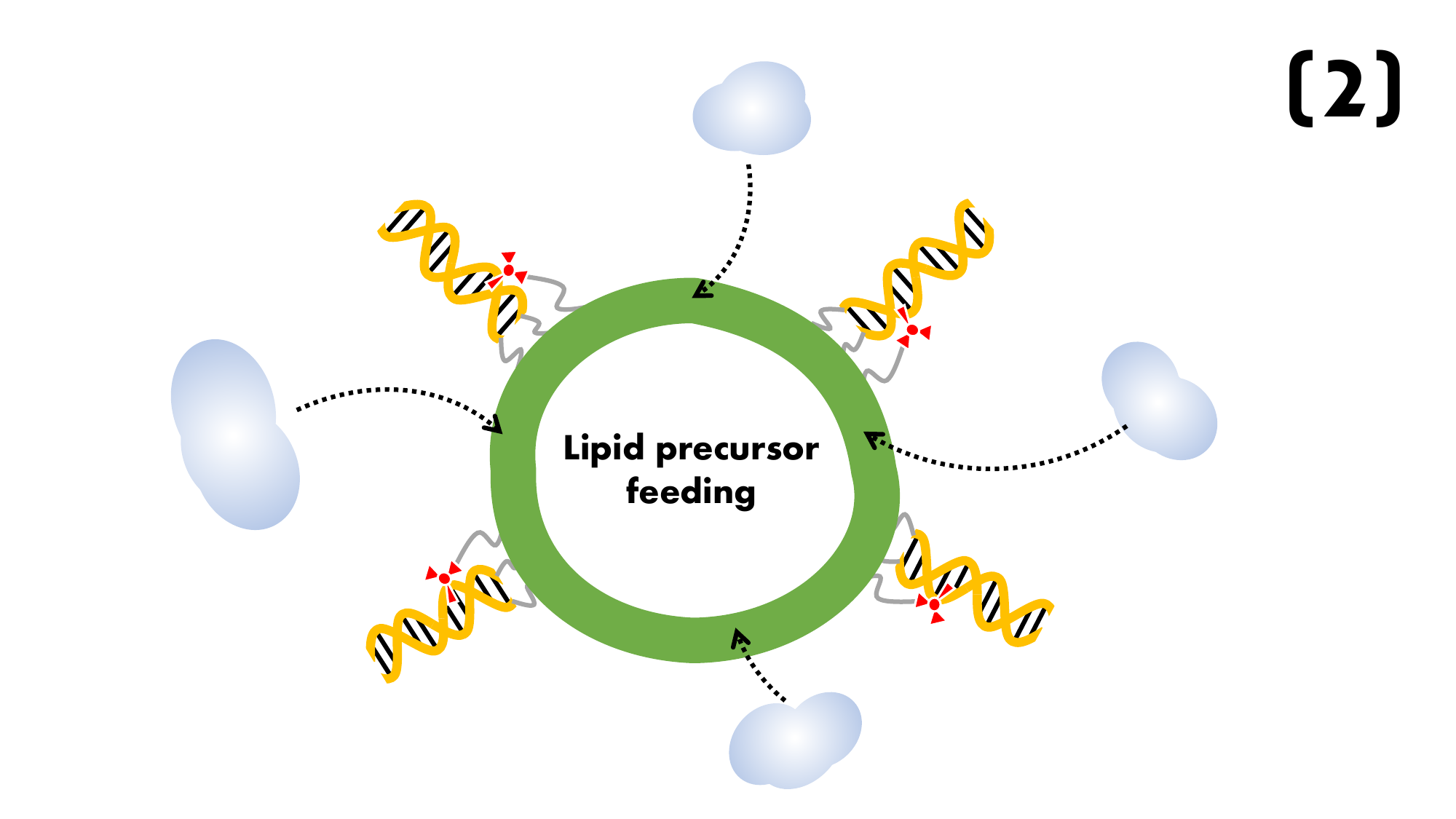}
\includegraphics[width=0.5\textwidth]{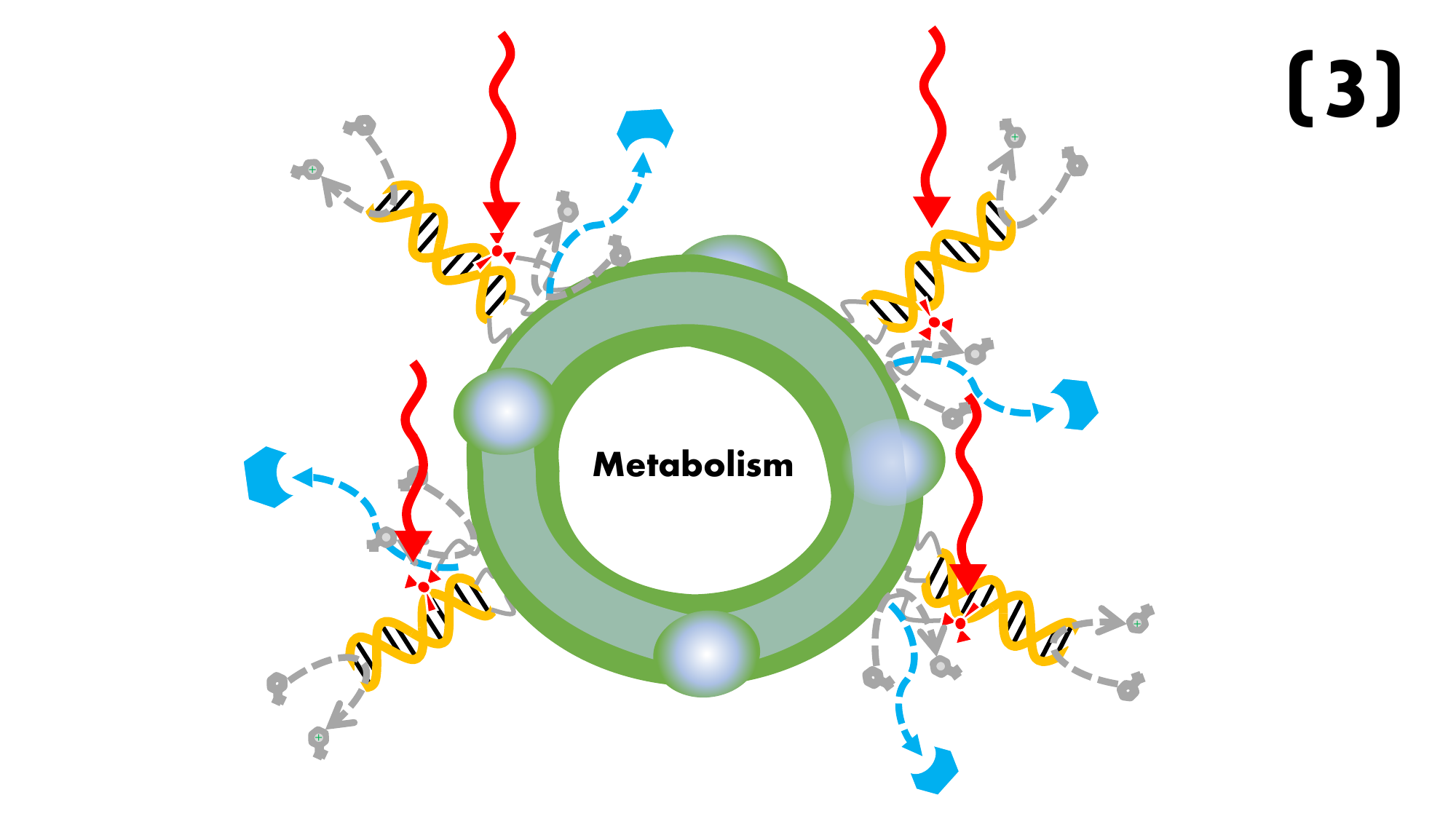}
\includegraphics[width=0.5\textwidth]{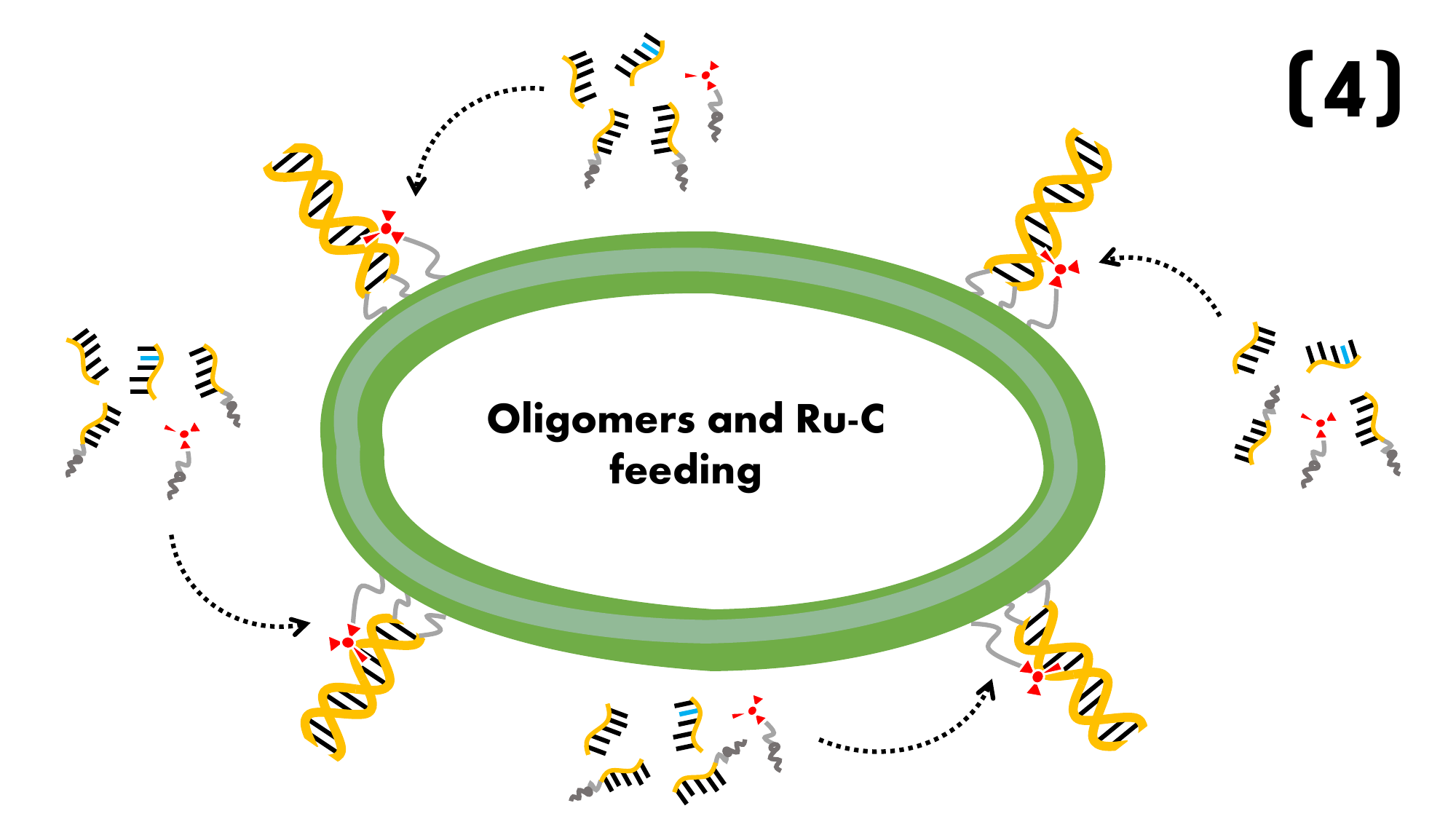}
\includegraphics[width=0.5\textwidth]{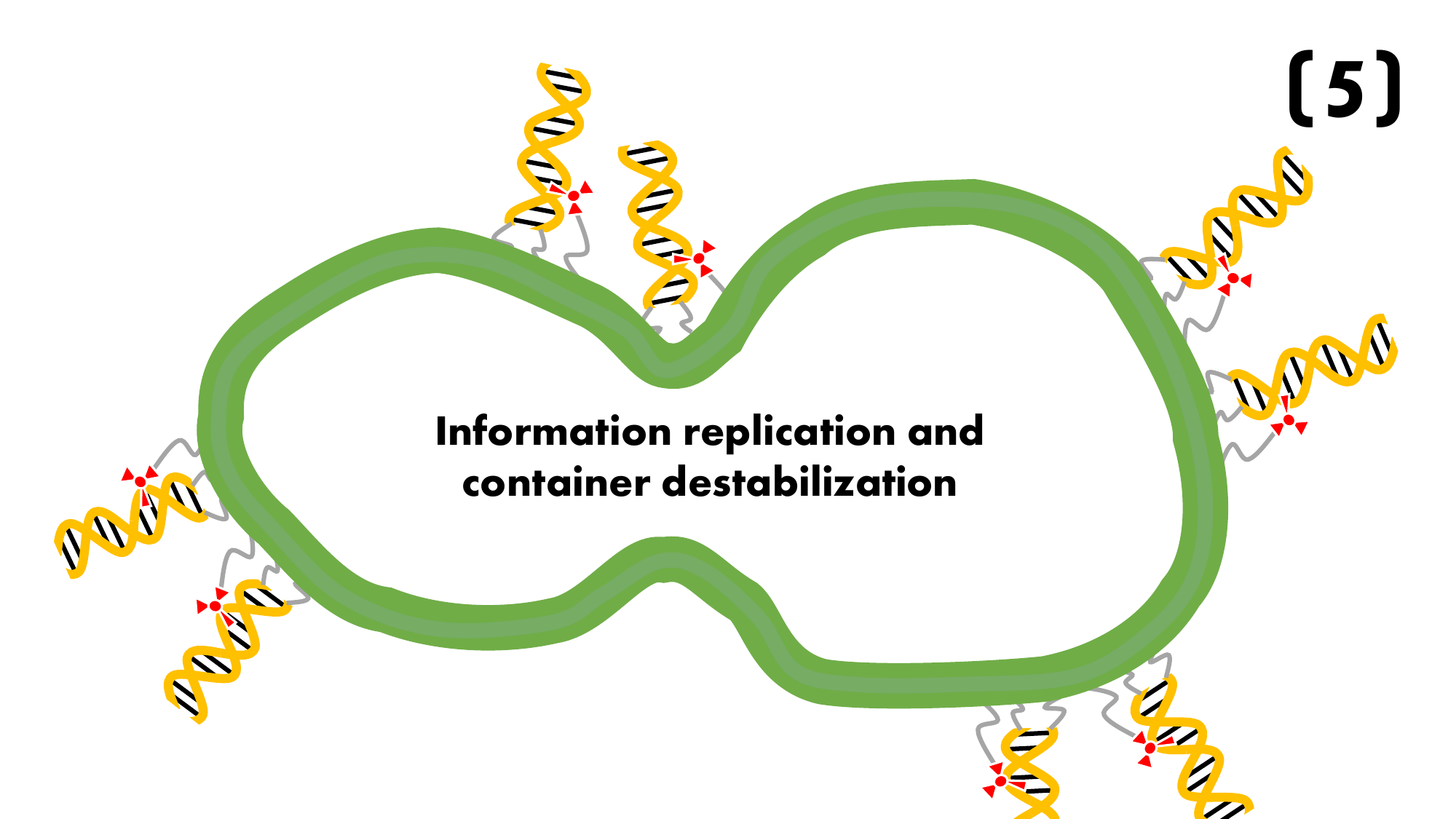}
\includegraphics[width=0.5\textwidth]{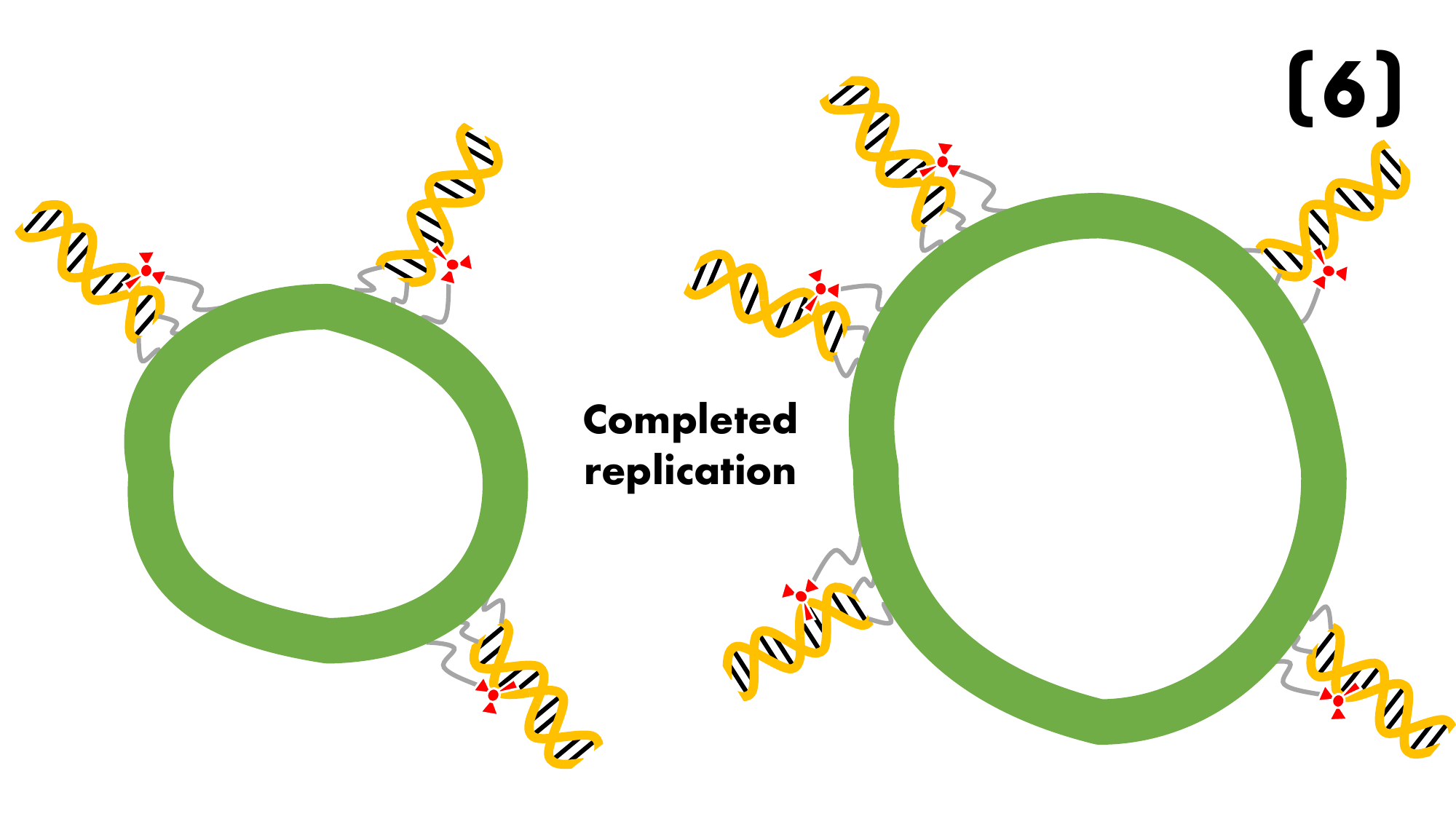}
\caption{Protocellular life-cycle: Panel (1) is a cartoon of the protocell with a fatty (decanoic) acid vesicle container decorated with anchored ruthenium complexes and DNA duplexes.  
Panel (2) shows droplet feeding of hydrophobic membrane precursors (picolineum esters). Panel (3) depicts absorption of the hydrophobic droplets into the membrane, where they are partly dissolved in the membrane, while the metabolism converts precursor lipids (picolinium ester) into lipids (decanoic acid). Panel (4) depicts ruthenium complex and the feeding of the DNA oligomer. Note that all precursors can be fed at once but are shown here as a two-step process for clarity. Panel (5) shows a completed (DNA) co-factor replication together with a completed fatty acid production, which results in membrane growth and vesicle destabilization. The original vesicle eventually breaks up and forms two new protocells, see Panel (6). 
}
\label{fig:life_cycle}
\end{figure}

 Further, it should be noted that the environment is highly regulated (e.g., temperature, pH, salts) and designed to support the protocellular life-cycle by delivering light energy as well as multiple necessary resource molecules including picolinium ester, ssDNA oligomers (with and without tails), Ru-complexes with tails, picolinium ester, dihydrophenyl glycine (hydrogen source). A summary discussion of the main processes in the protocellular life-cycle is shown in \autoref{fig:life_cycle}

%% file: 3-ChargeTransfer.tex
\section{Simulation Exploration of Co-factor Charge Transport}\label{sec:sim_explo}

Charge transport (CT) in DNA has been extensively studied since the early 1990s. The majority of DNA CT research has focused on hole transport (HT). Less research has been done on electron transport (ET) in DNA \cite{voityuk2005estimates}. DNA HT dynamics is usually associated with oxidized nucleobases in DNA. Oxidative damage occurs through the oxidization of guanines by reactive oxygen species leading to higher levels of 8-oxo-7,8-dihydro-2'-deoxyguanosine (oxoG), which is linked to mutagenesis and cancer \cite{loft1996cancer,roszkowski2011oxidative}. Therefore, electrochemistry-based sensors have been proposed for the detection of selected DNA sequences or mutated genes associated with human disease \cite{drummond2003electrochemical}. 

Our previous explorations of protocellular metabolic function have focused on direct ET from oxoG to a Ru-C, either covalently bound or in close spacial proximity \cite{declue2009nucleobase,maurer2011interactions,cape2012phototriggered,rasmussen2016generating,bornebusch2021reaction}). 
In the present work, the direct ET between oxoG and Ru-C is mediated by a combinatorial co-factor in the form of a DNA duplex. 
This causes CT to become a multi-step process, that both involves ET and HT. Electron transport from oxoG through DNA to Ru-C is equivalent to: (i) An initial ET from a nearby guanine within the DNA strand to the light activated Ru-C; (ii) Now the missing electron in the guanine (the hole) travels through the DNA strand HT until it meets an oxoG that donates the missing electron. To repair the now damaged oxoG, a proton donor (dihydrophenylglycine) is provided to the system as in our previous studies. 

Everything else being equal, the slower the CT rate within the DNA, the slower the resulting metabolic rate, because the CT rate modulates the rate at which photo-excited electrons can be used for digesting resource molecules and turning them into building blocks. 

 Previous studies have established that Ru-C can merge into the $\pi$ stack of DNA duplexes (intercalation), which can induce HT into DNA when photo-activated. It is known that Ru-C intercalates with the $\pi$ stack of DNA duplexes, yielding fast photoinduced transfer of a hole to DNA nucleobases, and such holes can migrate long distances (200 Å) through DNA \cite{mihailovic2006exploring,genereux2010mechanisms,williams2000variations}. It is also known that HT occurs in DNA strands containing oxoG, and that oxoG is a deep thermodynamic potential well for holes \cite{gasper1997intramolecular}. Thus, holes become `trapped' when reaching oxoG. 

\begin{figure}[H]
\centering
\includegraphics[scale=0.25]{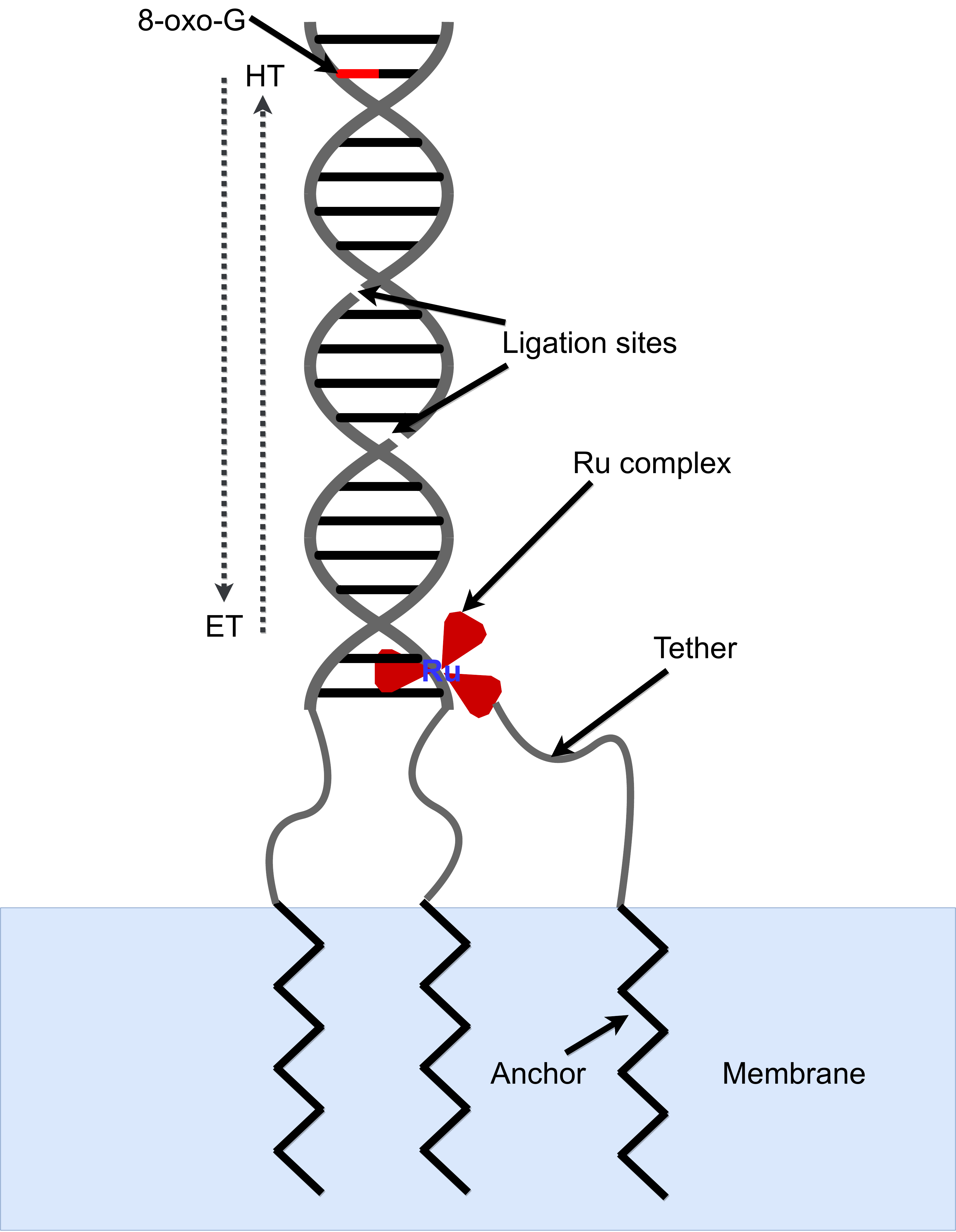}
\caption[Metabolism design with tethers]{Example of protocellular metabolism design. The co-factor containing oxoG and Ru-C are anchored to the protocellular container surface via tethers to amphiphilic anchors. The Ru-C intercalates with the DNA $\pi$ stack of the co-factor enabling CT between Ru-C and 8-oxo-G mediated by the nucleobases of the DNA strand, see text for details. A fast DNA HT means a fast metabolism, while a slow HT means a slow metabolism. Examples of DNA oligomer ligation sites are shown as they impacts the HT rates.}
\label{fig:tether}
\end{figure}

A design example of the protocellular metabolism, composed of the energy transducer and the co-factor, is shown in \autoref{fig:tether}.
Based on these experimental facts, we propose a protocell design in which DNA strands containing a single oxoG near one terminal is attached to the protocell container (membrane) via at least one amphiphilic anchor at the opposite terminal. Additionally, Ru-C (energy transducers) are anchored to the membrane via molecular tethers, thereby promoting intercalation with the $\pi$ stack of the DNA duplex.    

\subsection{Modelling charge transport reactions}\label{sec:CTdiffeq}

Because the efficiency of the DNA CT is critical for metabolic efficiency, we need to explore how the CT is influenced by the DNA base composition. We can model DNA CT as a system of multiple redox centers, which can be modeled as systems of coupled linear differential equations. A system with $N$ redox centers is described by $N$ equations of the following form:
\begin{equation}\label{eq:Pdot=KP}
    \dd{P_{i}(t)}{t} = \sum^N_{j=1} k_{ij} P_j(t) \,\,\,\,\,\,\, \textrm{where} \,\,\,\,\,\,\, i\in \{1,..,N\},
\end{equation}
where $P_{i}(t)$ describes the probability that the charge is located at the $i$th redox center at time $t$ and $k_{ij}$ are elements of a matrix containing the transition rates between redox centers: $k_{ij}$ being the transfer rate \emph{to} the $i$th \emph{from} the $j$th redox center. Assuming there is no loss of charge to the environment, the rate matrix should conserve charge, meaning $\sum_i k_{ij} = 0$.

We have chosen to restrict our kinetic simulations in the following sections to use previously calculated or experimentally measured CT rates.

We now use the rate matrix formalism to model CT in DNA. Because the transition rates depend on quantum tunneling rates, they decay exponentially as a function of the distance between redox centers. Models that only include thermally induced hole transport transitions (TIH) set entries far from the diagonal of the $k_{ij}$ matrix to zero, yielding a band matrix. If transitions to non-neighboring bases are allowed --- e.g., by including the super exchange (SE) mechanism --- the rate matrix will no longer be a simple band matrix, but instead have non-zero entries away from the band, as is the case for the rate matrix in Eq.~\ref{eq:AGAAGA} that represents a simple well-matched strand 5'-GAGAAGA-3':
\begin{equation} \label{eq:AGAAGA}
\hspace{-.5cm}
\textbf{K} =
\begin{bmatrix}
-k_{GA}             &   k_{AG}              &   0                   &    0                  &       0                   &       0\\
k_{GA}              &   -(2k_{AG}+k_{SE})  &   k_{GA}              &    0                  &       k_{SE}              &       0\\
0                   &   k_{AG}              &   -(k_{GA}+k_{AA})    &    k_{AA}             &       0                   &       0\\
0                   &   0                   &   k_{AA}              &    -(k_{AA}+k_{GA})   &       k_{AG}              &       0\\
0                   &   k_{SE}              &   0                   &    k_{GA}             &       -(2k_{AG}+k_{SE})   &       k_{GA} \\
0                   &   0                   &   0                   &    0                  &       k_{AG}              &       -k_{GA} \\
\end{bmatrix}
\end{equation}

For the construction of these models, the following assumptions are made: (i)~holes are localized at a single base; (ii)~HT transitions are always possible between neighboring bases; (iii)~holes only transfer between the bases with the lowest oxidation potential in each base pair ($E_{ox}^G<E_{ox}^C$ and $E_{ox}^A<E_{ox}^T$ \cite{bixon2002long}); (iv)~all SE transitions --- independent of bridge sequence --- are modelled as SE over a simple A/T bridge, i.e., the SE rate from G to GGG found by Bixon and Jortner is approximately equal to the SE rate from G to a single G\footnote{This assumption is based on the energy dependent nuclear Franck-Condon factor in Eq.~(\ref{eq:FE}), where Bixon and Jortner \cite{bixon2002long} use $\lambda=0.25$ eV and $\hbar\omega = 0.18$ eV, both of which are large compared to $\Delta G = -0.096$ eV.}; ($v$)~there is no charge loss from the strand to solution. Furthermore, in our simulation, we set a maximum bridge length for SE transitions of $n=9$ bp due to the rapid exponential decay of SE rates as a function of $n$. For all future simulations the charge is initially localized at the first base, i.e., $P_{1}(t=0)=1$ and $P_{i\neq1}(t=0)=0$.

\subsection{Charge transport simulations in a selection of strands}\label{sec:CTsims}

To obtain physically realistic co-factor CT rates we only simulate strands with known experimental or theoretical single-step HT $k_{ij}$ rates. These experimentally estimated rates are from Osakada et al. (2008) \cite{osakada2008kinetics}, while the theoretically estimated rates are from Bixon and Jortner (2002) \cite{bixon2002long}. The used theoretical rates computed by Bixon and Jortner, who use semi-classical Marcus theory are shown in \autoref{tab:rates}. In this table are also shown the rates from Osakada et al. (2008) that experimentally investigate charge transport through strands containing mismatches. From the charge transport they further deduce rates between certain redox sites (not necessarily neighboring base pairs) by fitting kinetic models to their results. As will become clear in Section \ref{sec:LIDA_kin} on co-factor replication, we have selected a group of strands containing both a bulge (an unpaired ``bulging'' base) and possibly mismatches (bases forming non-Watson–Crick pairs) \cite{santalucia2004thermodynamics}. A bulge occurs when one base is not is detached from the duplex structure somewhere in the middle of the strand, e.g., if one strand of in a well-matched duplex is one base longer then the other strand. 

Specifically, in our investigations, we assume the rate through the substrand containing the mismatch is independent of the sequence of the rest of the strand, so that rates across mismatches ($k_{GGt}$ and $k_{GGa}$) are implemented as a combined rate across multiple base pairs. Furthermore, the charge transport across the mismatch is modeled as being one-directional (no back-transfer) as by Osakada et al. (2008). The values of $k_{GGt}$ and $k_{GGa}$ are also shown in \autoref{tab:rates}.

Research by Barton and coworkers has found that conformational gating is highly regulating in DNA HT, with disruptions of the DNA $\pi$ stack generally yielding slower HT rates \cite{genereux2010mechanisms,bhattacharya2001influence}. This has motivated our estimate of HT rates through the bulge ($k_{AB}$ and $k_{BA}$) as being approximately one order of magnitude slower than HT rates through mismatches, as we view mismatches as a lesser disruption of the DNA $\pi$ stack compared to bulges.

\begin{table}[H]
\hspace{-1 cm}
\begin{tabular}{|l|l|l|l|}
\hline
Name            & Rate [s$^{-1}$]                     & Reference                                     & Description\\ \hline
$k_{AA}$        & $5\cdot10^{7}$                    & Bixon and Jortner (2002) \cite{bixon2002long}        & rate from A/T to A/T\\
$k_{GA}$        & $4\cdot10^{4}$                    & Bixon and Jortner (2002) \cite{bixon2002long}        & rate from G/C to A/T\\
$k_{AG}$        & $3\cdot10^{8}$                    & Bixon and Jortner (2002) \cite{bixon2002long}        & rate from A/T to G/C\\
$k_{GG}(n=1)$   & $1.2\cdot 10^{8\textbf{a}}$    & Bixon and Jortner (2002) \cite{bixon2002long}        & super-exchange rate from G/C to G/C \\
$k_{AoxoG}$        & $4\cdot10^{6}$                    & Estimation by authors                         & rate from A/T to oxoG/C\\
$k_{oxoGA}$        & $0^{\textbf{b}}$                  & Estimation by authors                         & rate from oxoG/C to A/T\\
$k_{GGt}$       & $1\cdot10^{4}$                    & Osakada et al. (2008) \cite{osakada2008kinetics}    & rate through substrand from G/C to G/T\\
$k_{GGa}$       & $9.5\cdot10^{3}$                  & Osakada et al. (2008)  \cite{osakada2008kinetics}   & rate through substrand from G/C to G/A\\
$k_{AB}$        & $1\cdot10^{3}$                    & Estimation by authors                         & rate from A/T to bulge\\
$k_{BA}$        & $1\cdot10^{3}$                    & Estimation by authors                         & rate from bulge to A/T\\ \hline
\end{tabular}
\caption[CT rates used in simulations]{$^{\textbf{a}}$The super-exchange rate between G/C and G/C through an A/T-bridge of length $n=1$. The super-exchange rates for A/T-bridges of length $n>1$ are defined as $k_{GG}(n) = k_{GG}(1) r^{n-1}$, where $r$ is determined by the electronic couplings in the bridge or alternatively a $\beta$-value specific for the bridge ($r=\exp(-\beta R_0)$). In our investigations, we use $\beta=0.7$ Å$^{-1}$ and $R_0=3.4$ Å; $R_0$ being the base separation in the DNA $\pi$ stack. In our simulations, we set a maximum bridge length of $n=9$ bp. $^{\textbf{b}}$Since oxoG is a thermodynamic trap, the rate out of oxoG is assumed to be very low compared to the rest of the used rates. $k_{oxoGA}$ is therefore set equal to zero for simplicity.}
\label{tab:rates}
\end{table}

Clearly, using only the elemental HT rates listed in \autoref{tab:rates} limits the possible strands we can simulate in a sequence of 18 bps.
Eqs.~(\ref{eq:sub1})-(\ref{eq:sub6}) below show the six used substrands from which the full 18 bp co-factor sequences are composed. 
From \autoref{tab:rates}, we have that (i) substrand 3 and 4 (mismatches) must be preceded by substrand 1 for the rate constants $k_{GG_a}$ and $k_{GG_t}$ to apply. Furthermore, (ii) the full strand must include one and only one bulge (substrand 6) near the center of the strand (see Section 4 for details). We choose to place it as the third substrand in our chosen set. Finally, (iii) the full strand must include one and only one 8-oxo-G (substrand 5). 
This results in a set of 160 strand sequences that are chosen as possible protocell co-factor candidates. 

\begin{align}
    \textrm{substrand 1:} \,\,\,\,\,\,      &\textrm{5'-AGA-3'}\label{eq:sub1}\\
                                            &\textrm{3'-TCT-5'}\nonumber\\
    \textrm{substrand 2:} \,\,\,\,\,\,      &\textrm{5'-AAA-3'}\\
                                            &\textrm{3'-TTT-5'}\nonumber\\
    \textrm{substrand 3:} \,\,\,\,\,\,      &\textrm{5'-AGA-3'}\\
                                            &\textrm{3'-TTT-5'}\nonumber\\
    \textrm{substrand 4:} \,\,\,\,\,\,      &\textrm{5'-AGA-3'}\\
                                            &\textrm{3'-TAT-5'}\nonumber\\
    \textrm{substrand 5:} \,\,\,\,\,\,      &\textrm{5'-A8oxoGA-3'}\\
                                            &\textrm{3'-TCT-5'}\nonumber\\
    \textrm{substrand 6:} \,\,\,\,\,\,      &\textrm{5'-AAA-3'}\label{eq:sub6}\\
                                            &\textrm{3'-TT-5'}\nonumber
\end{align}

The full 160 strands are shown in \autoref{tab:strands} in Section \ref{sec:5}. In addition, in that table, the conclusions from Sections~\ref{sec:sim_explo} and \ref{sec:4} are indicated by color coding of the different strands. 

For the protocell metabolism, the quantity of interest is the HT rate from Ru-C to oxoG (henceforth denoted $k_{CT}$). Note that this CT is coupled to the larger, full protocellular metabolism \cite{declue2009nucleobase,maurer2011interactions,cape2012phototriggered,rasmussen2016generating,bornebusch2021reaction}, which is further discussed in Section 3.3. 
For the following CT simulations, it is assumed that a Ru-C intercalates at the 5' end of the DNA strand, inducing a hole by fast HT into the first base of the 5' end. In principle, the hole could be induced at other sites instead. The hole then travels within the strand until it reaches oxoG, where it is trapped. A MATLAB program constructs the system of ODEs specific for each of the 160 strand sequences using the rate matrix formalism in Eq.~(\ref{eq:Pdot=KP}) and solves the resulting system numerically with \texttt{ode15s}. As previously stated, the hole is assumed to be initially localized at the first base, the initial conditions $P_1(t=0)=1$ and $P_{i\neq1}(t=0)=0$ are used.

\begin{figure}[H]
\centering
\hspace*{-1.8cm}
\includegraphics[scale=0.75]{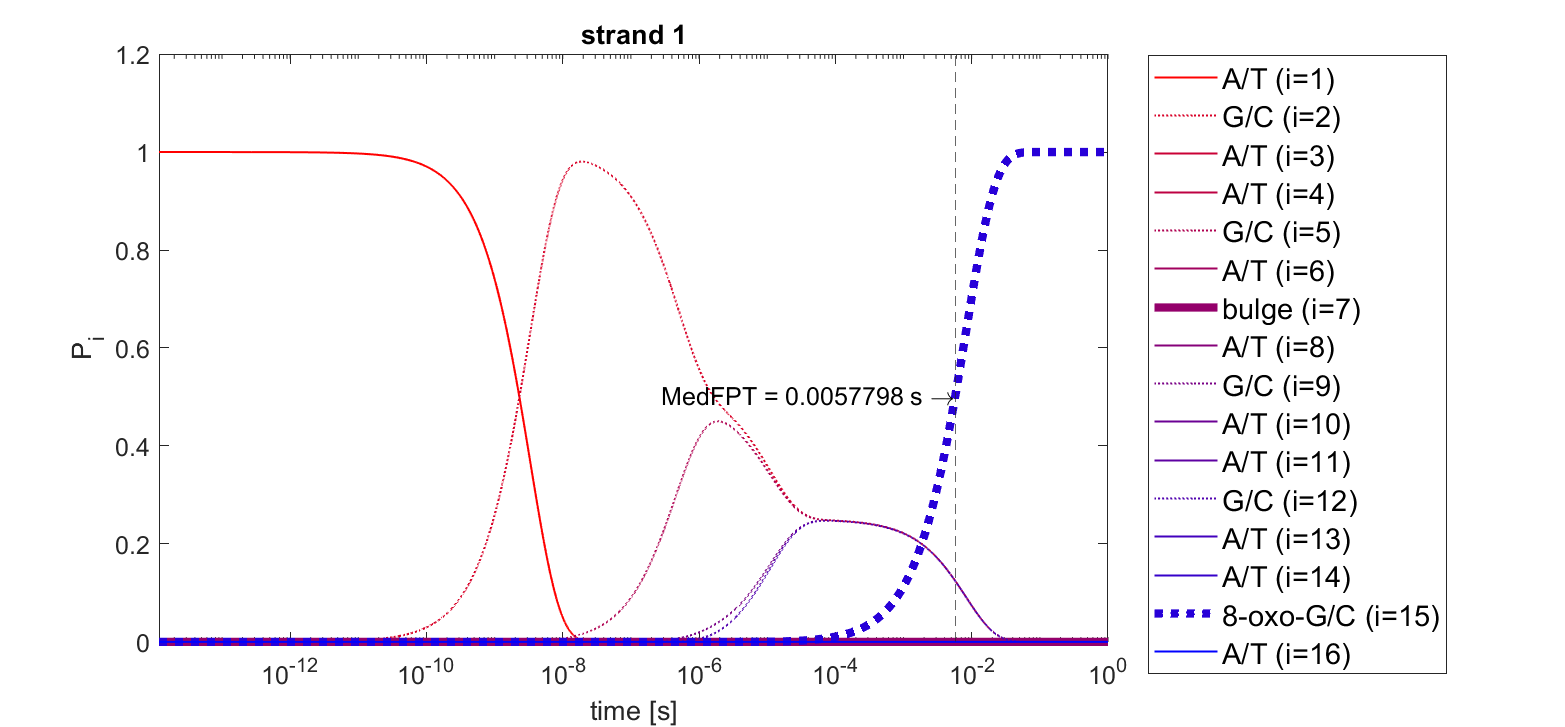}
\caption[Simulated CT in strand 1]{CT  simulation results for strand 1 to determine the median first passage time (MedFPT). The dynamics show the influence of the fast SE transition between G($i=5$) to G($i=9$) which circumvents the slow transition through the bulge.}
\label{fig:strand1}
\end{figure}

One way of quantifying transport in diffusion-to-target reactions is the Median First Passage Time (MedFPT), which describes the median time it takes for a diffusing particle to reach a target position for the first time. We will now use MedFPT to quantify the rate of HT through the co-factor to 8-oxo-G. Since the model assumes 8-oxo-G to be a complete trap for holes (no transitions out of 8-oxo-G), the mathematics of identifying the MedFPT simplifies to identifying the time at which $P_{8oxoG}(t)=0.5$, i.e., the time at which the probability of the hole being located at 8-oxo-G is 0.5. The inverse quantity $k_{CT}=1/\textrm{MedFPT}$ then defines the rate of HT to 8-oxo-G. To illustrate the typical emergent hole dynamics, \autoref{fig:strand1} shows the obtained temporal dynamics of CT for one of the simulated strands. 


\begin{figure}[H]
\centering
\includegraphics[width=0.6\textwidth]{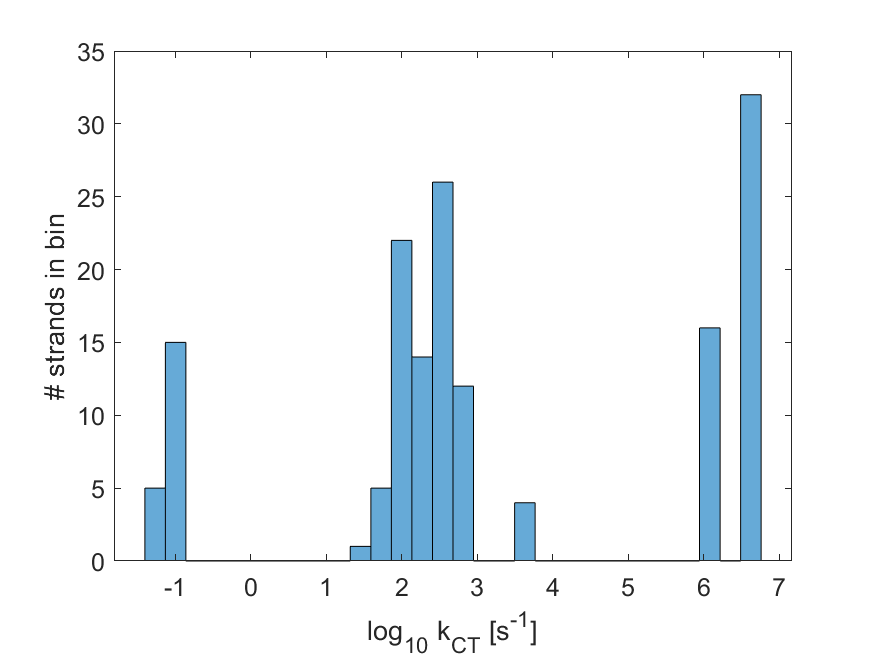}
\caption[Histogram of simulated CT rates]{Histogram showing the distribution of charge transfer rates to 8-oxo-G through the set of 160 strands. Note the three grouping of the CT rates and see text for details.}
\label{fig:CThist}
\end{figure}

The following tendencies emerge from the 160 strand simulations: Primarily, the fastest CT rates are observed when oxoG is located before the bulge, as expected. Second, SE and its distance dependence highly influence CT rates, especially when SE occurs over the bulge. When no SE transitions across the bulge are present (when NN guanines are separated by more than 9 bp), the slowest total CT rates are observed. Mismatches are less impeding than expected for CT rates in the current implementation, possibly due to mismatch CT rates being implemented as a single transition over multiple bases, which is comparable to the rate of multiple thermally induced hole (TIH) jumps across the well-matched strand. Finally, the assumption of no back-transfer across mismatches yields cases where the stopping condition ($P_{8oxoG}(t)=0.99$) is not fulfilled within the allotted simulation time, as charge becomes trapped away from oxoG. This assumption may also lead to mismatches increasing CT rates in cases where oxoG is located after mismatches, as charge is trapped by mismatches at close proximity to oxoG.  \autoref{fig:CThist} shows a histogram of all obtained CT rates for the 160 strands. 

We also calculated the Mean First Passage Time (MeanFPT) for the same charge transport matrices. To calculate the MeanFPTs, we defined $\tilde{\mathbf{K}}$ as the rate matrix in Eq.~\ref{eq:AGAAGA} except with the row and column corresponding to the absorbing 8-oxo-G site removed. The MeanFPT starting from initial site 1 is then given in terms of the matrix inverse as $\tau_\text{MeanFPT}= -\sum_i[\tilde{\mathbf{K}}^{-1}]_{i1}$~\cite{polizzi2016mean}. Generally, we find that MedFPT $\simeq 0.7 \times$ MeanFPT, except for Strands 71 and 73 where MedFPT $\simeq 0.04 \times$ MeanFPT. This difference is due to the special structure of these strands, as they both have the oxoG below the ligation site combined with a small transition probability for the hole to jump past the ligation site. Most holes move directly to the oxoG and get absorbed, while a few jump to the upper part of the strand where they get `stuck' diffusing along the upper part of the strand before eventually moving past the ligation site (bulge) and getting absorbed by the oxoG.

The CT rate data $k_{CT}$ shown in \autoref{fig:CThist} show three distinct groups: (i) strands with $k_{CT}<10^{-1}$ s$^{-1}$, strands with $10^{0}<k_{CT}<10^{4}$, and (iii) strands with $k_{CT}>10^{5}$ s$^{-1}$. The common factor for group (i) is the absence of a SE transition across the center bulge, which yields comparatively slower CT rates. Group (iii) is further divided into two subgroups corresponding to the 8-oxo-G being located in either the first or second substrand of the total strand, and the high observed CT rates are therefore simply due to the hole being initially located close to the 8-oxo-G. 

As discussed in Section 2, an 8-oxo-G location before the ligation site is not preferable as the metabolic processes in this case can proceed with the supply of only the lower-resource oligmer. Without a full replication process, the metabolism thereby loses its information control.  

Variations in group (ii) of strands with 1 s$^{-1}<k_{CT}<10^4$ s$^{-1}$ are due to an assortment of factors such as number of mismatches, mismatch locations, and distance dependence of SE transitions. A more complete correlation analysis remains a future task.

\subsection{Metabolic impact of DNA charge transfer}

As discussed in Section 2, DNA charge transfer is explored as a sequence-dependent process that could be part of the network of protocellular metabolic processes. 
The key issue for our investigations is to understand how the DNA charge transfer process influences the overall metabolic performance. 

\begin{figure}[H]
\centering
\includegraphics[width=1.0\textwidth]{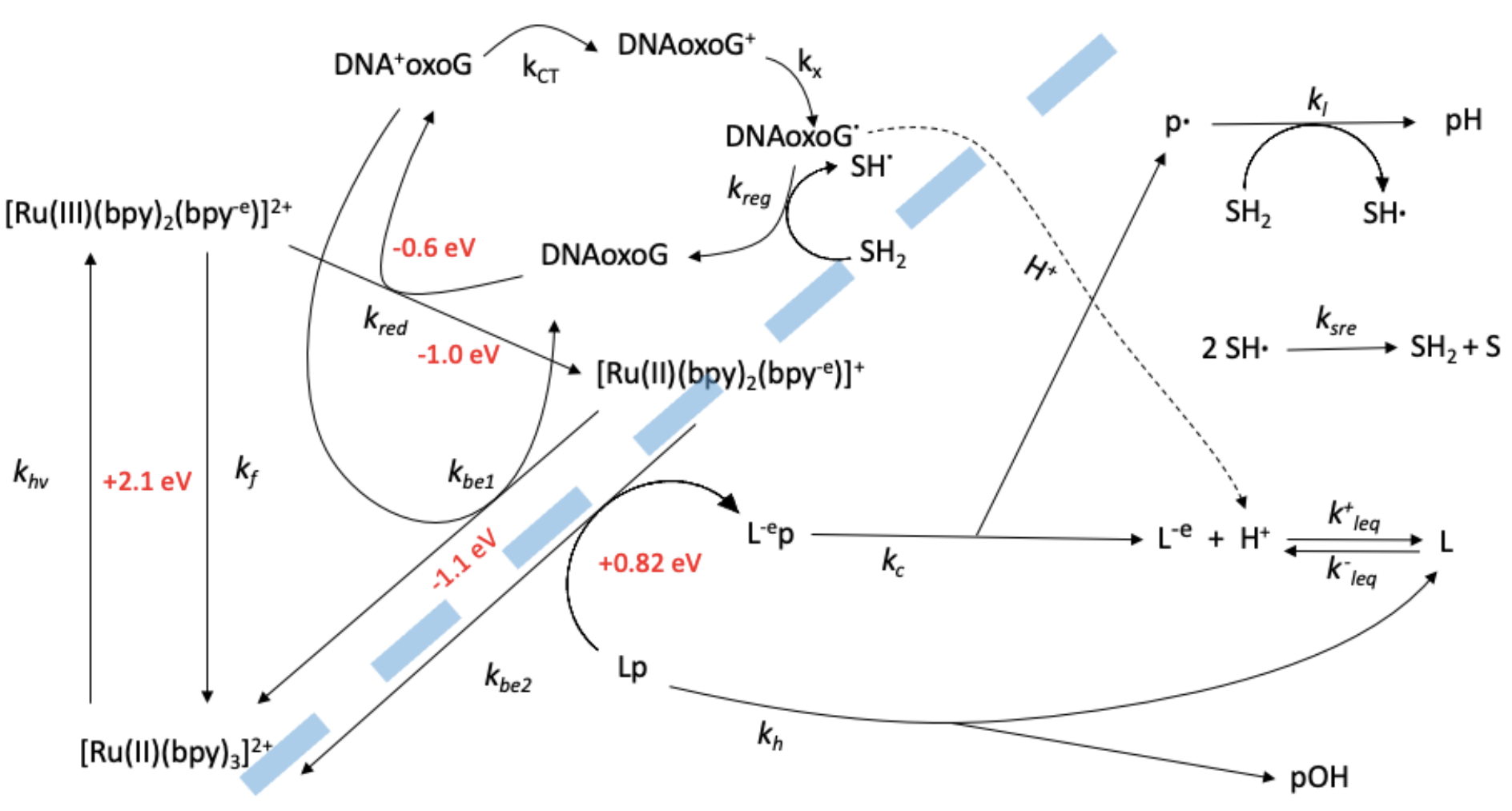}
\caption{Protocellular metabolic reaction network. See text for details. Note that we only need to consider the upper left part (above blue dotted line) of the reactions to evaluate the overall impact of the DNA charge transfer process. Presentation and discussion of the full network and its kinetics can be found in~\cite{declue2009nucleobase,bornebusch2021reaction,bornebusch2021thesis}.} 
\label{fig:MetabolicNetwork}
\end{figure}

DNA charge transfer impacts the rate by which the photo-activated electron can be utilized to transform resource molecules into building blocks. 
Therefore, for our purposes, it suffices to review the initial reactions in the metabolic reaction network, which are summarized in the upper left part of \autoref{fig:MetabolicNetwork} above the blue dotted line. 
A detailed discussion of the full metabolic network can be found in~\cite{declue2009nucleobase,bornebusch2021reaction,bornebusch2021thesis}, although this earlier work did not include the double stranded DNA charge transfer process (missing the step that includes the DNA$^+$oxoG complex).

In \autoref{fig:MetabolicNetwork}, the reaction constant $k_{h\nu} \simeq 0.5$/s determines the photo activation rate. The fluorescent rate constant is $k_{f} \simeq 3 \times 10^6$/s, while the bimolecular reduction rate constant is $k_{red} \simeq 3 \times 10^5$/M s. 
These rate constants are assumed to be constant in the following. 
Furthermore, if we assume that the back reaction for an electron already transferred from the DNA helix to the ruthenium complex can be ignored (no covalent binding between the two reactants), we can set $k_{be1} \simeq 0$. 

The circular reaction in the upper part of \autoref{fig:MetabolicNetwork} starts with the DNA duplex donating an electron to the ruthenium complex to become DNA$^+$oxoG. The rate of this reaction is determined by the bimolecular rate constant $k_{red} \simeq 3 \times 10^5/\textrm{M\,s}$  and the ruthenium complex concentration [RuC]. 
Next, a hole diffuses inside the DNA$^+$oxoG duplex toward the oxoguanine, whose rate constant $k_{CT}$  we estimated for a variety of strand combination in this Section and summarized in \autoref{fig:MetabolicNetwork}. 
The hole is eventually absorbed by the oxoguanine (DNAoxoG$^+$), which thereby loses a proton, as determined by the rate constant $k_x \simeq 10^8 $/s to become DNAoxoG$^\bullet$.
The oxoguanine, now missing a proton, is successively provided with a new proton from a sacrificial proton donor, SH$_2$ as dihydrophenylglycine, whose rate is determined by the bimolecular rate constant $k_{reg} \simeq 10^6$/M s and the concentration [SH$_2$]. This concludes the cycle and makes the DNA duplex ready to donate a new electron to the ruthenium complex. 

The reaction time $t_{circular}$ for complete this circular reaction is determined by the four involved reaction constants $k_{red}\text{[RuC]}, k_{CT}, k_x, k_{reg}\text{[SH$_2$]}$ and can be approximated by $t_{circular} \simeq 1/k_{red}\text{[RuC]} +1/k_{CT} + 1/k_{x} +1/k_{reg}\text{[SH$_2$]}$, so the resulting reaction rate constant can be approximated as 
\begin{equation}
\begin{aligned}
k_{circular} \simeq 
&\{k_{red}\text{[RuC]} \times k_{CT} \times k_x \times k_{reg}\text{[SH$_2$]}\}/   \\
&\{(k_{CT} \times k_x \times k_{reg}\text{[SH$_2$]}) + 
(k_{red}\text{[RuC]} \times k_x \times k_{reg}\text{[SH$_2$]}) + \\
&(k_{red}\text{[RuC]} \times k_{CT} \times k_{reg}\text{[SH$_2$]}) + 
(k_{red}\text{[RuC]} \times k_{CT} \times k_{x})\}.
\end{aligned}
\end{equation}
If we assume that $k_{CT} \ll k_{red}\text{[RuC]}, k_x, k_{reg}\text{[SH$_2$]}$, we have $k_{circular} \simeq k_{CT}$.
In this situation $k_{CT}$ dominates the ‘recharge’ time of the DNA co-factor, determining how fast it can again act as an electron donor for the ruthenium complex.
From the experiments in \cite{declue2009nucleobase} we have [RuC] = 1 mM and [SH$_2$] = 15 mM, which means that
$k_{red}\text{[RuC]} \simeq 3 \times 10^5/\text{Ms} \times 10^{-3}\text{M} \simeq$ 300/s and 
$k_{reg}\text{[SH$_2$]} \simeq 10^6/\text{s} \times 15 \times 10^{-3}\text{M} \simeq 1.5 \times 10^3$/s. 
Thus, $k_{CT} (=50\text{/s}) \ll k_{red}\text{[RuC]} (=300\text{/s})$ is only weakly fulfilled with these concentrations, as the ruthenium complex concentration is a secondary rate limiting factor for $k_{circular}$.

In any event, increasingly smaller $k_{CT}$ means an increasingly smaller rate of usable photo activated electrons determined by
\begin{equation}
k_{usable-photo} \simeq \frac{k_{h\nu} \times k_{CT} }{k_{h\nu} + k_{CT}} 
\end{equation}
This means that depending on whether $k_{CT}$ is the same as $k_{h\nu}$, or is 10 times larger, or is 100 times larger than $k_{h\nu}$, the amount of usable photo activated electrons will be reduced by about 50\%, 10\% and 1\% respectively. 
{\it In the following, we require that $k_{CT} > 100 \times k_{h\nu}$ = 50/s, so that the photo activation process is still the overall rate-limiting reaction.} 

Going back to \autoref{fig:CThist}, recall that the strands in group (ii) with the 8-oxo-G located above the ligation site are acceptable protocellular co-factor candidates. Furthermore, requiring that $k_{CT} > 50$/s should ensure co-factors that make viable protocells.  

The key questions is; how fast, or slow, is the resulting fatty acid production rate if we accept the above requirements? 
From previous experimental studies \cite{declue2009nucleobase,maurer2011interactions}, we measured an initial resulting fatty acid reaction rate constant of about $1.3 \times 10^{-5}$/s, which depends on the aggregate surface composition of picolinium ester (resource molecules) and fatty acid (products). 
More fatty acid membranes lower the reaction rate constant, presumably because the picolinium ester absorbed and integrated into the membrane is less accessible for the metabolic complex. There is, however, one more issue to consider, since a direct hydrolysis of the picolinium ester is also happening, which means that the production of fatty acid can also occur without the protocellular controlled photo-driven conversion. The hydrolysis rate constant for picolinium ester hydrolysis is measured up to about $10^{-6}$/s \cite{declue2009nucleobase}, about an order of magnitude lower than the catalyzed charge transport driven rate constant of about $1.3 \times 10^{-5}$/s, recall the discussion above. 

Assuming a rate constant of half the above and a 1:1 composition of picolinium ester and fatty acid, both at a concentration of 8 mM, we get a fatty acid production rate of about $0.65 \times 10^{-4}$/s $\times 8.0 \times 10^{-3}$M = $5.2 \times 10^{-7}$ M/s. 
Given this rate, $\tau_{double}$ is the time required to produce 8 mM more fatty acid from 8 mM picolinium ester concentration. Using $8 \times 10^{-3} \,\text{M} = \tau_{double} \times 5.2 \times 10^{-7}$ M/s yields $\tau_{double}$ = $1.54 \times 10^{4}$ s or about 4.3 hours.

%% file: 4-Replication.tex
\section{Simulation Exploration of Co-factor Replication}\label{sec:4}

Non-enzymatic molecular replication was introduced in Section~\ref{sec:Flint_design} based on lesion-induced DNA amplification (LIDA). This section explores LIDA for the protocellular information system through simulations of kinetic equations. The full kinetic system is introduced below in \autoref{fig:LIDA_fig} and Eqs.~(\ref{eq:first_LIDA_eq})-(\ref{eq:last_LIDA_eq})~
\cite{engelhardt2020thermodynamics,bornebusch2021reaction,bornebusch2021thesis}. 
The parameters (rate constants) of the kinetic system are computed for a selection of DNA strands using established research on the thermodynamics of DNA structural motifs. Finally, LIDA-based simulations are performed for the 160 unique DNA strands (\autoref{tab:strands}) in Appendix A already used for the CT simulations in Section~\ref{sec:sim_explo} above.

\subsection{The LIDA system for protocells}\label{sec:LIDA_sys}
Alladin-Mustan et al. (2015) reported an exponential amplification of DNA replication under isothermal conditions at room temperature by introducing destabilizing lesions into the DNA duplex (LIDA or Lesion Induced DNA Amplification), such as abasic sites and mismatches \cite{alladin2015achieving}. These lesions decrease the free energy of the duplex's hydrogen bonds to the point where thermal fluctuations at room temperature can deliver the necessary energy to cause dehybridization. Furthermore, Alladin-Mustan et al. (2015) showed that the optimal temperature for DNA amplification can be tuned by the type and number of lesions added.

\begin{figure}[H]
\centering
\includegraphics[scale=0.06]{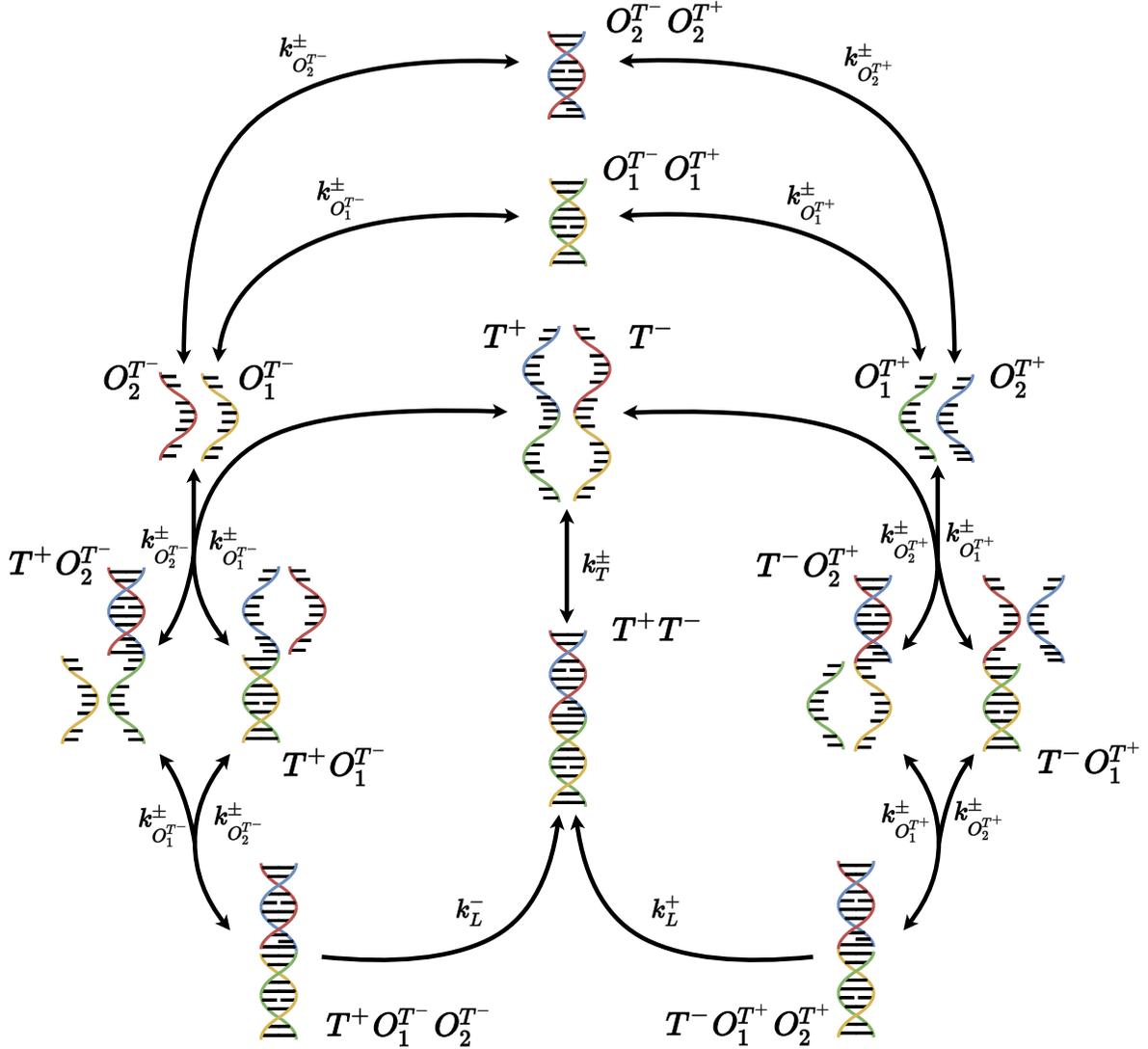}
\caption[LIDA reaction scheme]{The LIDA system consists of four DNA single stranded oligomers ($O^{T^+}_1$, $O^{T^+}_2$, $O^{T^-}_1$ and $O^{T^-}_2$) and two single stranded DNA templates ($T^+$ and $T^-$). Here, the reaction scheme of LIDA with $O^{T^+}_1$, $O^{T^+}_2$, $O^{T^-}_1$ and $O^{T^-}_2$ color coded as green, blue, yellow and red respectively, as well as the forward ($k^+$) and backward ($k^-$) reactions rates of all reactions. The oligomers $O^{T^+}_1$ and $O^{T^-}_1$ are near complementary (complementary except for mismatches and bulges; respectively illustrated by partially or fully unconnected nucleobases). Likewise, $O^{T^+}_2$ and $O^{T^+}_2$ are near complementary. These oligomers can thus form the short duplexes $O^{T^-}_2O^{T^+}_2$ and $O^{T^-}_1O^{T^+}_1$. Templates are longer single-stranded oligomers of the same sequence as two ligated short oligomers, i.e., ligating $O_1^{T^+}$ and $O_2^{T^+}$ forms $T^+$ and likewise ligating $O_1^{T^-}$ and $O_2^{T^-}$ forms $T^-$. The templates $T^+$ and $T^-$ are thus also near complementary. The oligomers $O^{T^\pm}_1$ and $O^{T^\pm}_2$ can hybridize with the templates $T^\mp$ forming four possible strands with sticky ends ($T^+O_2^{T^-}$, $T^+O_1^{T^-}$, $T^-O_1^{T^+}$ and $T^-O_2^{T^+}$). These four strands can further hybridize with oligomers that are near complementary to their sticky ends, forming the unligated strands $T^+O_1^{T^-}O_2^{T^-}$ and $T^-O_1^{T^+}O_2^{T^+}$. A reaction can irreversibly ligate these strands into the DNA duplex $T^+T^-$, which can dehybridize yielding new templates. The LIDA system thereby amplifies the concentration of templates while oligomers are present; in other words, LIDA makes DNA replication of a subset of strands possible without the complex mechanisms used in biological life.}
\label{fig:LIDA_fig}
\end{figure}

The total reaction scheme of the LIDA system is shown in \autoref{fig:LIDA_fig}, but can be summarized by the reaction equation
\begin{equation}
   O_1^- + O_2^- + O_1^+ + O_2^+\xrightarrow{\,\,\,T^+\,\textrm{or}\,T^-} T^+T^-
\end{equation}
meaning LIDA is the conversion of four short oligomers ($O_1^- + O_2^- + O_1^+ + O_2^+$
) into the full DNA strand in duplex. This reaction is catalyzed by either simplex ($T^+$ or $T^-$) of the full DNA strand\footnote{A non-catalyzed reaction via pseudo blunt-ended ligation may also occur, but much less frequent than the catalyzed reaction \cite{alladin2015achieving}. Furthermore, in the currently lab-achieved DNA amplification by LIDA, the system is also catalyzed by enzymes as noted in the Supporting Information for Alladin-Mustan et al. (2015) \cite{alladin2015achieving}.}, and since the duplex and simplex forms can be converted into each other via hybridization/dehybridization, the total system is an autocatalytic production of double stranded DNA from four single-stranded oligomers.

It should be noted that the 5' end of the oligomers need to be imidazole activated for the ligation to occur, and if such hydrolysis occurs the deactivated oligomer is no longer able to take part in a ligation reaction. 
However, it is known from experiments \cite{Kawamura_Maeda_2007} that the hydrolysis rate constant $k_{hyd-imp}$ of the imidazole activated oligomers is significantly lower than the ligation rate constant as well as the involved hybridization rates. Therefore, the imidazole hydrolysis process is not explicitly included in our LIDA simulations. 

In previous experimental work \cite{cape2012phototriggered} we have demonstrated that a picolil protection group can be attached to the 3' end of the oligomers, and that this protection group can be cleaved off using the same ruthenium-complex-based photo-activation process that generates fatty acids from picolin ester as discussed in previous Section~\ref{sec:sim_explo}. 
For simplicity, the details of this process are also not included in the LIDA simulation. 

The kinetic equations of the LIDA system are shown in Equations (\ref{eq:first_LIDA_eq})-(\ref{eq:last_LIDA_eq}). The equations are taken from \cite{bornebusch2021thesis} with new additional terms, due to the possibility that the near complementary oligomers form short duplexes ($O_1^{T^+}O_1^{T^-}$ and $O_2^{T^+}O_2^{T^-}$). This system of coupled differential equations is later used in Section \ref{sec:LIDAsims} for \textit{in silico} simulations of the LIDA system in order to obtain the replication dynamics of a selection of protocell co-factors.

\begingroup
\allowdisplaybreaks
\begin{align}
    \dd{\yI}{t} =& \kminusoItplus \yV + \kminusoItplus \yVII- \kplusoItplus \yXII \yI \label{eq:first_LIDA_eq}\\ &- \kplusoItplus \yVI \yI - \kplusoI \yIII \yI + \kminusoI \yXIV \nonumber \\  
    \dd{\yII}{t} =& \kminusoIItplus \yVI + \kminusoIItplus \yVII - \kplusoIItplus \yXII \yII\\ &- \kplusoIItplus \yV \yII - \kplusoII \yIV \yII + \kminusoII \yXV \nonumber \\  
    \dd{\yIII}{t} =& \kminusoItminus \yIX + \kminusoItminus \yXI - \kplusoItminus \yXIII \yIII\\ &- \kplusoItminus \yX \yIII - \kplusoI \yIII \yI + \kminusoI \yXIV \nonumber \\ 
    \dd{\yIV}{t} =& \kminusoIItminus \yX + \kminusoIItminus \yXI - \kplusoIItminus \yXIII \yIV\\ &- \kplusoIItminus \yIX \yIV - \kplusoII \yIV \yII + \kminusoII \yXV \nonumber \\ 
    \dd{\yV}{t} =& \kplusoItplus \yXII \yI + \kminusoIItplus \yVII - \kminusoItplus \yV\\ &- \kplusoIItplus \yV \yII \nonumber \\ 
    \dd{\yVI}{t} =& \kplusoIItplus \yXII \yII + \kminusoItplus \yVII - \kminusoIItplus \yVI\\ &- \kplusoItplus \yVI \yI \nonumber \\ 
    \dd{\yVII}{t} =& \kplusoItplus \yVI \yI + \kplusoIItplus \yV \yII - \kminusoItplus \yVII\\ &- \kminusoIItplus \yVII - \kl \yVII \nonumber \\ 
    \dd{\yVIII}{t} =& \kl \yVII + \kl \yXI + \ktplus \yXII \yXIII\\ &- \ktminus \yVIII \nonumber \\ 
    \dd{\yIX}{t} =& \kplusoItminus \yXIII \yIII + \kminusoIItminus \yXI - \kminusoItminus \yIX\\ &- \kplusoIItminus \yIX \yIV \nonumber \\ 
    \dd{\yX}{t} =& \kplusoIItminus \yXIII \yIV + \kminusoItminus \yXI - \kminusoIItminus \yX\\ &- \kplusoItminus \yX \yIII \nonumber \\ 
    \dd{\yXI}{t} =& \kplusoItminus \yX \yIII + \kplusoIItminus \yIX \yIV - \kminusoItminus \yXI\\ &- \kminusoIItminus \yXI - \kl \yXI \nonumber \\ 
    \dd{\yXII}{t} =& \kminusoItplus \yV + \kminusoIItplus \yVI + \ktminus \yVIII - \kplusoItplus \yXII \yI\\ &- \kplusoIItplus \yXII \yII - \ktplus \yXII \yXIII \nonumber \\ 
    \dd{\yXIII}{t} =& \kminusoItminus \yIX + \kminusoIItminus \yX + \ktminus \yVIII - \kplusoItminus \yXIII \yIII\\ &- \kplusoIItminus \yXIII \yIV - \ktplus \yXII \yXIII \nonumber \\ 
    \dd{\yXIV}{t} =& \kplusoI \yIII \yI - \kminusoI \yXIV\\ 
    \nonumber \\
    \dd{\yXV}{t} =& \kplusoII \yIV \yII - \kminusoII \yXV \label{eq:last_LIDA_eq} 
\end{align}
\endgroup

\subsection{Kinetic rates in LIDA}\label{sec:LIDA_kin}

All reactions in the LIDA kinetic system are of the two types: 
\begin{equation}
    A+B \xrightleftharpoons[k^-]{k^+} C\,\,\,\,\,\,\textrm{or}\,\,\,\,\, A\xrightleftharpoons[k^-]{k^+} B
\end{equation}
except for the irreversible ligation reaction, where the observed rate from \cite{cape2012phototriggered} is used. Equilibrium thermodynamics gives a relation between the standard Gibbs free energy $\Delta G^{\degree}$ and the equilibrium constant $K$:
\begin{equation}\label{eq:deltaG(K)}
    \Delta G^{\degree} = -RT \ln{K},
\end{equation}
where $R=1.987$ cal $\cdot$ K$^{-1} \cdot$ mol$^{-1}$ is the gas constant and $T$ is the temperature in Kelvin. Using the definition of the equilibrium constant $K = k^+/k^-$, we can obtain: 
\begin{equation}\label{eq:offrate}
    K = \frac{k^+}{k^-} = \exp\left(-\frac{\Delta G^{\degree}}{RT}\right) \Rightarrow k^- = k^+ \exp\left(\frac{\Delta G^{\degree}}{RT}\right)
\end{equation}
Thus, if we know the on-rate (forward rate constant) $k^+$, the temperature $T$, and the standard Gibbs free energy 
$\Delta G^\circ$, we can compute the off-rate (reverse rate constant) $k^-$. 
For simplicity, we assume all on-rates to be $2\cdot10^{7}$ s$^{-1}$ \cite{Christiensen_et_al_2001}, while 
$\Delta G^\circ$ is computed using the method and parameters introduced by SantaLucia and Hicks (2004) \cite{santalucia2004thermodynamics}. 
From this, we compute off-rates ($\kminusoItminus$, $\kminusoItplus$, $\kminusoIItminus$, $\kminusoIItplus$ and $\ktminus$) for LIDA simulations.

As mentioned, protocellular co-factors contain destabilizing mismatches to obtain faster off-rates, and specifically, the 160 investigated strands (see Table \ref{tab:strands}) contain G/A and G/T mismatches. Mismatch contributions to $\Delta G^{\degree}$ at 37$\degree$C are listed in Table 2 in \cite{santalucia2004thermodynamics}. For our LIDA simulations (in which the temperature is set to $26\degree$C), we must instead use $\Delta H^{\degree}$ and $\Delta S^{\degree}$ to compute $\Delta G^{\degree}_{26\degree}$ for G/T and G/A mismatches, which can be found in \cite{allawi1997thermodynamics,allawi1998nearest}. 

The investigated protocell genomes also contain a highly destabilizing bulge, which forms near the center of the templates in duplex ($T^+T^-$), replacing the abasic site as the main destabilizing element in the original LIDA system~\cite{alladin2015achieving}. This is required to obtain faster dehybidization rates for the $T^+T^-$ duplex ($k^-_{T}$). SantaLucia and Hicks (2004) state that bulge loops of size 1 contribute $\Delta G^{\degree}_{\textrm{bulge},37\degree}=4$ kcal$\cdot$mol$^{-1}$. A negative correction should be added to obtain $\Delta G^{\degree}_{\textrm{bulge},26\degree}$, but since we do not know the size of this correction, we simply use $\Delta G^{\degree}_{\textrm{bulge},37\degree}=4$ kcal$\cdot$mol$^{-1}$ in our simulations.

Dangling ends (DEs) also contribute to the hybridization free energy. Including DE contributions can have a significant effect on the LIDA replication dynamics, particularly when computing off-rates for single oligomer/templates duplexes ($T^-O^{T^+}_1$, $T^-O^{T^+}_2$, $T^+O^{T^-}_1$ and $T^+O^{T^-}_2$). DEs are predominantly stabilizing, while contributions from single nucleotide range from $+0.48$ to $-0.96$ kcal $\cdot$ mol$^{-1}$ depending on sequence and orientation (5'-3' or 3'-5') \cite{santalucia2004thermodynamics}. Nearly all of the DE contribution comes from the first DE, with additional nucleotides contributing less than 0.2 kcal $\cdot$ mol$^{-1}$ \cite{santalucia2004thermodynamics}. 

\begin{figure}[H]
\centering
\includegraphics[scale=0.75]{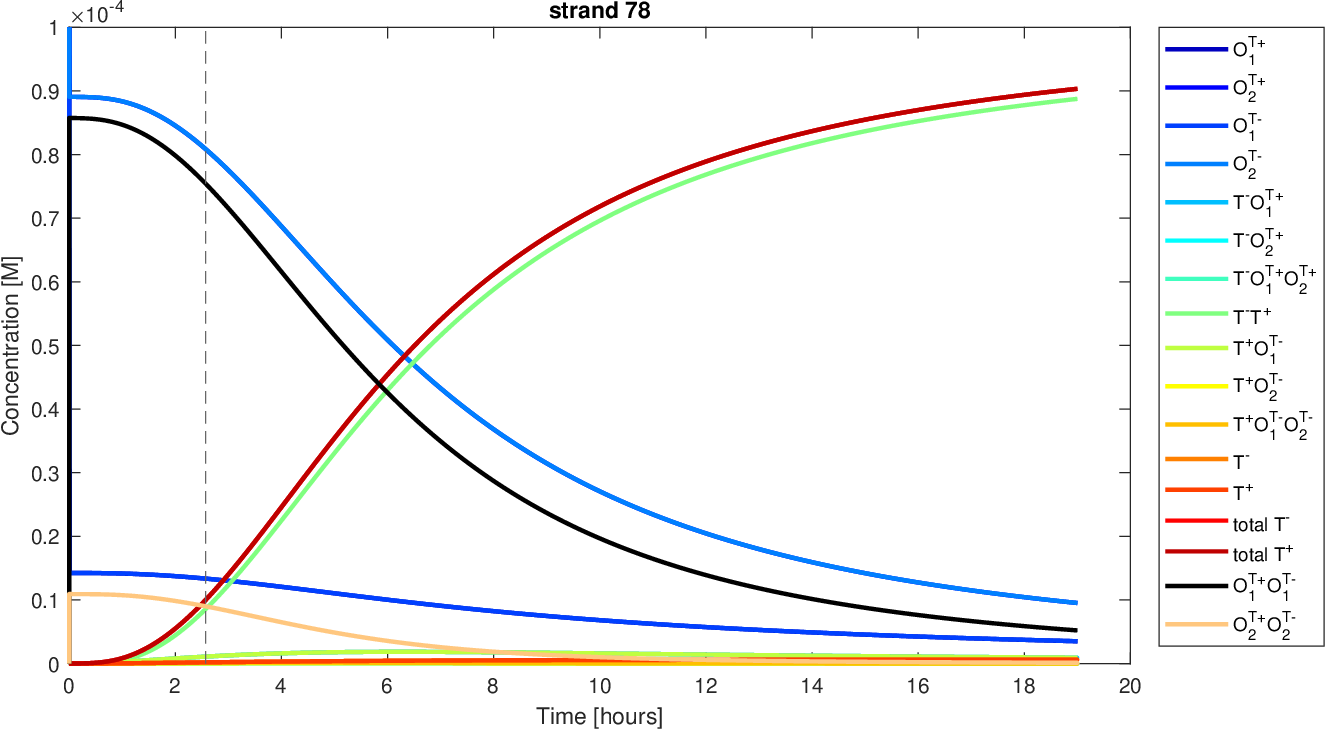}
\caption{Typical LIDA simulation where the graph color corresponds to the molecular complex indicated at the right side of the figure. Note how the two oligomer duplexes are formed at the onset, with $O^{T+}_1 O^{T-}_1$ (black line) in significantly higher concentration (higher hybridization energies) than the $O^{T+}_2 O^{T-}_2$ (beige line) duplex. These duplex concentrations thereafter decrease as the concentration of the total number of full plus-and-minus strands increases exponentially (red line) where they are mostly in a duplex configuration (green line, right below the red line). The light blue curve (above the black line) shows the kinetics of the free $O^{T+}_2$ and $O^{T-}_2$ oligomers, while the dark blue curve (above the beige line) shows the kinetics of the free $O^{T+}_1$ and $O^{T-}_1$ oligomers. The many intermediary template-oligomer complexes as well as the free plus and minus templates are at low concentrations throughout the simulation and therefore difficult to distinguish at this concentration resolution.}
\label{fig:typical_LIDA}
\end{figure}

For our implementation, we included a single correcting factor to the off-rates in the terms of dehybridization of the short oligomers from the strands with dangling ends, i.e., $T^-O^{T^+}_1$, $T^-O^{T^+}_2$, $T^+O^{T^-}_1$ and $T^+O^{T^-}_2$. SantaLucia and Hicks (2004) state that the average 5'-dangling end contributes $\Delta G^{\degree}_{5'}=-0.45$ kcal $\cdot$ mol$^{-1}$, while the average 3'-dangling end contributes $\Delta G^{\degree}_{3'}=-0.29$ kcal $\cdot$ mol$^{-1}$. We therefore chose the average of $\Delta G^{\degree}_{5'}$ and $\Delta G^{\degree}_{3'}$ as our correcting factor to the hybridization free energy: $\Delta G^{\degree}_{DE}=(-0.45-0.29)/2=-0.37$ kcal $\cdot$ mol$^{-1}$. Though a more rigorous calculation of the sequence dependence of DE contribution is possible, we use this single correcting factor to test the sensitivity of the LIDA kinetic system to DE contributions. When comparing LIDA simulations (as specified in Section \ref{sec:LIDAsims}) including and excluding DE contributions, we observe a decrease in replication times of protocellular co-factors via LIDA when DE contributions are included. A typical LIDA simulation is shown in Fig. \ref{fig:typical_LIDA}.

We note that there are inconsistencies between
the thermodynamic parameters computed using the method by SantaLucia and Hicks and the thermodynamic parameters computed by NUPACK \cite{bornebusch2021thesis}. All thermodynamic parameters used to compute off-rates for LIDA simulations in this paper are listed in Table \ref{tab:SLH_params}. 

\begin{table}[H]
\centering
\begin{tabular}{lll}
\hline
\multicolumn{1}{|l|}{NN base pair}         & \multicolumn{1}{l|}{$\Delta H^{\degree}$ {[}kcal$\cdot$mol$^{-1}${]}} & \multicolumn{1}{l|}{$\Delta S^{\degree}$ {[}cal$\cdot$mol$^{-1}\cdot$K$^{-1}${]}} \\ \hline
\multicolumn{1}{|l|}{AA/TT}                & \multicolumn{1}{l|}{-7.6}                           & \multicolumn{1}{l|}{-21.3}                                      \\
\multicolumn{1}{|l|}{TT/AA}                & \multicolumn{1}{l|}{-7.6}                           & \multicolumn{1}{l|}{-21.3}                                      \\
\multicolumn{1}{|l|}{AT/TA}                & \multicolumn{1}{l|}{-7.2}                           & \multicolumn{1}{l|}{-20.4}                                      \\
\multicolumn{1}{|l|}{TA/AT}                & \multicolumn{1}{l|}{-7.2}                           & \multicolumn{1}{l|}{-21.3}                                      \\
\multicolumn{1}{|l|}{GT/CA}                & \multicolumn{1}{l|}{-8.4}                           & \multicolumn{1}{l|}{-22.4}                                      \\
\multicolumn{1}{|l|}{GA/CT}                & \multicolumn{1}{l|}{-8.2}                           & \multicolumn{1}{l|}{-22.4}                                      \\
\multicolumn{1}{|l|}{AG/TC}                & \multicolumn{1}{l|}{-8.2}                           & \multicolumn{1}{l|}{-22.4}                                      \\
\multicolumn{1}{|l|}{TG/AC}                & \multicolumn{1}{l|}{-8.4}                           & \multicolumn{1}{l|}{-22.4}                                      \\ \hline
                                           &                                                     &                                                                 \\ \hline
\multicolumn{1}{|l|}{Mismatch type}        & \multicolumn{1}{l|}{$\Delta H^{\degree}$ {[}kcal$\cdot$mol$^{-1}${]}} & \multicolumn{1}{l|}{$\Delta S^{\degree}$ {[}cal$\cdot$mol$^{-1}\cdot$K$^{-1}${]}} \\ \hline
\multicolumn{1}{|l|}{AG/TA}                & \multicolumn{1}{l|}{-0.7}                           & \multicolumn{1}{l|}{-2.3}                                       \\
\multicolumn{1}{|l|}{AG/TT}                & \multicolumn{1}{l|}{1}                              & \multicolumn{1}{l|}{0.9}                                        \\
\multicolumn{1}{|l|}{AT/TG}                & \multicolumn{1}{l|}{-2.5}                           & \multicolumn{1}{l|}{-8.3}                                       \\
\multicolumn{1}{|l|}{AA/TG}                & \multicolumn{1}{l|}{-0.6}                           & \multicolumn{1}{l|}{-2.3}                                       \\ \hline
                                           &                                                     &                                                                 \\ \cline{1-2}
\multicolumn{1}{|l|}{Lesion type}          & \multicolumn{1}{l|}{$\Delta G^{\degree}_{37\degree}$ {[}kcal$\cdot$mol$^{-1}${]}}                             &                                                                 \\ \cline{1-2}
\multicolumn{1}{|l|}{bulge loop of size 1} & \multicolumn{1}{l|}{4}                              &                                                                 \\ \cline{1-2}
\end{tabular}
\caption[Thermodynamic parameters]{The thermodynamic parameters used to compute free energies of DNA hybridization for LIDA simulations. All parameters have been found in \cite{santalucia2004thermodynamics,allawi1997thermodynamics,allawi1998nearest} and is denoted using the same notation, where e.g. AA/TT corresponds to 5'-AA-3' hybridized with 3'-TT-5'.}
\label{tab:SLH_params}
\end{table}

\subsection{LIDA simulations for a selection of strands}\label{sec:LIDAsims}

By using DNA strand-specific kinetic off-rates from thermodynamic parameters and the kinetic equations of LIDA, we simulate co-factor replication for the same 160 DNA strands that we consider in Section \ref{sec:CTsims} when studying charge transport. This is done by splitting all strands into oligomers of equal length (9 bp or 3 substrands). For each strand, the hybridization free energies, Eq.~(\ref{eq:deltaG(K)}), are computed for the short duplexes ($O_1^{T-}O_1^{T+}$ and $O_2^{T-}O_2^{T+}$) and the templates in duplex ($T^+T^-$) using the method of SantaLucia and Hicks \cite{santalucia2004thermodynamics}. Note that the bulge only contributes to the free energy for $T^+T^-$. The thermodynamic contribution of 8-oxo-G is unknown, but Gasper and Schuster (1997) note that replacement of G by 8-oxo-G has little or no effect on the global duplex structure \cite{gasper1997intramolecular}. We therefore assume that 8-oxo-G can be treated as a G in calculations of hybridization free energy.

Inserting these free energies into Eq.~(\ref{eq:offrate}) with $T=26\degree$C yields off-rates, which can be implemented in the LIDA kinetic system, Eqs.~(\ref{eq:first_LIDA_eq})-(\ref{eq:last_LIDA_eq}). We assume that $k^-_{O_1^{T-}}=k^-_{O_1^{T+}}$ and $k^-_{O_2^{T-}}=k^-_{O_2^{T+}}$, although there may be discrepancies. The resulting kinetic systems are simulated using the numerical solver \texttt{ode15s} in MATLAB R2021b. 
If the simulation time reaches 100 hours, the simulation is stopped.

\begin{figure}[H]
\centering
\includegraphics[width=0.6\textwidth]{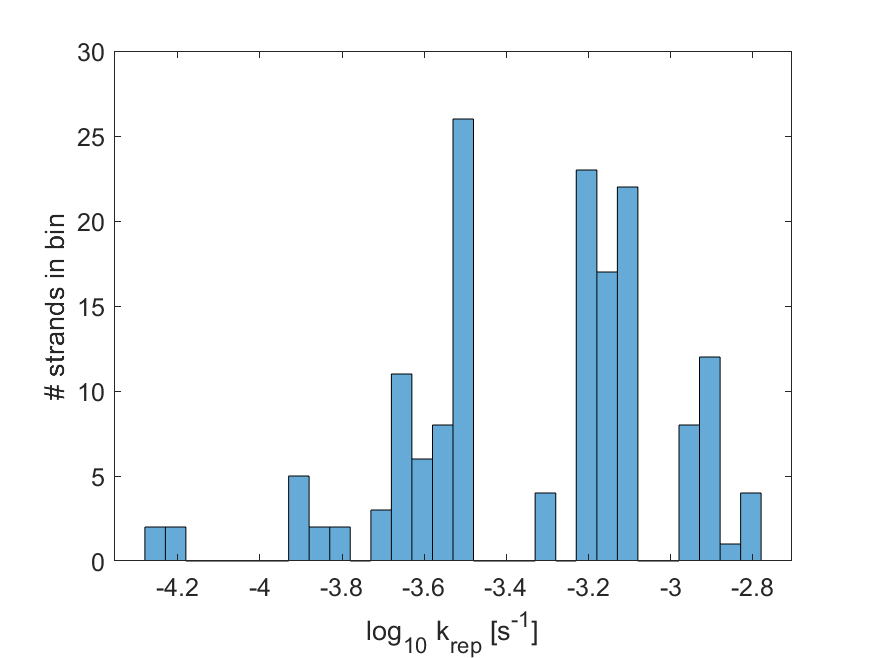}
\caption[Histogram of simulated replication times]{Histogram showing the distribution of LIDA simulated replication rates $k_{rep}$ for the set of 160 strands. Initial concentrations used in the LIDA simulations for the 160 strands are $[O_1^{T^+}]=[O_1^{T^-}]=[O_2^{T^+}]=[O_2^{T^-}]=100$ µM and $[T^-]=14$ nM with all other initial concentrations equal to zero M. We find typical doubling times for $k_{rep}$ between $10^{-4}$/s and $10^{-3}$/s where the corresponding doubling times are ${\tau}_{rep} = 6.70 \times 10^3$ s and $6.70 \times 10^2$ s respectively.
}
\label{fig:tau_rep_hist}
\end{figure}

We estimate the replication rate constant $k_{rep}$ by (i) measuring the time $\tau_{10\%}$ at which DNA amplification has reached 10\% of completion ($= 10\times 10^{-6}$ M) and (ii) assuming that the initial growth phase can be approximated as exponential. 
Then $k_{rep}$ can be estimated from $10\times 10^{-6}\,\text{M} = 14 \times 10^{-9}\,\text{M}\times \exp(k_{rep}\times \tau_{10\%})$ as $k_{rep} = \ln[(10/14)\times 10^3]/\tau_{10\%}.$
The doubling or replication time $\tau_{rep}$ can be estimated from $A\exp(k_{rep} \times \tau_{rep}) = 2A$ so $\tau_{rep} = \ln 2/k_{rep}$. 
\autoref{fig:tau_rep_hist} shows a histogram of the replication rates obtained for the 160 strands. 

The estimated replication rate constant $k_{rep}$ data in \autoref{fig:tau_rep_hist} show multiple semi-distinct groupings, with typical doubling times ranging between about 10 minutes and a couple of hours, with the slowest about 3.8 hours. Note that this is faster then the estimated vesicle doubling time of 4.3, we found in Section 3. 






Finally, we need to check the above estimated replication rate constants with the imidazole hydrolysis rate constant $k_{hyd-imp}$ that is measured to be between $4.78\times 10^{-6}$/s and $1.04\times 10^{-5}$/s at room temperature and depending on the ion concentration \cite{Kawamura_Maeda_2007}. 
DNA sequences with an estimated overall replication rate constant below $k_{hyd-imp}$ are not viable as the replication process will be inhibited by the decay of activated oligomers. 

Since most of our estimated replication rate constants satisfy $k_{rep} > 10^{-4}$/s, which is an order of magnitude larger than the largest reported hydrolysis rate constants $k_{hyd-imp} \sim 10^{-5}$/s, {\it as a first approximation we can assume that a protocellular replication is viable if we require $k_{rep} > 10^{-4}$/s.}

%% file: 5-ConnectingMetabolism.tex
\section{Connecting results from charge transport and replication simulations}
\label{sec:5}

The data on the replication rate and CT rates from the simulations performed in Sections \ref{sec:CTsims} and \ref{sec:LIDAsims} provide us with information about how the protocell fitness depends on the co-factor: Assuming everything else being equal, higher replication and/or CT rates mean higher fitness.

\begin{figure}[H]
\centering
\hspace*{-1cm}
\includegraphics[scale=0.7]{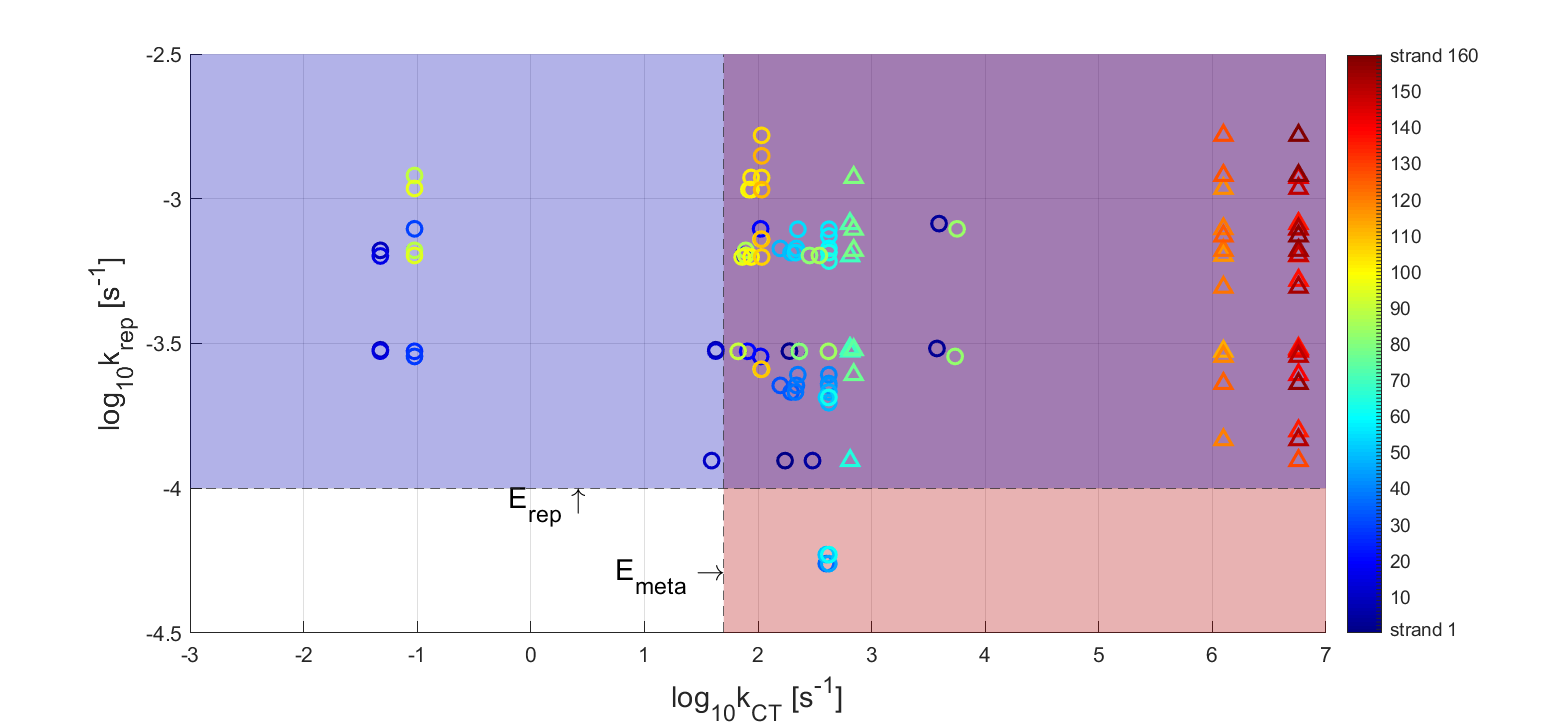}
\caption{The replication rate $k_{rep}$ versus the CT rate $k_{CT}$ for all 160 examined strands (see \autoref{tab:strands} for a listing of all strands). Color bar indicates which string corresponds to which data point; rings indicate that oxoG is above the ligation site, while triangles indicate that oxoG is below the ligation site. Everything else being equal, both the replication rate and the charge transfer rate have to be above some minimal value to ensure protocellular survival. This is indicated by the $E_{meta}$ and $E_{rep}$ values and their respectively pale red and pale blue areas. Thus, the purple area indicates both viable metabolic and replication rates. See text for details. 
}
\label{fig:E(x_meta+rep)}
\end{figure}

Obviously, a lower bound must exist for both the co-factor charge transfer and replication rates if the protocellular system should be able to grow and survive. If either the CT rate or the replication rate are too slow, a variety of degradation processes would exceed the production processes and the protocellular system would either not be able to form or it would disintegrate.  
Recall that from our analysis in Section \ref{sec:CTsims}, we require that $k_{CT} > 50$/s to ensure that the CT rate constant is well above the rate-limiting photo-activation rate constant. From our analysis in Section \ref{sec:LIDAsims}, we require that $k_{rep} > 10^{-4}$/s to ensure that the replication rate constant is well above the associated hydrolysis rates. These conditions ensure that charge-transport and replication processes can support viable protocells. 
Also, recall the discussion in Section \ref{sec:Flint_design}, where for simplicity we required that 8-oxo-G is located above the ligation site, which disqualifies strand numbers 65-80 and 113 to 160 (see \autoref{tab:strands} for details). Thus, 95 out of the 160 strands satisfy this condition.
In \autoref{fig:E(x_meta+rep)} we show the replication rate $k_{rep}$ versus the CT rate $k_{CT}$ for all 160 examined strands, where the `viable area' is indicated. 
In \autoref{tab:strands}, the 160 strands are listed where their replication and charge transfer properties are encoded using similar colors as used in \autoref{fig:E(x_meta+rep)}. 

It should be emphasized that as we combine the results from the CT and LIDA simulations, no interaction dynamics are taken into consideration. 
This includes interactions between the replication and the charge transfer processes, as well as the feeding, growth, and container division processes. 
Further, our simulations assume a constant environment (temperature, pH, salt, etc.). 
Thus, a combination of the results found under the CT and LIDA simulation can only be viewed as a simplified representation of what a full systems simulation would have given. 
This means that the viable area shown in the reaction rate diagram in \autoref{fig:E(x_meta+rep)} should be viewed as a simplified approximation.

\begin{table}[H]
\footnotesize
    \begin{tabular}{|l|l|}
        \hline
        Strand & \multicolumn{1}{|p{1.8cm}|}{\centering Substrand\\ sequence} \\ \hline
\rowcolor{p} 1 & 1  1  6  1  1  5 \\
\rowcolor{p} 2 & 1  1  6  1  2  5 \\
\rowcolor{p} 3 & 1  1  6  1  3  5 \\
\rowcolor{p} 4 & 1  1  6  1  4  5 \\
\rowcolor{p} 5 & 1  1  6  1  5  1 \\
\rowcolor{p} 6 & 1  1  6  1  5  2 \\
\rowcolor{b} 7 & 1  1  6  2  1  5 \\
\rowcolor{b} 8 & 1  1  6  2  2  5 \\
\rowcolor{b} 9 & 1  1  6  2  5  1 \\
\rowcolor{b} 10 & 1  1  6  2  5  2 \\
\rowcolor{b} 11 & 1  1  6  5  1  1 \\
\rowcolor{b} 12 & 1  1  6  5  1  2 \\
\rowcolor{g} 13 & 1  1  6  5  1  3 \\
\rowcolor{g} 14 & 1  1  6  5  1  4 \\
\rowcolor{b} 15 & 1  1  6  5  2  1 \\
\rowcolor{b} 16 & 1  1  6  5  2  2 \\
\rowcolor{p} 17 & 1  2  6  1  1  5 \\
\rowcolor{p} 18 & 1  2  6  1  2  5 \\
\rowcolor{p} 19 & 1  2  6  1  3  5 \\
\rowcolor{p} 20 & 1  2  6  1  4  5 \\
\rowcolor{p} 21 & 1  2  6  1  5  1 \\
\rowcolor{p} 22 & 1  2  6  1  5  2 \\
\rowcolor{b} 23 & 1  2  6  2  1  5 \\
\rowcolor{b} 24 & 1  2  6  2  2  5 \\
\rowcolor{b} 25 & 1  2  6  2  5  1 \\
\rowcolor{b} 26 & 1  2  6  2  5  2 \\
\rowcolor{b} 27 & 1  2  6  5  1  1 \\
\rowcolor{b} 28 & 1  2  6  5  1  2 \\
\rowcolor{b} 29 & 1  2  6  5  1  3 \\
\rowcolor{b} 30 & 1  2  6  5  1  4 \\
\rowcolor{b} 31 & 1  2  6  5  2  1 \\
\rowcolor{b} 32 & 1  2  6  5  2  2 \\
\rowcolor{p} 33 & 1  3  6  1  1  5 \\
\rowcolor{p} 34 & 1  3  6  1  2  5 \\
\rowcolor{g} 35 & 1  3  6  1  3  5 \\
\rowcolor{r} 36 & 1  3  6  1  4  5 \\
\rowcolor{p} 37 & 1  3  6  1  5  1 \\
\rowcolor{p} 38 & 1  3  6  1  5  2 \\
\rowcolor{p} 39 & 1  3  6  2  1  5 \\
\rowcolor{p} 40 & 1  3  6  2  2  5 \\
\hline \end{tabular} \hspace{0.25cm} \begin{tabular}{|l|l|} \hline Strand & \multicolumn{1}{|p{1.8cm}|}{\centering Substrand\\ sequence} \\ \hline
\rowcolor{p} 41 & 1  3  6  2  5  1 \\
\rowcolor{p} 42 & 1  3  6  2  5  2 \\
\rowcolor{p} 43 & 1  3  6  5  1  1 \\
\rowcolor{p} 44 & 1  3  6  5  1  2 \\
\rowcolor{g} 45 & 1  3  6  5  1  3 \\
\rowcolor{r} 46 & 1  3  6  5  1  4 \\
\rowcolor{p} 47 & 1  3  6  5  2  1 \\
\rowcolor{p} 48 & 1  3  6  5  2  2 \\
\rowcolor{p} 49 & 1  4  6  1  1  5 \\
\rowcolor{p} 50 & 1  4  6  1  2  5 \\
\rowcolor{r} 51 & 1  4  6  1  3  5 \\
\rowcolor{p} 52 & 1  4  6  1  4  5 \\
\rowcolor{p} 53 & 1  4  6  1  5  1 \\
\rowcolor{p} 54 & 1  4  6  1  5  2 \\
\rowcolor{p} 55 & 1  4  6  2  1  5 \\
\rowcolor{p} 56 & 1  4  6  2  2  5 \\
\rowcolor{p} 57 & 1  4  6  2  5  1 \\
\rowcolor{p} 58 & 1  4  6  2  5  2 \\
\rowcolor{p} 59 & 1  4  6  5  1  1 \\
\rowcolor{p} 60 & 1  4  6  5  1  2 \\
\rowcolor{r} 61 & 1  4  6  5  1  3 \\
\rowcolor{p} 62 & 1  4  6  5  1  4 \\
\rowcolor{p} 63 & 1  4  6  5  2  1 \\
\rowcolor{p} 64 & 1  4  6  5  2  2 \\
\rowcolor{p} \textcolor{darkgray}{65} & \textcolor{darkgray}{1  5  6  1  1  1} \\
\rowcolor{p} \textcolor{darkgray}{66} & \textcolor{darkgray}{1  5  6  1  1  2} \\
\rowcolor{g} \textcolor{darkgray}{67} & \textcolor{darkgray}{1  5  6  1  1  3} \\
\rowcolor{g} \textcolor{darkgray}{68} & \textcolor{darkgray}{1  5  6  1  1  4} \\
\rowcolor{p} \textcolor{darkgray}{69} & \textcolor{darkgray}{1  5  6  1  2  1} \\
\rowcolor{p} \textcolor{darkgray}{70} & \textcolor{darkgray}{1  5  6  1  2  2} \\
\rowcolor{p} \textcolor{darkgray}{71} & \textcolor{darkgray}{1  5  6  1  3  1} \\
\rowcolor{g} \textcolor{darkgray}{72} & \textcolor{darkgray}{1  5  6  1  3  2} \\
\rowcolor{p} \textcolor{darkgray}{73} & \textcolor{darkgray}{1  5  6  1  4  1} \\
\rowcolor{g} \textcolor{darkgray}{74} & \textcolor{darkgray}{1  5  6  1  4  2} \\
\rowcolor{p} \textcolor{darkgray}{75} & \textcolor{darkgray}{1  5  6  2  1  1} \\
\rowcolor{p} \textcolor{darkgray}{76} & \textcolor{darkgray}{1  5  6  2  1  2} \\
\rowcolor{p} \textcolor{darkgray}{77} & \textcolor{darkgray}{1  5  6  2  1  3} \\
\rowcolor{p} \textcolor{darkgray}{78} & \textcolor{darkgray}{1  5  6  2  1  4} \\
\rowcolor{p} \textcolor{darkgray}{79} & \textcolor{darkgray}{1  5  6  2  2  1} \\
\rowcolor{p} \textcolor{darkgray}{80} & \textcolor{darkgray}{1  5  6  2  2  2} \\
\hline \end{tabular} \hspace{0.25cm} \begin{tabular}{|l|l|} \hline Strand & \multicolumn{1}{|p{1.8cm}|}{\centering Substrand\\ sequence} \\ \hline
\rowcolor{p} 81 & 2  1  6  1  1  5 \\
\rowcolor{p} 82 & 2  1  6  1  2  5 \\
\rowcolor{p} 83 & 2  1  6  1  3  5 \\
\rowcolor{p} 84 & 2  1  6  1  4  5 \\
\rowcolor{p} 85 & 2  1  6  1  5  1 \\
\rowcolor{p} 86 & 2  1  6  1  5  2 \\
\rowcolor{p} 87 & 2  1  6  2  1  5 \\
\rowcolor{b} 88 & 2  1  6  2  2  5 \\
\rowcolor{b} 89 & 2  1  6  2  5  1 \\
\rowcolor{b} 90 & 2  1  6  2  5  2 \\
\rowcolor{p} 91 & 2  1  6  5  1  1 \\
\rowcolor{p} 92 & 2  1  6  5  1  2 \\
\rowcolor{g} 93 & 2  1  6  5  1  3 \\
\rowcolor{g} 94 & 2  1  6  5  1  4 \\
\rowcolor{b} 95 & 2  1  6  5  2  1 \\
\rowcolor{b} 96 & 2  1  6  5  2  2 \\
\rowcolor{p} 97 & 2  2  6  1  1  5 \\
\rowcolor{p} 98 & 2  2  6  1  2  5 \\
\rowcolor{p} 99 & 2  2  6  1  3  5 \\
\rowcolor{p} 100 & 2  2  6  1  4  5 \\
\rowcolor{p} 101 & 2  2  6  1  5  1 \\
\rowcolor{p} 102 & 2  2  6  1  5  2 \\
\rowcolor{p} 103 & 2  2  6  2  1  5 \\
\rowcolor{p} 104 & 2  2  6  2  2  5 \\
\rowcolor{p} 105 & 2  2  6  2  5  1 \\
\rowcolor{p} 106 & 2  2  6  2  5  2 \\
\rowcolor{p} 107 & 2  2  6  5  1  1 \\
\rowcolor{p} 108 & 2  2  6  5  1  2 \\
\rowcolor{p} 109 & 2  2  6  5  1  3 \\
\rowcolor{p} 110 & 2  2  6  5  1  4 \\
\rowcolor{p} 111 & 2  2  6  5  2  1 \\
\rowcolor{p} 112 & 2  2  6  5  2  2 \\
\rowcolor{p} \textcolor{darkgray}{113} & \textcolor{darkgray}{2  5  6  1  1  1} \\
\rowcolor{p} \textcolor{darkgray}{114} & \textcolor{darkgray}{2  5  6  1  1  2} \\
\rowcolor{p} \textcolor{darkgray}{115} & \textcolor{darkgray}{2  5  6  1  1  3} \\
\rowcolor{p} \textcolor{darkgray}{116} & \textcolor{darkgray}{2  5  6  1  1  4} \\
\rowcolor{p} \textcolor{darkgray}{117} & \textcolor{darkgray}{2  5  6  1  2  1} \\
\rowcolor{p} \textcolor{darkgray}{118} & \textcolor{darkgray}{2  5  6  1  2  2} \\
\rowcolor{p} \textcolor{darkgray}{119} & \textcolor{darkgray}{2  5  6  1  3  1} \\
\rowcolor{p} \textcolor{darkgray}{120} & \textcolor{darkgray}{2  5  6  1  3  2} \\
\hline \end{tabular} \hspace{0.25cm} \begin{tabular}{|l|l|} \hline Strand & \multicolumn{1}{|p{1.8cm}|}{\centering Substrand\\ sequence} \\ \hline
\rowcolor{p} \textcolor{darkgray}{121} & \textcolor{darkgray}{2  5  6  1  4  1} \\
\rowcolor{p} \textcolor{darkgray}{122} & \textcolor{darkgray}{2  5  6  1  4  2} \\
\rowcolor{p} \textcolor{darkgray}{123} & \textcolor{darkgray}{2  5  6  2  1  1} \\
\rowcolor{p} \textcolor{darkgray}{124} & \textcolor{darkgray}{2  5  6  2  1  2} \\
\rowcolor{p} \textcolor{darkgray}{125} & \textcolor{darkgray}{2  5  6  2  1  3} \\
\rowcolor{p} \textcolor{darkgray}{126} & \textcolor{darkgray}{2  5  6  2  1  4} \\
\rowcolor{p} \textcolor{darkgray}{127} & \textcolor{darkgray}{2  5  6  2  2  1} \\
\rowcolor{p} \textcolor{darkgray}{128} & \textcolor{darkgray}{2  5  6  2  2  2} \\
\rowcolor{p} \textcolor{darkgray}{129} & \textcolor{darkgray}{5  1  6  1  1  1} \\
\rowcolor{p} \textcolor{darkgray}{130} & \textcolor{darkgray}{5  1  6  1  1  2} \\
\rowcolor{p} \textcolor{darkgray}{131} & \textcolor{darkgray}{5  1  6  1  1  3} \\
\rowcolor{p} \textcolor{darkgray}{132} & \textcolor{darkgray}{5  1  6  1  1  4} \\
\rowcolor{p} \textcolor{darkgray}{133} & \textcolor{darkgray}{5  1  6  1  2  1} \\
\rowcolor{p} \textcolor{darkgray}{134} & \textcolor{darkgray}{5  1  6  1  2  2} \\
\rowcolor{p} \textcolor{darkgray}{135} & \textcolor{darkgray}{5  1  6  1  3  1} \\
\rowcolor{p} \textcolor{darkgray}{136} & \textcolor{darkgray}{5  1  6  1  3  2} \\
\rowcolor{p} \textcolor{darkgray}{137} & \textcolor{darkgray}{5  1  6  1  4  1} \\
\rowcolor{p} \textcolor{darkgray}{138} & \textcolor{darkgray}{5  1  6  1  4  2} \\
\rowcolor{p} \textcolor{darkgray}{139} & \textcolor{darkgray}{5  1  6  2  1  1} \\
\rowcolor{p} \textcolor{darkgray}{140} & \textcolor{darkgray}{5  1  6  2  1  2} \\
\rowcolor{p} \textcolor{darkgray}{141} & \textcolor{darkgray}{5  1  6  2  1  3} \\
\rowcolor{p} \textcolor{darkgray}{142} & \textcolor{darkgray}{5  1  6  2  1  4} \\
\rowcolor{p} \textcolor{darkgray}{143} & \textcolor{darkgray}{5  1  6  2  2  1} \\
\rowcolor{p} \textcolor{darkgray}{144} & \textcolor{darkgray}{5  1  6  2  2  2} \\
\rowcolor{p} \textcolor{darkgray}{145} & \textcolor{darkgray}{5  2  6  1  1  1} \\
\rowcolor{p} \textcolor{darkgray}{146} & \textcolor{darkgray}{5  2  6  1  1  2} \\
\rowcolor{p} \textcolor{darkgray}{147} & \textcolor{darkgray}{5  2  6  1  1  3} \\
\rowcolor{p} \textcolor{darkgray}{148} & \textcolor{darkgray}{5  2  6  1  1  4} \\
\rowcolor{p} \textcolor{darkgray}{149} & \textcolor{darkgray}{5  2  6  1  2  1} \\
\rowcolor{p} \textcolor{darkgray}{150} & \textcolor{darkgray}{5  2  6  1  2  2} \\
\rowcolor{p} \textcolor{darkgray}{151} & \textcolor{darkgray}{5  2  6  1  3  1} \\
\rowcolor{p} \textcolor{darkgray}{152} & \textcolor{darkgray}{5  2  6  1  3  2} \\
\rowcolor{p} \textcolor{darkgray}{153} & \textcolor{darkgray}{5  2  6  1  4  1} \\
\rowcolor{p} \textcolor{darkgray}{154} & \textcolor{darkgray}{5  2  6  1  4  2} \\
\rowcolor{p} \textcolor{darkgray}{155} & \textcolor{darkgray}{5  2  6  2  1  1} \\
\rowcolor{p} \textcolor{darkgray}{156} & \textcolor{darkgray}{5  2  6  2  1  2} \\
\rowcolor{p} \textcolor{darkgray}{157} & \textcolor{darkgray}{5  2  6  2  1  3} \\
\rowcolor{p} \textcolor{darkgray}{158} & \textcolor{darkgray}{5  2  6  2  1  4} \\
\rowcolor{p} \textcolor{darkgray}{159} & \textcolor{darkgray}{5  2  6  2  2  1} \\
\rowcolor{p} \textcolor{darkgray}{160} & \textcolor{darkgray}{5  2  6  2  2  2} \\
\hline
    \end{tabular}
    \caption[Substrand sequences of simulated strands]{All 160 strands used for simulations and their substrand sequences. The nucleotide sequences of different substrands are shown in \autoref{eq:sub1}-\ref{eq:sub6}. Pale red cells contain strands that satisfy the condition for sufficiently fast CT and pale blue cells contain strands that satisfies the condition for sufficiently fast replication (see Section \ref{sec:FI}). Purple cells contain strands that satisfy both of the aforementioned conditions, just as in \autoref{fig:E(x_meta+rep)}.  Black text signifies a strand sequence containing an 8-oxo-G after the ligation site (rings in \autoref{fig:E(x_meta+rep)}) while grey text means the 8-oxo-G is before the central bulge (triangles in \autoref{fig:E(x_meta+rep)}). Red cells contain strands for which either the LIDA or CT simulation did not finish within the allotted time. Note that the number of viable strands, indicated by purple cells and black text, is 63.}
\label{tab:strands}
\end{table}

\subsection{Functional characterization of co-factors}

\subsubsection{Direct rate combination} 

To further explore the viability transition discussed above, we can in principle rank the combinatorial co-factors $F(k_{rep}, k_{CT} , ...)$ such that their combined replication and charge transfer (plus subsequent fatty acid production) rates are increasing along an axis. Again note that a reliable combined co-factor ’fitness’ ranking cannot yet be estimated without a complete protocell-environment simulation. 
However, we can still study co-factor rankings based on the results of the CT and LIDA simulations as they are of theoretical interest.

\begin{figure}[H]
\centering
\includegraphics[width=1\textwidth]{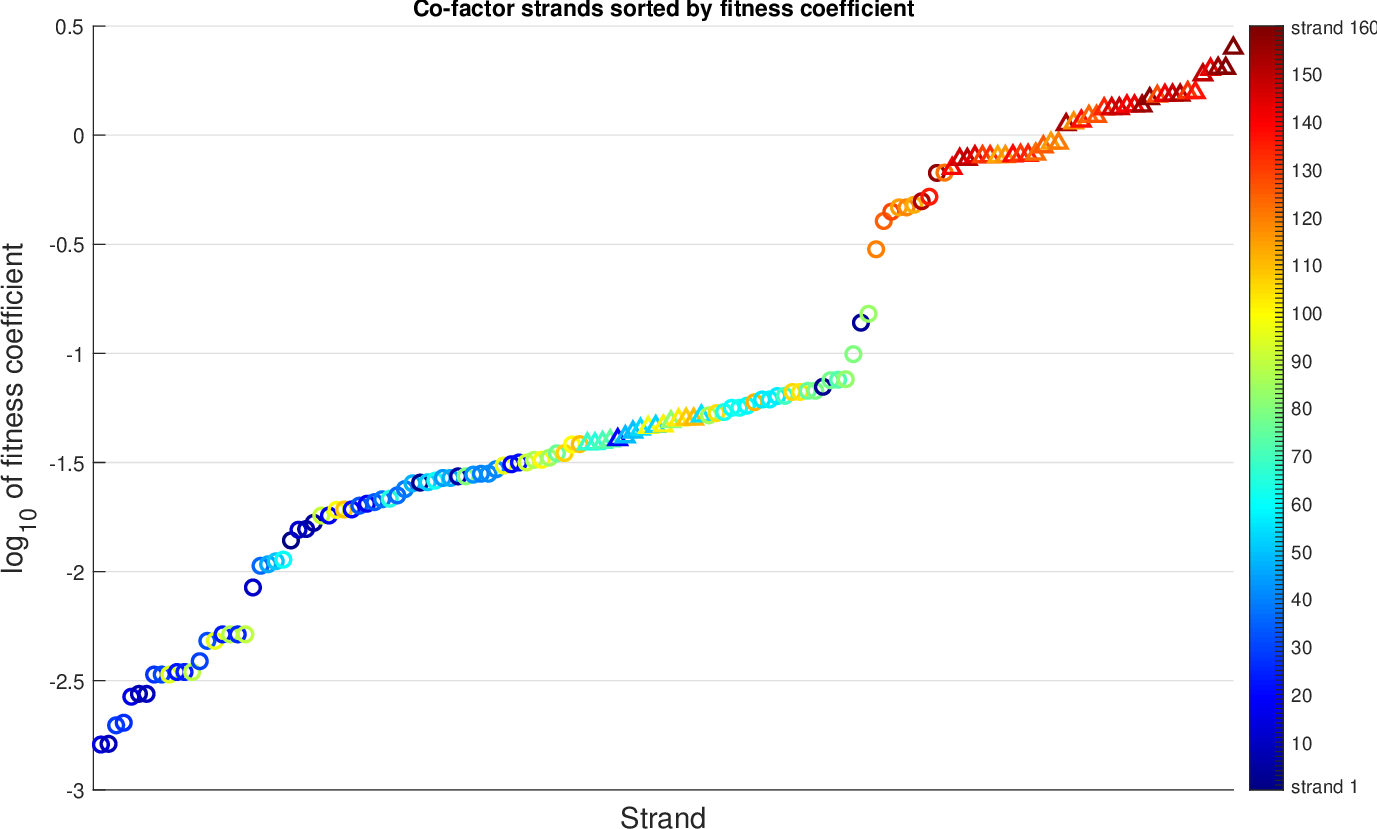}
\caption[Fitness coefficients of sorted co-factors]{Ansatz for a combined metabolic and replication growth rate from \cite{rocheleau2007emergence} depicted as the sorted sum of ${2\over3}\log_{10}[k_{rep}]$ +  ${1\over3}\log_{10}[k_{CT}]$. Note the three growth rate regimes (blue to blue/green to red) dominated by the clustered charge transport rates. Also note that no metabolism-replication interaction are included in our simulations, so the graph is a simple (superposition) approximation of the protocellular fitness based on co-factor for the 160 examined strands. The color bar shows which strand corresponds to which data point; rings indicate that oxoG is above the ligation site, while and triangles indicate the oxoG is located below the ligation site. See text for details.}
\label{fig:fitcoeff}
\end{figure}

A simplistic method to approximate the combined fitness impact of the charge transfer (CT) and the replication rates is to simply multiply the two rates. This can be done by combining the replication rate data and the CT rate data as the sum of the $\log_{10}$ of replication rate and the CT rate for each strand. 
There are, however, problems with a direct multiplication. 
Rocheleau et al. (2007) \cite{rocheleau2007emergence} studied a similar protocellular system that combined information replication and container growth. They showed that due to non-trivial couplings between interacting processes, the combined protocellular growth factor is proportional to $[k_{rep}^2 k_{CT}]^{1/3}$ rather than $(k_{rep} k_{CT})$.
This is shown when the coupled equations (4.1)-(4.6) in \cite{rocheleau2007emergence} are solved analytically. 
Although the coupling between our charge transfer rate and our co-factor replication rate is different from the coupled aggregated template and container replication rates discussed in \cite{rocheleau2007emergence}, the two systems are closely related and we believe that the results from \cite{rocheleau2007emergence} provide a reasonable ansatz for a combined fitness estimate given our current knowledge of these systems. 
This fitness estimate is shown in \autoref{fig:fitcoeff}, which can be interpreted as a simple approximation for evaluating the fitness of a protocellular co-factor. 
Note, however, that the growth law found in  \cite{rocheleau2007emergence} is derived under restricted assumptions and defining a more realistic fitness coefficient would require an expanded and more detailed model that is outside the scope of this work. 

The sorted fitness coefficient data shows three groups similar to the clustered CT rate distribution, with added smoothness arising from the more continuous replication rate distribution. The three groups are mainly explained by the same factors as the groups from the CT rate data, recall Section \ref{sec:sim_explo} and \autoref{fig:CThist}, as the CT rates vary across more orders of magnitude than the replication rates.


\subsubsection{Functional information}\label{sec:FI}

An alternative way to explore the viability of the presented protocellular system is to estimate the functional information of the involved co-factors. 

The concept of functional information was first introduced by Szostak (2003) to define and quantify the information content of biopolymer sequences \cite{szostak2003functional}. Hazen et al. (2007) later used functional information as a more general measure of complexity \cite{hazen2007functional}. Hazen et al. (2007) argue that common for all complex systems is the ability to alter their environment in one or more ways, which they refer to as functions. For a system with a specified function ($x$) and a set of configurations $M$, each configuration $m\in M$ of the system has a ``degree of function'' $E_x(m)$ defined by its ability to perform the function $x$. For a specific value of $E_x$, only a subset of the total configurations has the same-or-higher degree of function, $E_x(m)\geq E_x$. In mathematical terms, the size of this set is given by $\omega=|\{m \in M : E_x(m) \ge E_x \}|$. With the size of the total set of configurations $\Omega=|M|$, we can define the probability of a random configuration having degree of function $E_x(m)\geq E_x$ as
\begin{equation}
    P_x(E_x) = \frac{\omega}{\Omega}.
\end{equation}
Functional information is then defined as
\begin{equation}\label{eq:FE}
    I[E_x] = -\log_2 P_x(E_x) = -\log_2\left(\frac{\omega}{\Omega}\right).
\end{equation}

As Hazen et al. (2007) concludes, a rigorous analysis of the functional information of a system with respect to a specified function $x$ requires knowledge of two attributes: ({i}) all possible configurations $M$ of the system and ({ii}) the degree of function $x$ for every configuration \cite{hazen2007functional}. The difficulty of obtaining knowledge of these two attributes clearly scales with the complexity of the system.

The concept of a function obviously has a connection to the concept of evolutionary fitness, as functions contribute to the fitness of an evolving entity. This is true for systems at multiple levels of complexity; e.g., the degree of function of a metabolic pathway of a resource molecule utilized by a microorganism, or the degree of function of the beak shape of a finch allowing consumption of a specific food source. In protocellular populations, fitness can be decomposed into two main functions: ({i}) metabolic function, e.g., the metabolic rate of lipid precursors into usable lipids ($E_{meta}$), which is regulated by the rate of charge transport through co-factors and ({ii}) replicative function, e.g., the replication rate of co-factors through LIDA ($E_{rep}$). The distribution of degrees of these functions in protocell populations therefore gives rise to a fitness distribution, and thus selection dynamics. As mentioned, a full analysis of these functions of the protocell information system would require complete knowledge of the possible configurations $m\in M_{proto}$ and their corresponding degrees of function $E_{meta}(m)$ and $E_{rep}(m)$.

$M_{proto}$ is hard to define. An upper bound could be every possible configuration of the combinatorial co-factor, i.e., every possible DNA duplex of length 18 bp which can be formed using A, T, G, C and 8-oxo-G while also allowing mismatches, bulges and variations of oligomer lengths. The size $\Omega_{proto}$ of the set $M_{proto}$ is then clearly a very large number; too big to ever permit a brute-force analysis of the functional information. 

Our investigation of protocellular functional information is limited to the set of sequences used for which CT and LIDA simulations have been performed.
As discussed above, based on the results from Sections \ref{sec:sim_explo} and \ref{sec:4}, we calculated that minimum values of $k_{meta} > 50$/s and $k_{rep} > 10^{-4}$/s are needed to support a viable protocells, so we can now estimate the probabilities of randomly choosing a strand from the 160 tested strands that satisfies each of these conditions.

\begin{equation}
    P(k_{CT} > 50 /\textrm{s}) = \frac{[\textrm{\# strands that satisfy $k_{CT} > 50$/s}]}{[\textrm{total \# strands}]}
\end{equation}
\begin{equation}
    P(k_{rep} > 10^{-4}\textrm{/s}) = \frac{[\textrm{\# strands that satisfy $k_{rep} > 10^{-4}$/s}]}{ [\textrm{total \# strands}]}
\end{equation}
\begin{equation}
    P(k_{CT} > 50\textrm{/s \& } k_{rep} > 10^{-4}\textrm{/s}) = \frac{[\textrm{\# strands that satisfy $k_{CT} > 50$/s \& $k_{rep} > 10^{-4}$/s}]}{[\textrm{total \# strands}]}.
\end{equation}

Needed compositional functional information in strands for viable protocells:

\begin{equation}
    I(\textrm{CT is good}) = -\ln [ P(k_{CT} > 50\textrm{/s}) ]
\end{equation}
\begin{equation}
    I(\textrm{rep is good}) = -\ln [ P(k_{rep} > 10^{-4}\textrm{/s}) ]
\end{equation}
\begin{equation}
    I(\textrm{CT \& rep is good}) = -\ln [ P(k_{CT} > 50 /\textrm{s \& } k_{rep} > 10^{-4}\textrm{/s}) ].
\end{equation}

\autoref{tab:strands} indicates the total number of protocells with co-factor functions greater than $E_{meta}(m)\ge 50\text{/s}$ and $E_{rep}(m)\ge 10^{-4}\text{/s}$ yields $\omega_{meta+rep} = 63$. We can now compute the total functional information needed for a viable protocell if a co-factor strand is randomly picked from from the ensemble of 160 strands, 
\begin{align}
    I[E_{meta+rep}] &= -\log_2 (F(E_{meta+rep})) \\
    &= -\log_2 \left(\frac{\omega_{meta+rep}}{\Omega_{proto}}\right) = -\log_2 \left(\frac{63}{160}\right) \approx 1.34 \,\textrm{bits}.
\end{align}
Thus, 1.34 bits is the estimated functional information required for a protocell to be viable when its co-factor is randomly drawn from the limited available set of 160 co-factors. 









\begin{figure}[H]
\centering
\includegraphics[width=0.48\textwidth]{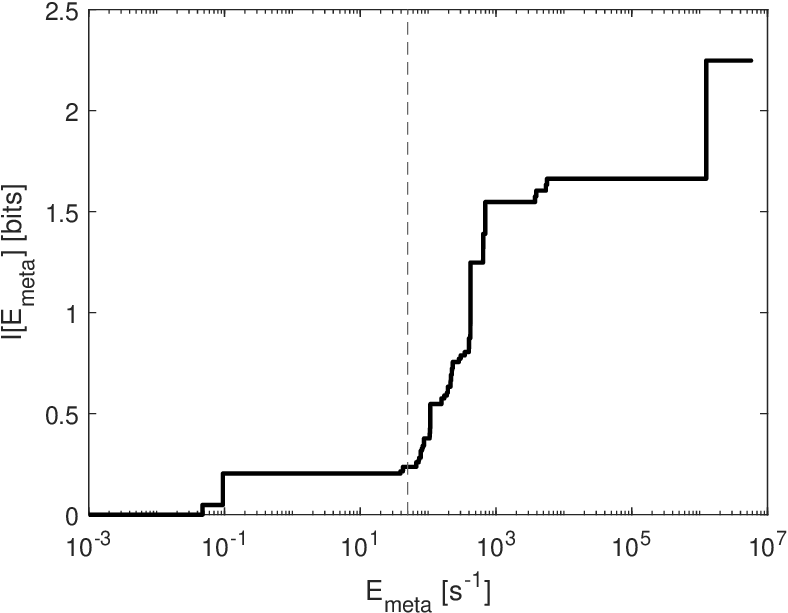}
\includegraphics[width=0.48\textwidth]{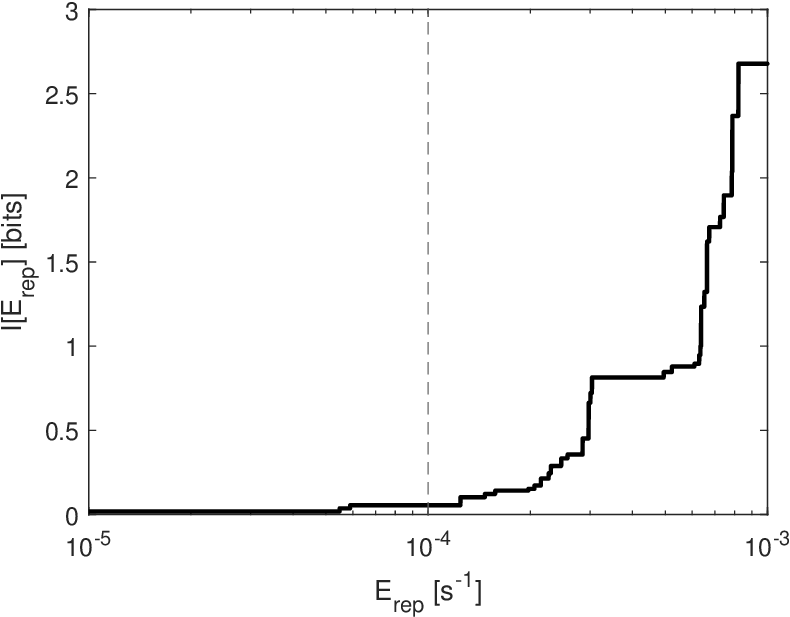}
\caption[Metabolic functional information]{(Left) The functional information $I[E_{meta}]$ of co-factors as a function of $E_{meta}$ and (Right) $I[E_{rep}]$ as a function of $E_{rep}$. Both show an increase in functional information with $E_x$ as fewer co-factors have degrees of function $\geq E_x$. The vertical dashed lines indicate $E_{meta}=50$/s and $E_{rep}=10^{-4}$/s respectively. Note that the maximum functional information differs for the two functions, as the number of identical fastest replicators and charge-transfer strands differs. For instance, for the charge transfer function 32 strands have an identical charge transfer rate of about $4 \times 10^6$/s, yielding about 2.32 bits.}
\label{fig:protoFI}
\end{figure}

We can also calculate the corresponding functional information of the co-factors as they are ranked according to their metabolic and replication rate constants, which is shown in \autoref{fig:protoFI}.

In \autoref{fig:protoFI}, functional information of both  charge transport and  replication reflect, in different ways, the same qualitative shape of the estimated protocellular growth rates shown in \autoref{fig:fitcoeff}. 
From \autoref{tab:strands}, we see that the CT threshold $k_{CT} > 50$/s (and 8-oxo-G above the ligation site) is satisfied for 67 strands, thus $\omega_{meta} \simeq 67$, while the replication threshold $k_{rep} > 10^{-4}$/s (and 8-oxo-G above ligation site) is satisfied for 86 strands, thus $\omega_{rep} \simeq 86$ . This corresponds to functional information thresholds of 1.26 bits and 0.89 bits for charge transport and replication functions, respectively. Note as a reference that the theoretical maximal possible functional information for this co-factor system is $-\log_2(1/160)\, \text{bits} \simeq 7.30 \,\text{bits}$.

%% file: 6-Discussion.tex
\section{Discussion}\label{sec:disc}



The presented work consists of four main parts:  (i) conceptual discussion of the protocellular functionalities and design; (ii) simulations of DNA charge transport between Ru-C and oxoG; (iii) simulations of lesion-induced DNA amplification; and (iv) a combination of the results from (ii) and (iii). In this section, we critically discuss the applied assumptions and approximations. 


{\bf About (i)}:
In Section 2 we present and discuss a conceptual picture of our protocellular design and functionalities, as a background for our later analysis of the two critical co-factor properties: replication and charge transfer. 
Our protocell system is also discussed in \cite{rasmussen2003bridging,rocheleau2007emergence,declue2009nucleobase,cape2012phototriggered}
\cite{rasmussen2016generating}, while alternative protocellular designs are presented and discussed in \cite{adamala2016collaboration,rasmussen2008protocells,rasmussen2008roadmap,schwille2018maxsynbio,pols2019synthetic,sato2022expanding}. 

{\bf About (ii)}: See Section 3. It is known that certain Ru-C merges into the DNA base stack (intercalation) and is able to oxidize guanines after photoexcitation \cite{mihailovic2006exploring,genereux2010mechanisms}. To induce this intercalation the Ru-C and DNA duplex are usually tethered. Without direct tethering a similar intercalation should be achieved by attaching both the Ru-C and two of the DNA duplex to hydrophilic anchors via carbon tethers. The goal is to only have charge transfer between the RuC and the oxoG through a duplex, such that the DNA replication process that creates more duplexes positively influence the fatty acid production process, thus establishing a catalytic coupling between the two. 


Our choice of kinetic modeling for the CT simulations is based on its low computational cost. More refined methods, e.g., full quantum-mechanical simulations, would have drastically increased the computational cost of the simulations and thereby also drastically limited the number of DNA co-factors that could be simulated in the allotted time. Thus, it would not have been possible to compare the CT capabilities of multiple co-factor sequences. 

To use ODEs means that basepair-to-basepair CT rates have to be known for each base pair CT interaction.  Thus, we consulted the literature where such rates are reported, either from direct measurements or from ab initio calculations.
As it turns out not all base pair CT combinations rates are listed in the available literature, we could only simulate a limited set of duplex strings. Also, several CT rates had to be estimated by the authors as no previous research was found (e.g., on CT rates through bulges).

It is not known to the authors how feasible super exchange (SE) transitions are through disruptions of the DNA $\pi$ stack such as mismatches, bulges, and possibly 8-oxo-G.  
Barton and coworkers found that conformational gating to be highly regulating in DNA hole transport (HT), with disruptions of the DNA $\pi$ stack generally yielding slower HT rates \cite{genereux2010mechanisms,bhattacharya2001influence}.
For simplicity, we assume that SE transitions across mismatches, bulges, and possibly 8-oxo-G, are feasible and comparable to SE transitions through a well-matched A/T bridge. Note that the validity of this assumption is significant for model accuracy. If SE transitions across bulges are not feasible or very slow compared to thermally induced hopping (TIH) rates through bulges, the resulting CT rate across the full duplex strand becomes increasingly dependent on the (currently rough) estimate of the TIH rate through bulges, which are expected to yield significantly slower full duplex CT rates when it involves bulges or mismatches. If SE across a bulge turns out not to be possible, one could probably utilize co-factors with the 8-oxo-G located just below the ligation site and still support co-factor replication. See Section 3 for details.   

In our simulations, for simplicity we assume charges are contained in DNA, i.e., there is no charge transport between DNA strands and the solvent. Note that solvent interactions have been argued to be a significant factor in CT through mismatched DNA duplexes \cite{giese2000influence}. This is expected to decrease the CT rates in DNA duplex due to the partial loss of charges to the solvent. 

Finally, we have conducted our CT simulations by assuming the HT starts from the first base pair up to the oxoG location. More correctly, our CT simulation should have started between the intercalated Ru-C and the nearest guanines within the DNA duplex. This simplification probably underestimates CT rates.

{\bf About (iii)}: See Section 4. Reaction kinetic seems to be the natural choice for investigating the LIDA based replication dynamics for multiple (>100) strands. We have made several simplification for  these kinetic LIDA simulations. 

In our simulations we assume the involved DNA template strands and precurser oligomers interact in bulk solution. They are in fact vesicle-anchored DNA duplex strands, where the supplied oligomer resources consist of two oligomers with and two oligomers without membrane anchors. More realistic kinetic rates could therefore be obtained if the relevant rates were adjusted to fit the DNA anchoring. Preliminary simulation investigations of such a system indicate that the relative replication rates differences between different sequences are conserved, although the overall replication rate is expected to be faster mainly due to a faster collision rates in 2D compared to 3D \cite{marco_MS_2023}.

Further, our LIDA kinetic equations do not include pseudo-blunt-end ligation, which is a process by which non-template-directed ligation is known to occur in the LIDA system \cite{alladin2015achieving}. The inclusion of pseudo-blunt-end ligation becomes important for predicting simulation dynamics of LIDA when the initial template concentrations are extremely low, as amplification can be observed in the physical system even when the initial template concentration is zero.  However, the kinetic rate of blunt-end ligation is estimated to be orders of magnitude smaller than the rates of template-directed ligation. Thus, pseudo-blunt-end ligation can therefore be assumed to have only minor influence on the dynamics of LIDA when there is an initial template concentration of a few nM.

Finally, we assume a constant and sequence-independent on-rate, which simplifies the calculations, although we are aware that this is always true.  For example, sequences with many base pairs repeats are known to initially mismatch followed by slower sliding and re-hybridization processes that eventually leads to correct hybridization state. These sliding and re-hybridization processes slow down the resulting hybridization process. However, assuming constant on-rates, we only need to estimate the off-rates based on the fundamental relation between the free energy of DNA hybridization and the equilibrium constant, which is equivalent to the fraction between the on-rate and the off-rate. As the off-rates are orders of magnitude slower than the on-rates, an assumption about constant on-rates will presumably not significantly influence the results. 

{\bf About (iv)}:
See Section 5. As already previously emphasized,
no interaction dynamics between the charge transport and the replication dynamics nor with the environment are taken into consideration. Thus, a combination of the simulation results found under (ii) and (iii) is a simplified (superposition), but still interesting, representation of what a more complex simulation would provide.

Given the above assumptions, we observe that viable protocells require minimum thresholds for the reaction rate constants for both co-factor replication and charge transport. Thus, protocells cannot exist with production rates below a certain threshold. These threshold rates are estimated by requiring that the overall production rate, an aggregation of the replication and the charge transfer rates, are higher than the involved degradation rates, see Sections 5 and 5.1.1. 

To assess how likely it is to pick a viable co-factor at random from the available ensemble of the 160 investigated strands, we use functional information as defined by \cite{hazen2007functional}.
Estimation of the functional information is generally difficult,but in our situation we can make functional information estimates based on the rate estimates in Section 3 and 4. For details see Section 5.1.2.

%% file: 7-Conclusion.tex
\section{Conclusion}


The main result of our investigation emerge as we compare the estimated $k_{CT}$ and $k_{rep}$ values with the associated observed degradation rates, which indicate that protocellular viability cannot be rejected for 63 of the 160 tested co-factors, given the assumptions that underpin our estimations. 

The work is based on an operational definition of minimum life that can be supported by four interconnected functionalities: a metabolic energy transducer, an informational co-factor, and a container, which must all be situated in an appropriate environment, see \cite{rasmussen2003bridging,rasmussen2008protocells,rasmussen2016generating}. 
We investigate the molecular requirements of a co-factor-modulated energy transducer, assuming the co-factor is a short DNA duplex containing an 8-oxo-guanine that is able to replicate and where the energy transducer is a ruthenium complex. 
We present simulation results for charge transfer (CT) and replication abilities for 160 DNA co-factors each composed of 18 base pairs \cite{rasmussen2003bridging,rasmussen2008protocells,declue2009nucleobase,maurer2011interactions,cape2012phototriggered,rasmussen2016generating}. 

CT simulations are performed for DNA duplexes using charge transfer kinetics, based on an ODE model where the transition rate constants $k_{CT}$ are determined from both theoretical calculations and empirical studies in the literature. A few CT rates had to be guesstimated by the authors (e.g., charge transfer across a bulge). 
Typical CT rate constants $k_{CT}$ range from $10^{-1}$/s to $10^{6}$/s for the 160 investigated DNA duplex strands, where the resulting $k_{CT}$ is estimated from the first base of the 18 bp long strand to an 8-oxo-G located somewhere along the duplex strand. 

The observed metabolic fatty acid production rate in a system with a direct ruthenium to 8-oxo-G charge transfer (without an intermediate DNA charge transfer) is measured to be about $1.3 \times 10^{6}$/s and a corresponding fatty acid (and vesicle) doubling time of about 4.3 hours\cite{declue2009nucleobase}
\cite{maurer2011interactions} \cite{bornebusch2021reaction}
\cite{bornebusch2021thesis}. 
If we require the DNA CT to slow down this rate no more that $1\%$, we require that $k_{CT} > 50$/s.

  
Replication dynamics of the combinatorial DNA co-factor is estimated by lesion-induced DNA amplification (LIDA) simulations using reaction kinetics ODEs, where reaction constants are determined directly from thermodynamic calculations and observed non-enzymatic ligation rate constant \cite{cape2012phototriggered}. The obtained overall replication constants $k_{rep}$ typically range from just below $10^{-4}$/s to above $10^{-3}$/s, which means co-factor doubling times from below 20 minutes to many hours. 
Oligomer degradation rate constants are observed to be up to about $10^{-5}$/s (imidazole hydrolysis), so we require that $k_{rep} > 10^{-4}$/s to ensure a viable co-factor replication. 

The protocellular viability is further quantified using the concept of functional information, where the needed minimal information in a co-factor is estimated.
Assuming a co-factor duplex is randomly selected from the sample of 160 investigated duplexes 1.26 bits are required to satisfy the charge transport rate constant requirement, 0.89 bits are required to satisfy the replication rate requirement, while both requirements need 1.34 bits of information. 

